\documentclass[twocolumn]{aastex61}
\hypersetup{citecolor=cyan,linkcolor=blue}

\usepackage{gensymb}

\shorttitle{[C{\sc ii}], UV and dust in LAEs at $z\approx6-7$}
\shortauthors{Matthee et al.}

\begin{document}

\title{Resolved UV and [CII] structures of luminous galaxies within the epoch of reionisation}

\correspondingauthor{J. Matthee}
\email{mattheej@phys.ethz.ch}

\author[0000-0003-2871-127X]{J. Matthee} \thanks{Zwicky Fellow} 
\affil{Department of Physics, ETH Z\"urich, Wolfgang-Pauli-Strasse 27, 8093 Z\"urich, Switzerland}
\affil{Leiden Observatory, Leiden University, PO\ Box 9513, NL-2300 RA, Leiden, The Netherlands}

\author{D. Sobral}
\affiliation{Department of Physics, Lancaster University, Lancaster, LA1 4YB, UK}

\author{L. A. Boogaard}
\affiliation{Leiden Observatory, Leiden University, PO\ Box 9513, NL-2300 RA, Leiden, The Netherlands}

\author{H. R\"ottgering}
\affiliation{Leiden Observatory, Leiden University, PO\ Box 9513, NL-2300 RA, Leiden, The Netherlands}

\author{L. Vallini}
\affiliation{Leiden Observatory, Leiden University, PO\ Box 9513, NL-2300 RA, Leiden, The Netherlands}

\author{A. Ferrara}
\affiliation{Scuola Normale Superiore, Piazza dei Cavalieri 7, I-56126 Pisa, Italy}
\affiliation{Kavli Institute for the Physics and Mathematics of the Universe (WPI), University of Tokyo, Kashiwa 277-8583, Japan}

\author{A. Paulino-Afonso}
\affiliation{Max-Planck-Institut f\"ur Extraterrestrische Physik, Postfach 1312, Giessenbachstr., 85748 Garching, Germany}

\author{F. Boone}
\affiliation{Universit\'e de Toulouse; UPS-OMP; IRAP; Toulouse, France}

\author{D. Schaerer}
\affiliation{Observatoire de Gen\`eve, Universit\' e Gen\`eve, 51 Ch. des Maillettes, 1290 Versoix, Switzerland}
\affiliation{Universit\'e de Toulouse; UPS-OMP; IRAP; Toulouse, France}

\author{B. Mobasher}
\affiliation{Department of Physics and Astronomy, University of California, Riverside, CA 92521, USA}



\begin{abstract}
\noindent We present new deep ALMA and {\it HST}/WFC3 observations of MASOSA and VR7, two luminous Ly$\alpha$ emitters (LAEs) at $z=6.5$, for which the UV continuum level differ by a factor four. No IR dust continuum emission is detected in either, indicating little amounts of obscured star formation and/or high dust temperatures. MASOSA, with a UV luminosity M$_{1500}=-20.9$, compact size and very high Ly$\alpha$ EW$_{0}\approx145$ {\AA}, is undetected in [C{\sc ii}] to a limit of L$_{\rm [CII]}<2.2\times10^7$ L$_{\odot}$ implying a metallicity $Z\lesssim0.07 Z_{\odot}$. Intriguingly, our {\it HST} data indicates a red UV slope $\beta=-1.1\pm0.7$, at odds with the low dust content. VR7, which is a bright (M$_{1500}=-22.4$) galaxy with moderate color ($\beta=-1.4\pm0.3$) and Ly$\alpha$ EW$_0 = 34${\AA}, is clearly detected in [C{\sc ii}] emission (S/N=15). VR7's rest-frame UV morphology can be described by two components separated by $\approx1.5$ kpc and is globally more compact than the [C{\sc ii}] emission. The global [C{\sc ii}]-UV ratio indicates $Z\approx0.2 Z_{\odot}$, but there are large variations in the UV-[C{\sc ii}] ratio on kpc scales. We also identify diffuse, possibly outflowing, [C{\sc ii}]-emitting gas at $\approx 100$ km s$^{-1}$ with respect to the peak. VR7 appears assembling its components at a slightly more evolved stage than other luminous LAEs, with outflows already shaping its direct environment at $z\sim7$. Our results further indicate that the global [C{\sc ii}]-UV relation steepens at SFR $<30$ M$_{\odot}$ yr$^{-1}$, naturally explaining why the [C{\sc ii}]-UV ratio is anti-correlated with Ly$\alpha$ EW in many, but not all, observed LAEs. 
\end{abstract}


\section{Introduction} \label{sec:intro}
The advent of the Atacama Large Millimetre Array (ALMA) has enabled the first detailed studies of the interstellar medium (ISM) in various kinds of star-forming galaxies and quasar hosts within the epoch of reionization at $z>6$ \citep[e.g.][]{Wang2013,Maiolino2015,Willott2015,Inoue2016,Decarli2017,Dodorico2018}. ALMA also opened the opportunity of resolving internal structure at kiloparsec scales \citep[e.g.][]{Jones2017,Matthee2017ALMA,Carniani2018,Hashimoto2018Dragons}. Currently, ALMA observations of galaxies at $z>6$ focus on measuring the dust far-infrared (FIR) continuum and FIR fine-structure lines [C{\sc ii}]$_{\rm 158 \mu m}$ and [O{\sc iii}]$_{\rm 88 \mu m}$. The [C{\sc ii}] line partly traces H{\sc ii} regions and neutral gas (i.e. photo-dissociating regions; PDRs; \citealt{Vallini2013}), while the [O{\sc iii}] line only traces H{\sc ii} regions. Combined with detailed photoionisation modelling, these lines can be used to constrain gas properties such as the metallicity and ionisation state \citep[e.g.][]{Olsen2017,Vallini2017}. 

While the dust continuum is typically undetected in galaxies selected through their UV or Ly$\alpha$ emission (e.g. \citealt{Capak2015,Bouwens2016,Pentericci2016,Matthee2017ALMA}, but see \citealt{Bowler2018,Hashimoto2018Dragons,Tamura2018} for counter examples), the fine-structure lines are now routinely detected. Due to the low metallicity and higher ionisation state of the ISM in high-redshift galaxies, the [O{\sc iii}] line may be the strongest far infrared emission line \citep[e.g.][]{Ferkinhoff2010,Inoue2016}, unlike at low-redshift, where [C{\sc ii}] is typically the strongest \citep[e.g.][]{Stacey1991,Wolfire2003,Herrera-Camus2015}. Indeed, recent studies have shown [O{\sc iii}] to be more luminous than [C{\sc ii}] in objects for which both lines are constrained \citep{Inoue2016,Carniani2017,Hashimoto2018Dragons,Walter2018}.

However, even though being potentially less luminous in high-redshift galaxies, the [C{\sc ii}] line is more easily observed for the vast majority of galaxies currently known at $6\lesssim z\lesssim7$. This is because [C{\sc ii}] redshifts into favourable frequencies in ALMA bands 6 and 7 at $z>5$, while this only happens at $z\gtrsim8$ for [O{\sc iii}] (with the exception of a few narrow redshift windows around $z\sim7$, e.g. \citealt{Carniani2017}, where the sensitivity to [C{\sc ii}] is still a factor $\approx2$ better). Therefore, practically, observations of the [C{\sc ii}] line are still the least expensive for measuring systemic galaxy redshifts (c.f. Ly$\alpha$) and studying ISM properties and kinematics.

Some early ALMA observations of galaxies at $z\approx5-6$ \citep[e.g.][]{Capak2015} indicated relatively luminous [C{\sc ii}] emission relative to the UV star formation rate (SFR), compared to the observed relation in the local Universe \citep{DeLooze2014}. Several other searches at $z\approx6-7$ either found a moderate [C{\sc ii}] deficit \citep{Pentericci2016,Bradac2017} or report stringent upper limits on the [C{\sc ii}] luminosity which would place these distant galaxies far below the local relation \citep[e.g.][]{Ouchi2013,Ota2014}. Such deficit could potentially be due to the prerequisite of a known Ly$\alpha$ redshift, which leads to a bias of observing young, metal-poor systems \citep[e.g.][]{Trainor2016}. New observations of additional galaxies \citep[e.g.][]{Matthee2017ALMA,Smit2017,Carniani2018,Hashimoto2018Dragons} and re-analysis of earlier ALMA data \citep{Carniani2018Himiko} indicate there is more scatter at high-redshift \citep{Carniani2018} instead of a strongly preferred low or high [C{\sc ii}]-UV ratio. The [C{\sc ii}]-UV ratio correlates with the relative strength of Ly$\alpha$ emission \citep{Carniani2018,Harikane2018}. These measurements are nonetheless complicated by the observation that [C{\sc ii}] is often offset spatially from the UV emission and multiple components of [C{\sc ii}] emission can be associated to UV components \citep[e.g.][]{Matthee2017ALMA}.

In this paper, we present new ALMA and {\it HST} observations of two very luminous LAEs at $z=6.5$, named VR7 and MASOSA \citep{Sobral2015,Matthee2017SPEC}, combined with a multi-wavelength analysis of archival data. While these galaxies have similar Ly$\alpha$ luminosity, the UV continuum is very different, leading to a factor $\approx5$ difference in the equivalent width (EW). What causes these differences in their Ly$\alpha$ EWs? Are the SFRs, dust content and/or ages of the galaxies different? Are there any differences in their metallicities? To answer these questions, we focus on characterising the properties of the ISM (metallicity, kinematics, dust content) and the stellar populations (star formation rate and age), and compare these to other galaxies observed at $z\approx6-7$. 

Our targets resemble the Ly$\alpha$ luminosity of the well-known LAEs Himiko and CR7 \citep{Ouchi2013,Sobral2015} which are clearly resolved into several components separated by $\approx4$ kpc in both the rest-frame UV and in [C{\sc ii}] emission \citep[e.g.][]{Matthee2017ALMA,Carniani2018,Sobral2019}. Do these luminous LAEs also consist of several components, which could indicate mergers are ubiquitous in the strongest Ly$\alpha$ emitters? We address this with high spatial resolution observations in both rest-frame UV and [C{\sc ii}] emission.

The structure of this paper is as follows. We present the new ALMA and {\it HST}/WFC3 observations and their data reduction in \S $\ref{sec:observations}$. \S $\ref{sec:properties}$ presents the rest-frame UV and Ly$\alpha$ properties of both galaxies and compares them to the observed galaxy population at similar lookback time. We then investigate the FIR continuum data from ALMA to constrain the galaxies' dust continuum emission in \S $\ref{sec:FIR}$. In \S $\ref{sec:CII}$ we present integrated measurements of the [C{\sc ii}] line, while we present the structure in VR7, resolved in both the spatial and spectral dimension in \S $\ref{sec:resolvedVR7}$. In \S $\ref{sec:discussion}$, we discuss the nature of MASOSA and which physical properties determine the [C{\sc ii}]/UV ratio in high-redshift galaxies. Finally, we summarise our results in \S $\ref{sec:conclusions}$. We use a $\Lambda$CDM cosmology with $\Omega_{\Lambda} = 0.70$, $\Omega_{\rm M} = 0.30$ and H$_0 = 70$ km s$^{-1}$ Mpc$^{-1}$ and a \cite{Salpeter1995} initial mass function. Throughout the paper, UV and IR luminosities are converted to SFR following \cite{Kennicutt1998}.

\section{Observations \& Data reduction} \label{sec:observations}
\subsection{ALMA} 
We performed ALMA observations in band 6 of our targets aiming at detecting [C{\sc ii}] line emission and FIR dust-continuum emission through ALMA program 2017.1.01451.S in Cycle 5. We observed in four spectral windows: two centred around redshifted [C{\sc ii}]$_{\rm 158 \mu m}$ emission at $z\approx6.5$ ($\approx252$ GHz) and two around 235 GHz, each with a bandwidth of 1875 MHz and 7.8 MHz ($9-10$ km s$^{-1}$) resolution. Both galaxies were observed with 43 antennas in configuration C43-4, allowing baselines from 15 to 783 m, leading to a natural resolution of $\approx0.4''$. Each galaxy was observed for three executions that consisted of 49 minutes integration time on target, leading to a total on target exposure of 147 minutes.

MASOSA \citep{Matthee2015,Sobral2015} was observed during three executions on March 23-24 2018 with precipitable water vapour (PWV) column of 0.9, 1.9 and 2.0mm, respectively. The quasar J1058+0133 was used as atmospheric bandpass and flux calibrator, while quasar J0948+0022 was used as phase calibrator. VR7 \citep{Matthee2017SPEC} was observed under good conditions (PWV column of 1.0, 1.0 and 0.7 mm, for each execution, respectively) on 6 and 7 September 2018. The quasars J2148+0657 and J2226+0052 were used for atmospheric bandpass and flux calibrator, and phase calibrator, respectively.

The data have been reduced and calibrated using {\sc Casa} version 5.1.1-5 following the standard pipeline procedures. The final imaging was performed using the {\sc clean} task, with different choices for weighting and tapering depending on the specific scientific question. In general, using natural weighting, we measure a background rms = 6 $\mu$Jy beam$^{-1}$ in the continuum and a background rms = $0.08-0.10$ mJy beam$^{-1}$ in 18 km s$^{-1}$ channels.  

Our analysis and reduction strategy was as follows: we first reduced the data with natural weighting and a UV tapering of 650 k$\lambda$ (corresponding to an on-sky FWHM of $0.3''$) and by averaging over two velocity channels, leading to a data-cube with a beam FWHM $\approx 0.7\times0.7''$ and $\approx 18$ km s$^{-1}$ velocity resolution. The tapering was used to optimise detectability of extended emission, while the typical [C{\sc ii}] line with FWHM$\approx150$ km s$^{-1}$ would still be well resolved. This reduction was used for initial inspection. For detections, we then re-imaged the data with higher spatial resolution (through briggs weighting with robust parameter 0.5 and without tapering) and/or without channel-averaging, with specific parameters chosen optimised for the science question (and motivated in each relevant section independently and summarised in Table $\ref{tab:reductions}$). For non-detections, we test the robustness of not-detecting the line/continuum by changing the reduction method, and use the tapered reduction motivated by the properties of similar galaxies.


\begin{table}
\centering
\caption{Overview of the different ALMA reductions used in various parts of our analysis.}
\begin{tabular}{lrr}
Reduction & Beam size & Used in Section \\ \hline
VR7 & & \\
Natural, 650 k$\lambda$ taper & $0.64''\times0.61''$  & \S $\ref{sec:FIR}$, \S $\ref{sec:CII}$, \S $\ref{sec:PV}$  \\
Briggs, Robust=0.5 & $0.47''\times0.40''$ & \S $\ref{sec:resolvedVR7}$ \\
MASOSA & & \\
Natural, 650 k$\lambda$ taper & $0.68''\times0.60''$ & \S $\ref{sec:FIR}$, \S $\ref{sec:CII}$ \\
\hline

\label{tab:reductions}
\end{tabular}
\end{table}

\subsection{{\it HST}/WFC3}
We present {\it HST}/WFC3 observations of MASOSA observed as part of {\it HST} program 14699 (PI: Sobral) on March 6 2018. Observations were performed for two orbits with the F110W and F160W filters. During each orbit, four exposures of $\approx 650$ s were taken following the standard WFC3/IR dither pattern. This results in total exposure times of 5.2ks. We acquire flat-fielded and calibrated {\sc flt} images from the {\sc STScI} server and fix the astrometric solution to the ground-based near-infrared data (that is fixed to GAIA DR2 astrometry). Then, after masking bad pixels and cosmic rays, we median combine individual exposures to a final image with 0.064$''$ pixel scale with bilinear interpolation using {\sc Swarp} \citep{Bertin2010}. We also apply this reduction strategy to the {\it HST}/WFC3 data on VR7 (originating from the same {\it HST} program) that was presented earlier in \cite{Matthee2017SPEC}. Compared to the previous reduction, the spatial resolution is increased slightly as the image is better sampled. We have verified that the integrated flux and sensitivity are consistent within $5$ \%.

We use unsaturated, high S/N detections of stars to measure a resolution of FWHM=0.25, 0.28$''$ in the F110W and F160W images, respectively. The depth is estimated by measuring the standard deviation of the total counts in 1000 apertures with 0.8$''$ diameter placed in empty sky regions. We measure a 3$\sigma$ limit of F110W$=27.2$ and F160W$=26.6$ AB magnitude after correcting for aperture losses.

\section{The properties of targeted galaxies} \label{sec:properties}
\begin{figure}
\includegraphics[width=8.6cm]{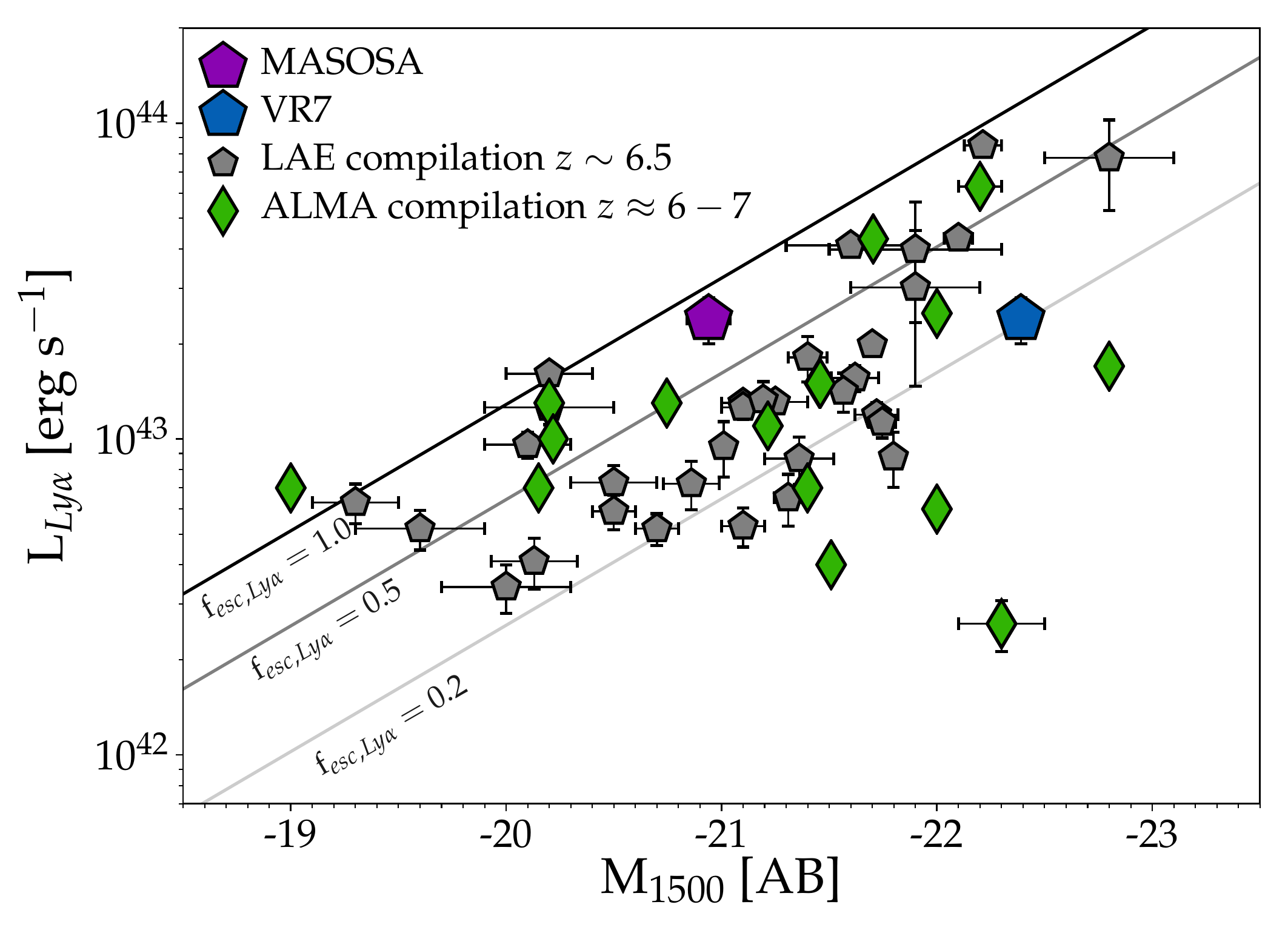}
\caption{The observed Ly$\alpha$ and UV luminosities of MASOSA and VR7 compared to the galaxy population at $z\approx6-7$. Grey pentagons show a compilation of LAEs from \citet{Curtis-Lake2012,Jiang2013,Matthee2017SPEC,Shibuya2017,Matthee2018}, while green diamonds show galaxies at $z\approx6-7$ targeted by previous ALMA programs as compiled in Appendix $\ref{appendix:compilation}$). Solid lines indicate the Ly$\alpha$ escape fraction assuming $\xi_{\rm ion} = 10^{25.4}$ Hz erg$^{-1}$ (e.g. \citealt{BouwensXION}), negligible Lyman-continuum escape fraction and no dust or IGM attenuation. This $\xi_{\rm ion}$ implies an intrinsic Ly$\alpha$ EW=146 {\AA} (\citealt{SM2019}). VR7 is among the most UV luminous galaxies targeted by ALMA so far, while MASOSA is among the galaxies with highest Ly$\alpha$ escape fraction.  \label{fig:MUV_LYA}}  
\end{figure}

In this section, we present the general properties of the targeted galaxies based predominantly on Ly$\alpha$ measurements and {\it HST}/WFC3 data. The measurements are summarised in Table $\ref{tab:global_properties}$. Our targets have been selected on their high Ly$\alpha$ luminosity ($>2\times L^{\star}$; e.g. \citealt{Matthee2015,Konno2018}), a confirmed spectroscopic redshift $z\sim6.5$ and their observability with ALMA. The Ly$\alpha$ luminosities of our targets are relatively high compared to the majority of earlier ALMA observations, but not extreme (Fig. $\ref{fig:MUV_LYA}$).
 
MASOSA, at $z=6.541\pm0.001$ \citep{Sobral2015}, is located in the COSMOS field, such that there is {\it Spitzer}/IRAC data available through the SPLASH program (in particular in the [3.6] and [4.5] bands; \citealt{Steinhardt2014}). VR7, at $z=6.534\pm0.001$ \citep{Matthee2017SPEC}, is located in the CFHTLS-W4/SA22 field where no deep {\it Spitzer}/IRAC data is available. Upper limits on rest-frame UV lines besides Ly$\alpha$ based on X-SHOOTER observations (i.e. C{\sc iv}, He{\sc ii}, C{\sc iii}]) are presented for VR7 in \cite{Matthee2017SPEC}. The non-detections of high-ionisation lines and the narrow Ly$\alpha$ line-widths ($<400$ km s$^{-1}$) indicate that our targets are not powered by a strong active galactic nucleus (AGN; e.g. \citealt{Sobral2018}.)

\subsection{Ly$\alpha$ luminosity and rest-frame equivalent width}
We measure the Ly$\alpha$ flux using a combination of narrow-band (NB921) and broadband ($z$) imaging from Subaru/Suprime-Cam as described in \cite{Matthee2015}. The transmission of the NB921 filter is not a perfect top-hat (which is assumed in the flux calculation in \citealt{Matthee2015}). Therefore, we correct the line-flux for the actual transmission at the specific wavelength where Ly$\alpha$ is observed which is known thanks to spectroscopy. Our flux measurements are therefore not dependent on (assumptions on) slit losses or the flux calibration of spectra. VR7 and MASOSA have an almost identical integrated Ly$\alpha$ luminosity, while their UV continuum luminosity differs by a factor $\approx4$ (see the next subsection). Measurements of the UV continuum luminosity and the UV slope are used to constrain the continuum around Ly$\alpha$, and to estimate the EW. The Ly$\alpha$ EW of VR7 is moderate (EW$_0=34\pm4$ {\AA}), while the EW of MASOSA is extremely high (albeit with large uncertainties; EW$_{0}=145^{+50}_{-43}$ {\AA}). The emission of both Ly$\alpha$ lines consist of a single, clearly asymmetric red peak, with FWHM ranging from 340-390 km s$^{-1}$ \cite{Sobral2015,Matthee2017SPEC}. 

 \begin{figure}
\includegraphics[width=8.6cm]{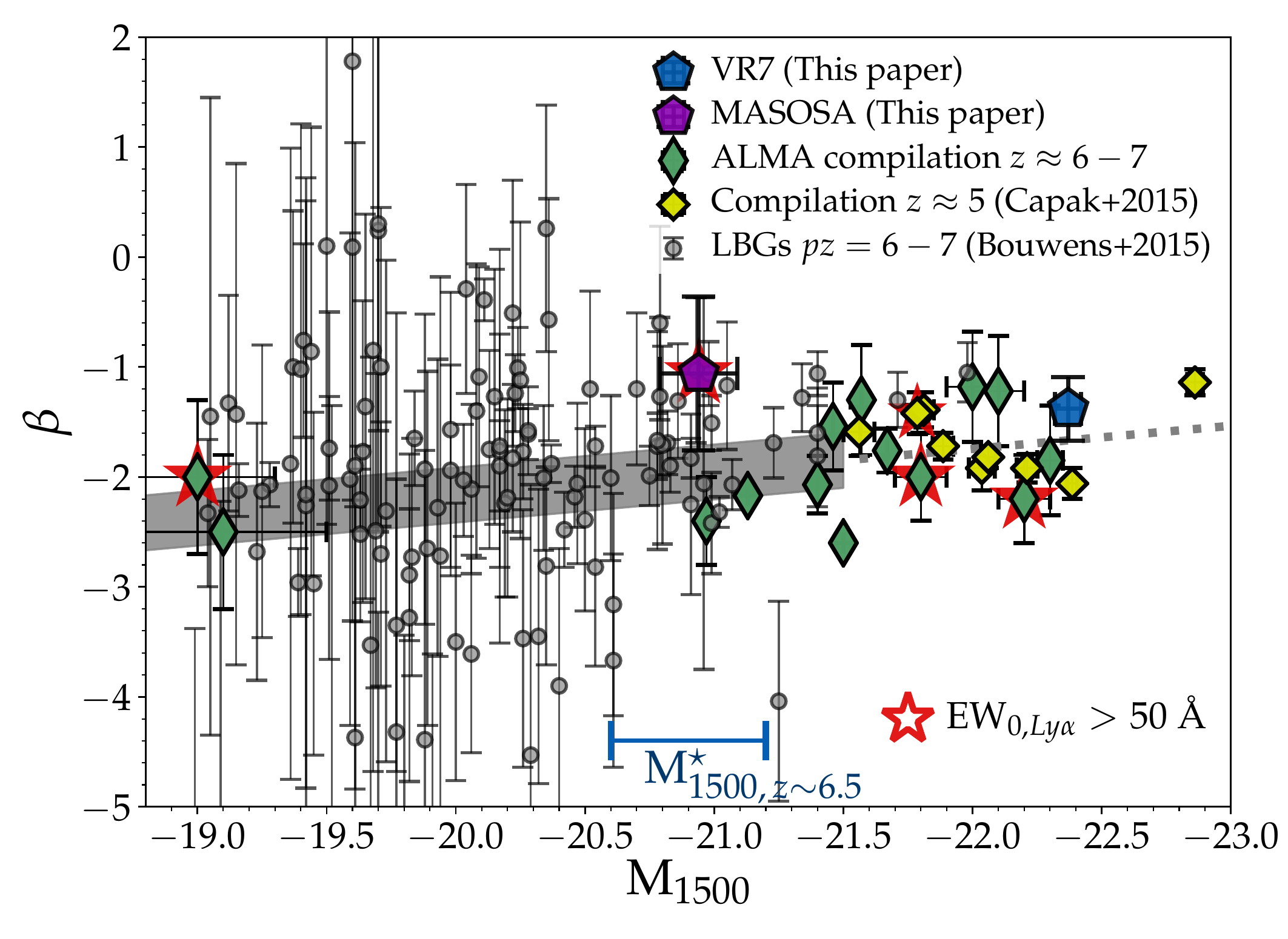}
\caption{The observed UV slope $\beta$ and UV luminosities of MASOSA and VR7 compared to galaxy compilations at $z\approx5$ (from \citealt{Capak2015}) and $z\approx6-7$ observed with ALMA and to the general Lyman-break population from \citet{Bouwens2014}. Red stars highlight sources with the Ly$\alpha$ EW$_{0}>50$ {\AA}. The shaded region shows the fitted relation (\citealt{Bouwens2014}) and its uncertainty, while we show its extrapolation to higher UV luminosities with a dotted line.  \label{fig:MUV_BETA}} 
\end{figure}

\subsection{UV luminosity and colors} \label{sec:UVprops}
We measure the rest-frame UV luminosity using the {\it HST}/WFC3 F110W and F160W data. Measurements are performed with 1.2$''$ diameter apertures to account for the objects sizes and include corrections for missing encapsulated flux based on the WFC3 manual. We correct the F110W photometry for the contribution from Ly$\alpha$ emission based on the measured Ly$\alpha$ luminosity and measure the UV slope $\beta$ using the F110W-F160W colors following \cite{Ono2010}. 

For MASOSA, we measure F110W$=25.74^{+0.13}_{-0.11}$ and F160W$=25.58^{+0.23}_{-0.20}$. These measurements translate into an absolute UV luminosity M$_{1500} = -20.94^{+0.14}_{-0.13}$ and a red UV slope $\beta=-1.06^{+0.68}_{-0.72}$ ($-0.14$ magnitude brighter and $\beta_{\rm obs}=-1.49^{+0.56}_{-0.58}$ if we would not correct for the Ly$\alpha$ contribution). The potential red color is intriguing given the high Ly$\alpha$ EW, as the Ly$\alpha$ escape fraction is typically lower for red galaxies \citep[e.g.][]{Matthee2016}. We discuss this in more detail in \S $\ref{sec:discuss_MASOSA}$. MASOSA has {\it Spitzer}/IRAC magnitudes $[3.6]=24.0\pm0.1$ and $[4.5]=24.6\pm0.2$ \citep{Laigle2016}, resulting in a relatively blue color $[3.6]-[4.5]=-0.6\pm0.2$ similar to other LAEs at $z\approx6.5$ \citep[e.g.][]{Harikane2018}.

VR7 is very luminous in the rest-UV, with F110W$=24.35^{+0.05}_{-0.05}$ and F160W$=24.28^{+0.08}_{-0.07}$, which translates in a UV luminosity M$_{1500}=-22.37^{+0.05}_{-0.05}$. VR7 is also relatively red, with $\beta=-1.38^{+0.29}_{-0.27}$, yet in agreement with the observed relation between UV slope and UV magnitude extrapolated to the luminosity of VR7 \citep{Bouwens2012}, see Fig. $\ref{fig:MUV_BETA}$. Due to its lower EW, the correction for the Ly$\alpha$ line has less impact ($-0.04$ magnitude and $\beta_{\rm obs}=-1.49^{+0.26}_{-0.26}$).

 \begin{figure}
\hspace{-0.2cm}\includegraphics[width=8.9cm]{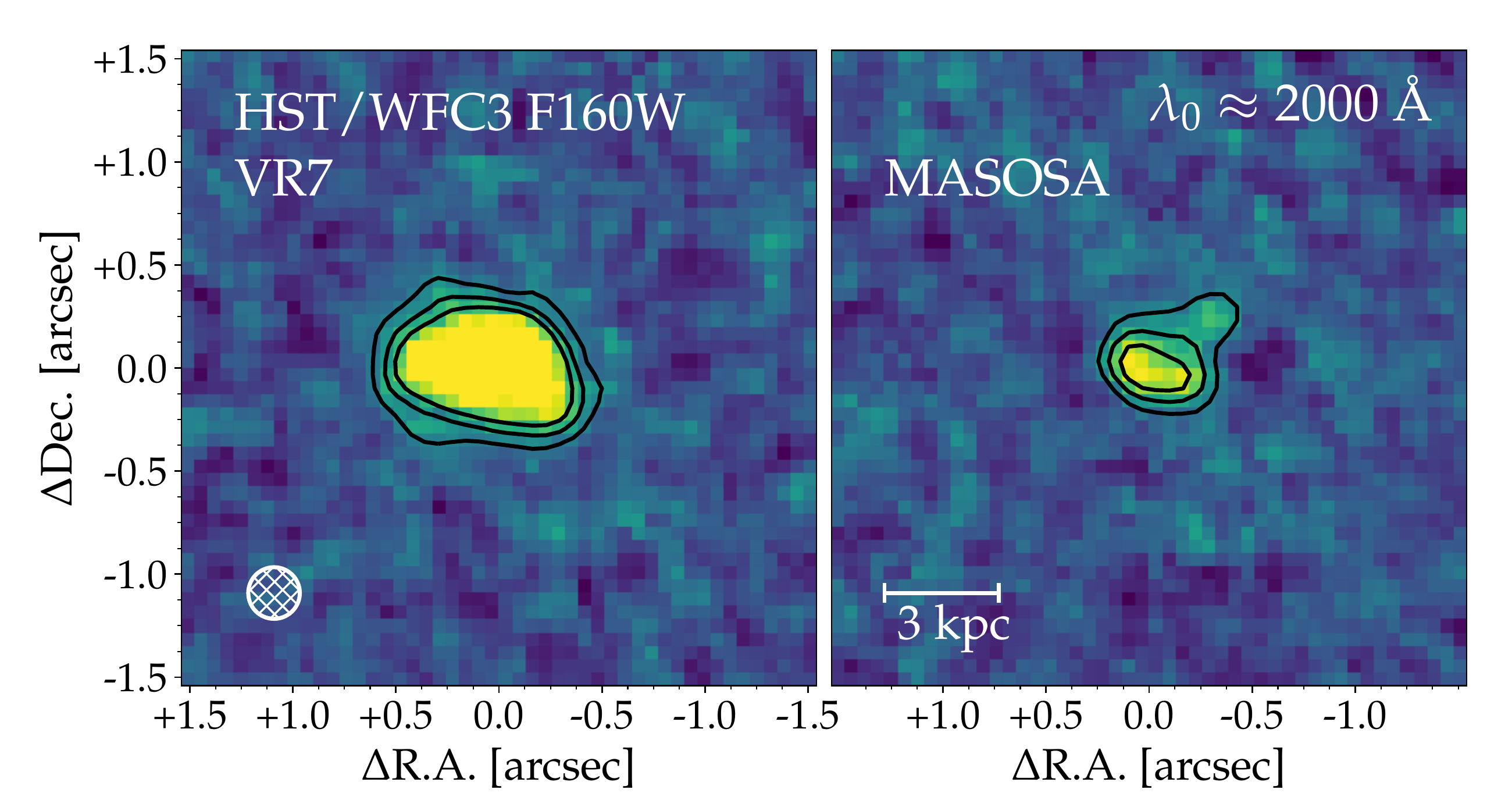}
\caption{{\it HST}/WFC3 F160W thumbnail images of VR7 (left) and MASOSA (right) with matched scale and contrast. Contour levels are at 3, 5, 7$\sigma$. VR7's UV emission (rest-frame wavelength $\lambda_0\approx2000$ {\AA}) is more luminous and more extended than MASOSA. MASOSA is compact, but has an indication of a second component which is only seen in F160W.  \label{fig:thumbs}} 
\end{figure}

\subsection{Rest-frame UV sizes} \label{sec:UVsize}
We measure the de-convolved UV-sizes of MASOSA and VR7 by modelling the light distribution with an exponential profile using the {\sc imfit} software package \citep{Erwin2015}. The PSF image is created by averaging the (normalised) light-profiles of four nearby non-saturated stars and can be modelled accurately with a Moffat profile with FWHM=0.25$''$. As the morphology in the F110W filter may be affected by Ly$\alpha$ emission, we measure morphology in the F160W filter (see Fig. $\ref{fig:thumbs}$; Appendix $\ref{app:size}$ for details on the fitting). For MASOSA, we measure $r_{\rm eff} = 1.12^{+0.44}_{-0.19}$ kpc (corresponding to FWHM=$0.41^{+0.16}_{-0.07}$' arcsec, so marginally resolved in the data) with an ellipticity $0.12^{+0.33}_{-0.12}$ and a position angle $27\pm27 \degree$, meaning that the UV-light is almost spherically symmetric. We note that we would measure a slightly smaller size when allowing the S\'ersic index to vary ($r_{\rm eff} \approx 0.9$ kpc for $n\approx0.1$) or when using the F110W filter ($r_{\rm eff} = 0.87\pm0.05$ kpc, indicating relatively compact Ly$\alpha$ emission). While the F160W image hints towards a faint second clump, a two-component exponential does not provide a better fit to the data.

VR7 is well resolved and a single exponential model results in $r_{\rm eff} = 1.56\pm0.05$ kpc with PA=$80\pm2$ and ellipticity $0.55\pm0.02$. Relaxing the S\'ersic index results again in a slightly smaller size ($r_{\rm eff} = 1.52\pm0.05$ kpc) with $n=0.4\pm0.08$. In Appendix $\ref{app:size}$ we show that the {\it HST} data from VR7 is better described by two slightly smaller exponential components with a separation of $\approx0.35''$ (1.9 kpc), but we list the single component fit in Table $\ref{tab:global_properties}$ for consistency with integrated [C{\sc ii}] and UV luminosity measurements.

\begin{figure*}
\begin{tabular}{cc}
\includegraphics[width=8.6cm]{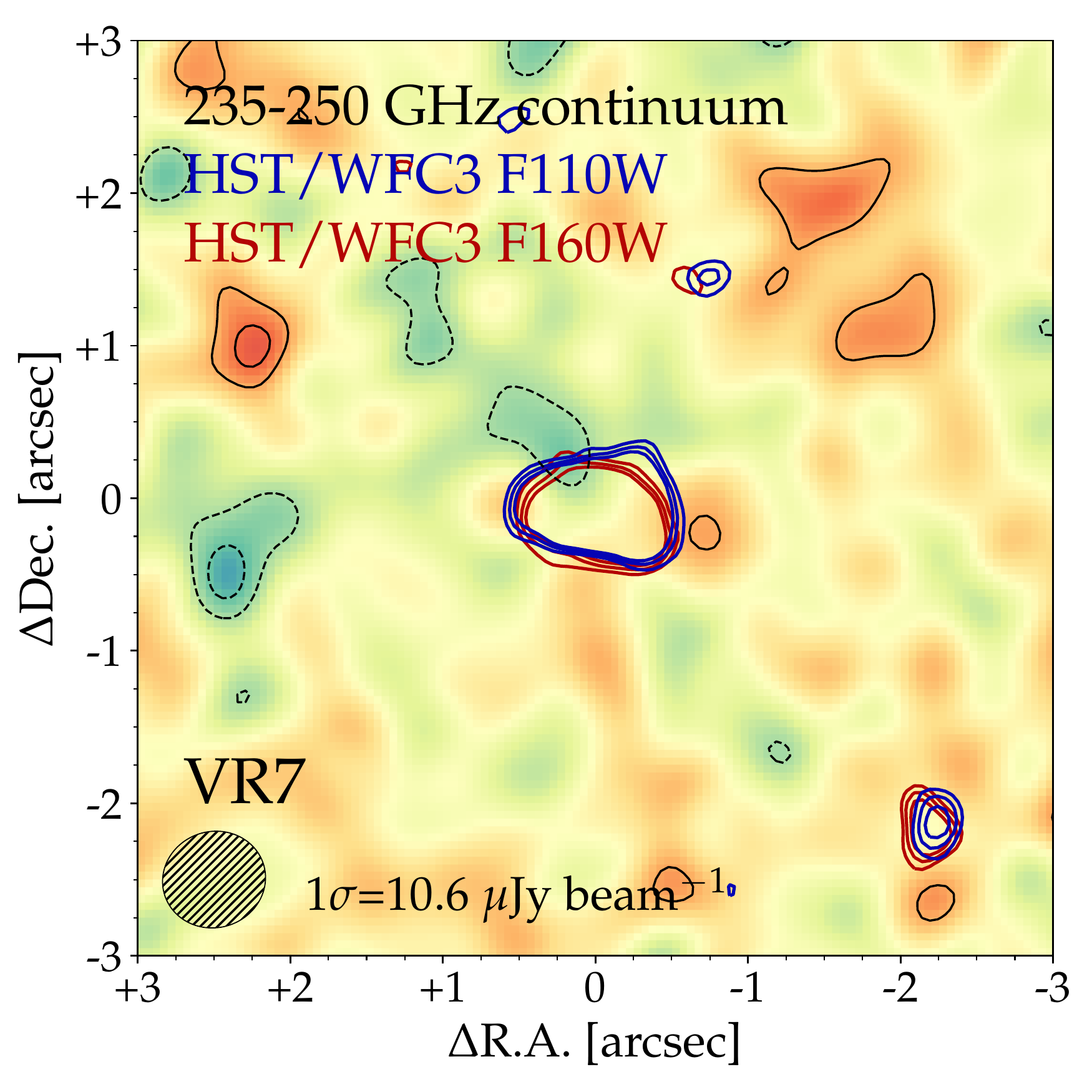}&
\includegraphics[width=8.6cm]{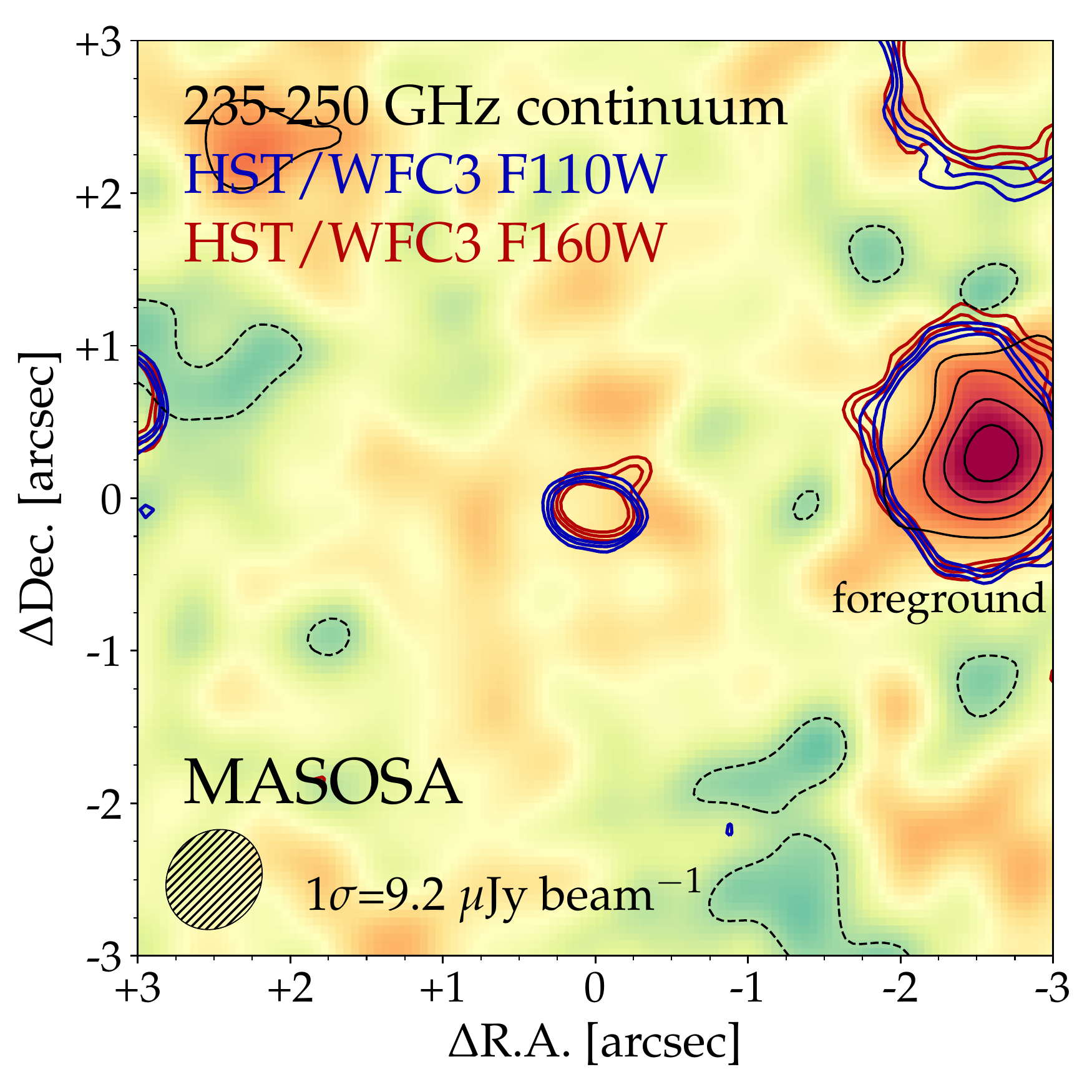}\\
\end{tabular}
\caption{FIR continuum of VR7 (left) and MASOSA (right) in primary beam-corrected data with natural weighting and 0.3$''$ taper (beam major-axes of $\approx0.7''$). The black lines show the $\pm2,3,4 \sigma$ contours, where the 1$\sigma$ level ranges from 9.2-10.6$\mu$Jy beam$^{-1}$. Dark blue contours illustrate the morphologies in the {\it HST}/WFC3 F110W band, while dark red shows the F160W band. Both bands trace the rest-frame UV emission. No FIR continuum emission is detected at the locations of VR7 or MASOSA, independent of the data-reduction method. \label{fig:FIR_zooms}}
\end{figure*} 

\subsection{How do these galaxies compare to the general galaxy population?}
We compare the UV and Ly$\alpha$ luminosities of MASOSA and VR7 to other LAEs identified at $z\sim6.5$ and to other UV-selected galaxies at $z\sim6-7$ observed with ALMA at similar rest-frame frequency (see Appendix $\ref{appendix:compilation}$). The latter sample mostly comprises galaxies for which the redshifts have previously been confirmed with (strong) Ly$\alpha$ emission, as ALMA observations require that the redshift is precise to within $\Delta z \approx0.05$, which is smaller than the typical uncertainty of photometric redshifts. A few notable exceptions include the galaxy B14-65666 at $z=7.15$ with faint Ly$\alpha$ emission \citep{Hashimoto2018Dragons} and two Lyman-break galaxies at $z\approx6.8$ for which the photometric redshift could reliably be estimated \citep{Smit2017}, see Appendix $\ref{appendix:compilation}$ for a complete overview.

As shown in Fig. $\ref{fig:MUV_LYA}$, VR7 is among the most luminous galaxies within the epoch of re-ionisation ever observed with ALMA. MASOSA has an $\approx$M$_{1500}^{\star}$ luminosity and is among the objects with the highest Ly$\alpha$ EW. We note that the spatial resolution of the ALMA observations presented in this work is a factor $\sim2$ higher than most previous observations, allowing more detailed investigation of dynamics and kinematics. The UV sizes of MASOSA and VR7 are comparable to the sizes of other galaxies with similar luminosity \citep[e.g.][]{Shibuya2015,Bowler2017}.

\section{The FIR continuum} \label{sec:FIR}
The ALMA data in the four spectral windows are combined to create 235-250 GHz (1.2 mm) continuum images centred on both objects. These continuum images constrain the FIR continuum luminosity. For VR7, we mask the frequencies where [C{\sc ii}] is detected (see \S $\ref{sec:CII}$), while for MASOSA we use the full frequency range. Several foreground objects are identified in the continuum maps, which can all be associated to objects in either {\it HST}, {\it Spitzer} or VLT/MUSE data (Matthee et al. in prep). These objects lie in the foreground of our targets because of detections in images with filters blue-wards of the Lyman-break at $z\approx6.5$. The foreground objects confirm that the relative astrometry between the ALMA and {\it HST} data and ground-based imaging is accurate to within $\approx0.1-0.2''$. The general properties of these foreground objects and estimates of their redshifts are summarised in Table $\ref{tab:foreground_properties}$. 

\subsection{Upper limits at $\lambda_0\approx160 \mu$m}
No FIR continuum emission is detected at the locations of VR7 or MASOSA (Fig. $\ref{fig:FIR_zooms}$). This result does not depend on the weighting or tapering applied when imaging the visibility data from ALMA, meaning that it is unlikely that flux is resolved out. As source-sizes are $\approx0.6-1.0''$ (major axis) in the rest-frame UV, we use ALMA continuum images constructed with natural weighting and 650 k$\lambda$ tapering (synthesized beam-FWHM $\approx0.7''$) to measure physically motivated and conservative upper limits. For VR7, we measure a 1$\sigma$ limiting $f_{\lambda_0=160 \mu {\rm m}} < 10.6 \mu$Jy beam$^{-1}$, while for MASOSA the 1$\sigma$ limit is $f_{\lambda_0=160 \mu {\rm m}} < 9.2 \mu$Jy beam$^{-1}$. 

\begin{figure}
\includegraphics[width=8.6cm]{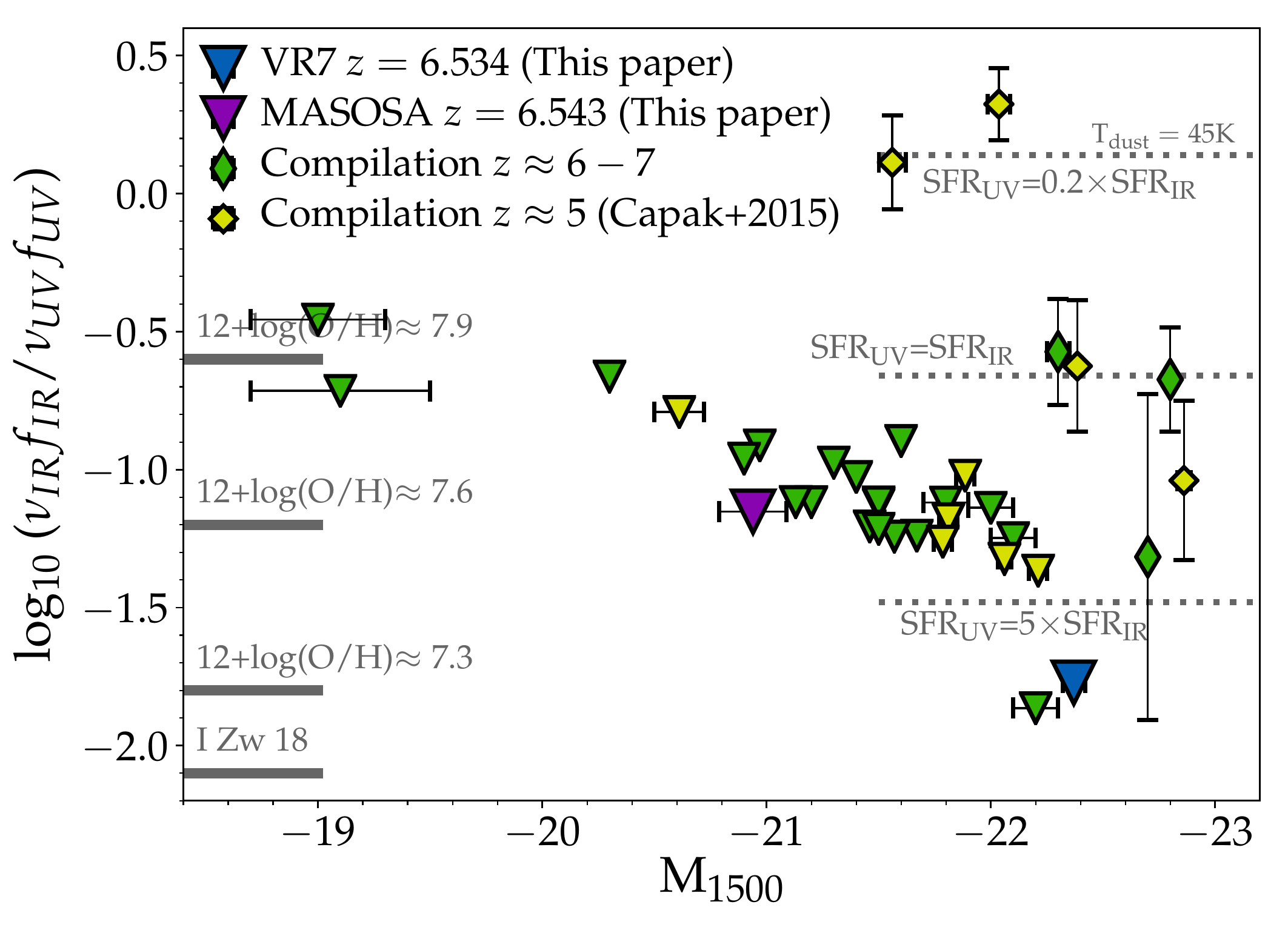}
\caption{UV continuum luminosity versus the ratio of the UV ($\lambda_0=0.15\mu$m) and IR ($\lambda_0=160\mu$m) flux densities for the galaxies observed in this paper and compilations of galaxies observed with ALMA at $z\approx5$ and $z\approx6-7$. Dotted grey lines indicate the 160$\mu$m/1500{\AA} ratio at which the unobscured SFR are a factor 0.2, 1, 5 times the obscured SFR, assuming a dust temperature of 45 K and \citet{Kennicutt1998} conversions. We indicate the typical gas-phase metallicities measured in galaxies in the local Universe with comparable flux density ratios (based on a compilation by \citealt{Maiolino2015}) and the upper limit in the extremely low-metallicity galaxy I Zw 18, but stress that the flux ratio is not exclusively metallicity sensitive. \label{fig:IRUV}}
\end{figure}

We compare our measured upper limits on the continuum flux density around $\lambda_0=160\mu$m with those measured in other UV-selected galaxies at $z\approx6-7$ in Fig. $\ref{fig:IRUV}$. We show IR flux densities, instead of (temperature dependent) IR luminosities, but note that we corrected the flux density for CMB heating assuming a dust temperature $T=45$K, leading to a median correction of $\approx0.03$ dex \citep{daCunha2013}. We also indicate the expected 1500{\AA}/160$\mu$m flux density ratios for local galaxies with extremely low metallicities as inferred from a compilation of galaxies in the local Universe by \cite{Maiolino2015}. For comparison, we show the upper limiting flux density ratio for low-metallicity dwarf galaxy I Zw 18, that we determine using measurements from {\it GALEX}/NUV \citep{GildePaz2007} and {\it Herschel}/PACS F$_{160}$ \citep{RemyRuyer2015}. This upper limit is comparable to the upper limit for VR7.
The IR continuum sensitivity in both MASOSA and VR7 is higher than other galaxies of comparable UV luminosity. VR7 is the most-UV luminous galaxy for which the continuum around $\lambda_0=160\mu$m is currently undetected. This extremely low 160$\mu$m/1500{\AA} ratio indicates that VR7 has a lower dust-to-gas ratio and/or higher dust temperature, compared to other luminous galaxies. The only other galaxy with even stronger limits on the 160$\mu$m/1500{\AA} flux density ratio is the luminous LAE CR7 \citep{Matthee2017ALMA}.

\subsection{IR luminosity and total SFR} \label{sec:IR_SFR}
In order to use these measurements to derive upper limits on the total IR luminosity (L$_{\rm IR}$), one has to make assumptions on the distribution of dust temperatures that combined result in a modified blackbody spectral energy distribution (SED), as an upper limit at a single frequency can not constrain the shape of this SED. Moreover, at $z\sim7$ the temperature of the cosmic microwave background is $\approx 20$K, which contributes to the heating of dust grains \citep{daCunha2013}. The choice of (luminosity weighted) dust temperature strongly impacts the integrated IR luminosity. For example, an increase in the dust temperature from 25 to 45 K decreases the limiting L$_{\rm IR}$ by a factor $\approx 4$ when the continuum is constrained around $\lambda_0\approx160 \mu$m \citep[e.g.][]{Schaerer2015}.

While dust temperatures in typical star-forming galaxies in the $z<2$ Universe are $\approx 35$ K \citep[e.g.][]{RemyRuyer2013}, it is plausible that the dust temperature is higher at higher redshifts, in particular in galaxies with hard ionising sources (i.e. early generations of stars) and a low metallicity \citep[e.g.][]{CenKim2014,Maiolino2015}. While direct constraints on the dust temperatures in high-redshift galaxies require large investments of ALMA time \citep[e.g.][]{Bouwens2016}, \cite{Faisst2018} show that the luminosity-weighted dust temperature in high$-z$ analogues at low-redshift may be as high as 70 K. \cite{Behrens2018} furthermore simulate that intense radiation fields in young galaxies could even result in dust temperatures as high as 90 K.

\begin{figure}
\includegraphics[width=8.7cm]{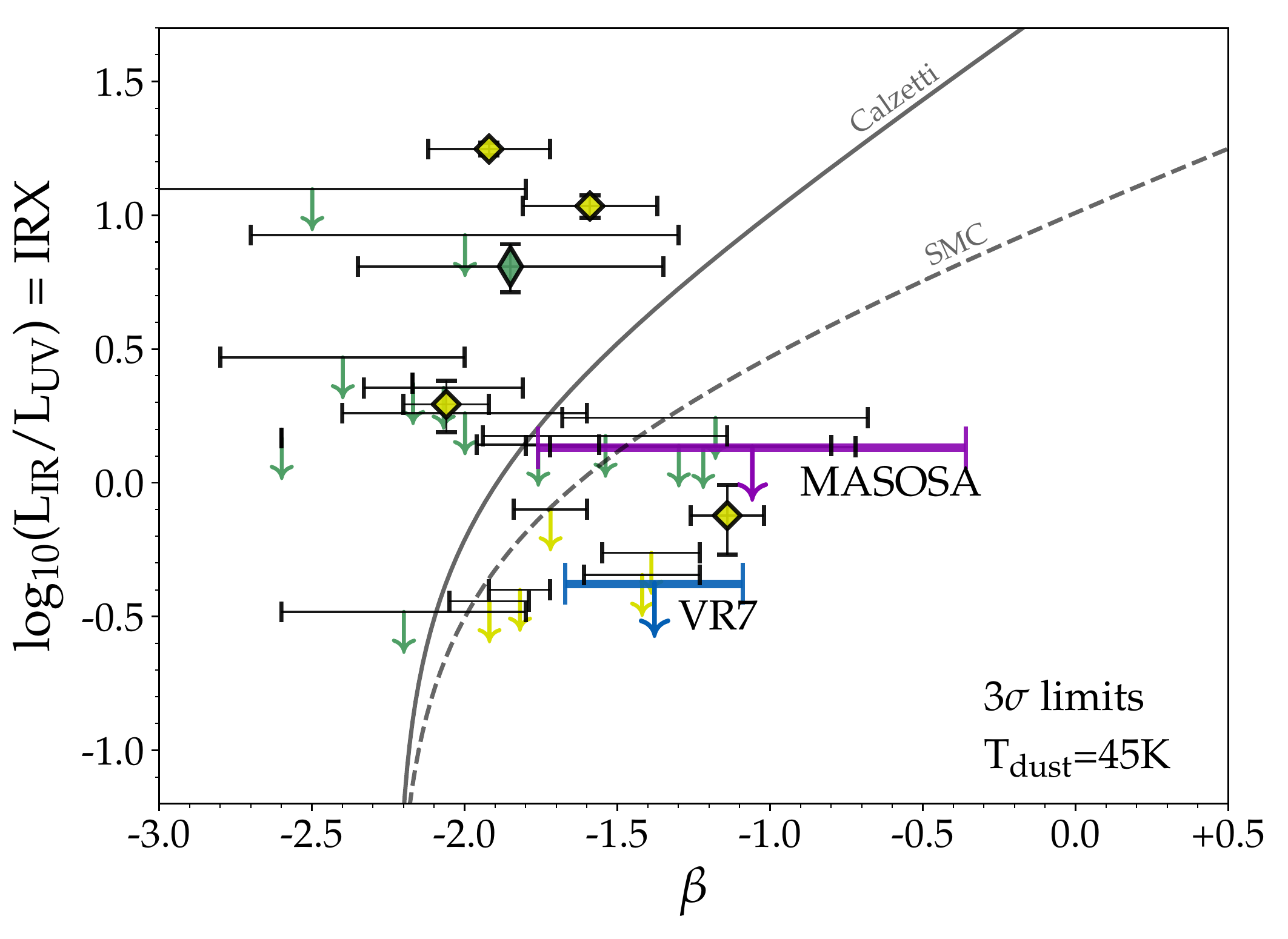}
\caption{The relation between the IR/UV luminosity ratio (IRX) and the observed UV slope $\beta$ for our observed galaxies (MASOSA in purple and VR7 in blue) and compilations of galaxies at $z\approx5$ (yellow and $z\approx6-7$ (green). IR luminosity has been calculated assuming a dust temperature T$=45$K, and upper limits are shown at the 3$\sigma$ level. The solid and dashed line shows the relation for the commonly used Calzetti and SMC extinction curves. \label{fig:IRXbeta}}
\end{figure} 

With these caveats and considerations in mind, we convert our upper limits for VR7 to a limiting L$_{\rm IR, T=45K} < 2.6\times10^{10}$ L$_{\odot}$ and L$_{\rm IR, T=45K} < 2.3\times10^{10}$ L$_{\odot} (1\sigma)$ for MASOSA integrating an optically thin grey body between 8-1000$\mu$m (see Table $\ref{tab:global_properties}$ and Appendix $\ref{appendix:compilation}$ for details). We assume a dust temperature 45 K and power law exponent $\beta=1.5$ and correct for the heating from the CMB following \cite{Ota2014}, although note these corrections are only a factor $\approx0.93$ and are less important for higher dust temperatures (see Appendix $\ref{appendix:compilation}$ for more details). The assumed dust temperature is similar to the measured dust temperature in local low-metallicity galaxy I Zw 18 \citep{RemyRuyer2015}. Assuming a dust temperature of 35 K would result in L$_{\rm IR, T=35K} < 1.3\times10^{10}$ L$_{\odot}$ and L$_{\rm IR, T=35K} < 1.2\times10^{10}$ L$_{\odot}$ for both sources, respectively.

\begin{figure*}
\begin{tabular}{cc}
\includegraphics[width=8.6cm]{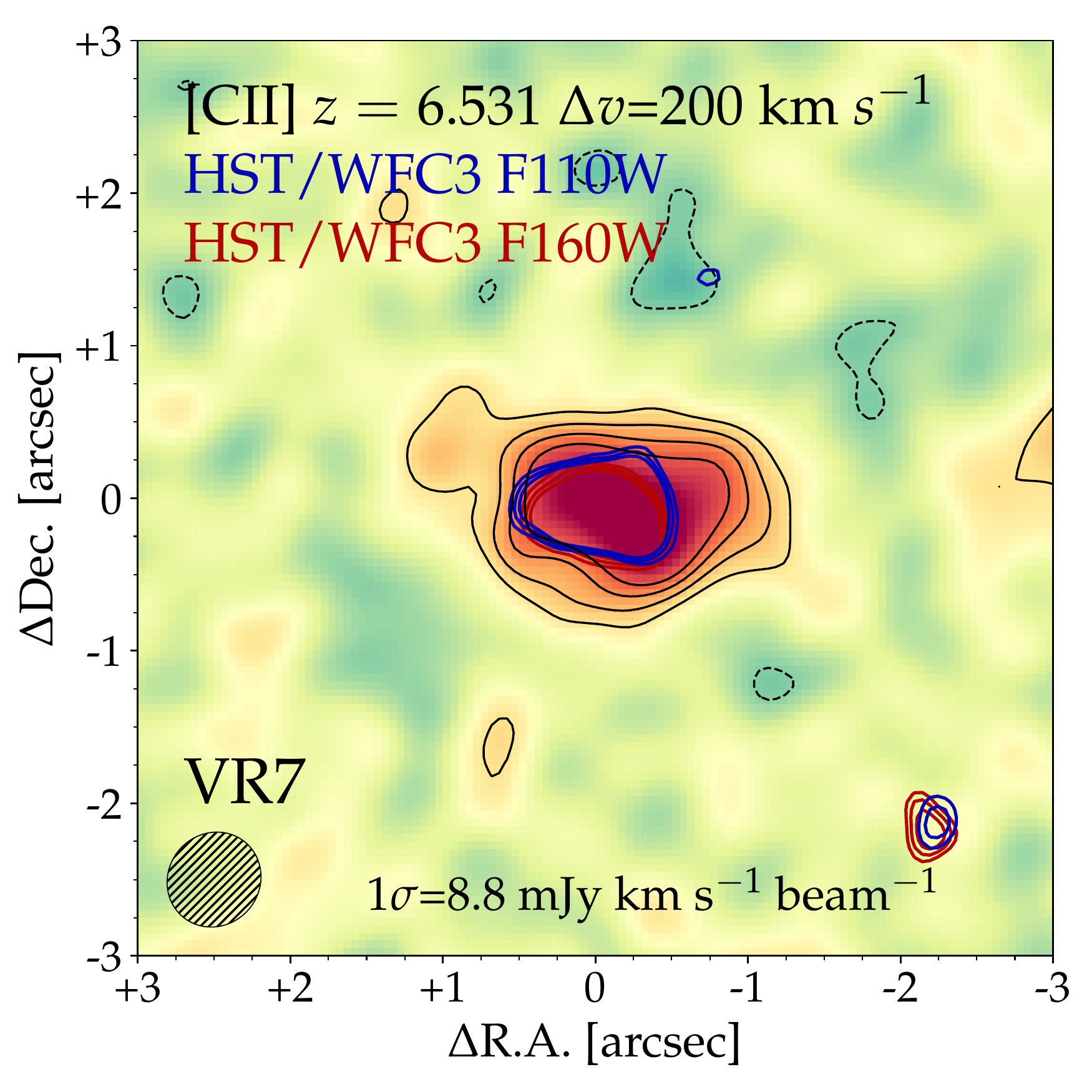}&
\includegraphics[width=8.6cm]{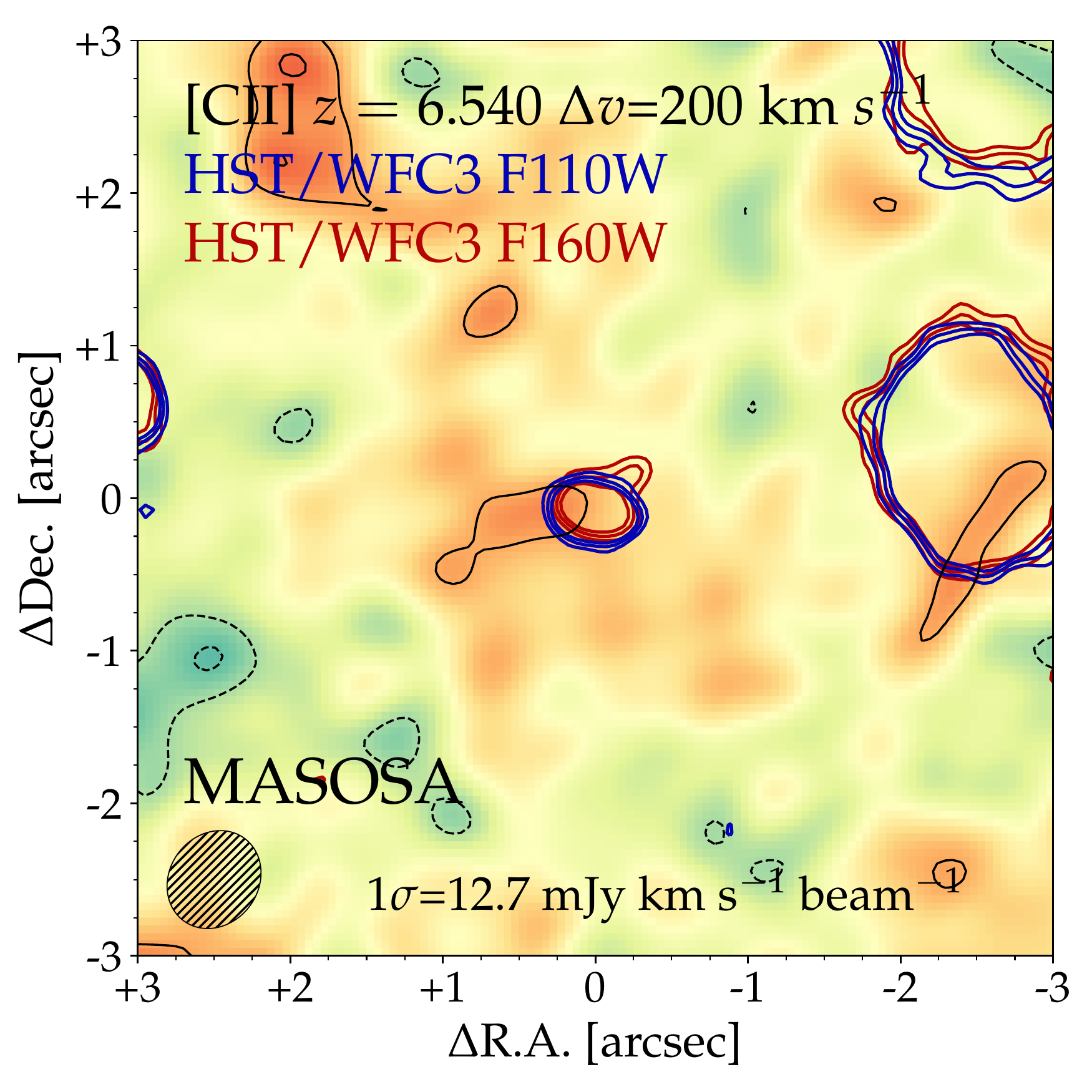}
\end{tabular}
\caption{As Fig. $\ref{fig:FIR_zooms}$, but now collapsing frequencies around the [C{\sc ii}] line with VR7 on the left and MASOSA on the right. Black contours show the $\pm 3,4,5 \sigma$ level, where $\sigma=8.8-12.7$ mJy km s$^{-1}$ beam$^{-1}$. Frequencies are collapsed with a width of 200 km s$^{-1}$ around a central frequency that is blue-shifted by 200 km s$^{-1}$ from the Ly$\alpha$ redshift, as motivated in \S $\ref{sec:CII}$. [C{\sc ii}] emission is clearly detected in VR7 (with peak S/N=9.4 and resolved over multiple beams), while we do not detect [C{\sc ii}] emission from MASOSA (see \S $\ref{sec:MASOSAlimit}$). Note that the PSF-FWHM of the {\it HST} imaging is a factor $\approx2.5$ smaller than of this ALMA reduction.}\label{fig:CII_detection_zooms}
\end{figure*}

As the observed UV continuum slope correlates with the level of dust attenuation \citep[e.g.][]{Meurer1999}, it is expected that the IR/UV luminosity ratio (the IRX ratio) depends on the UV slope, with the relation being dependent on the shape of the attenuation curve and the dust geometry \citep[e.g.][]{Faisst2018} and the intrinsic stellar SED \citep{Reddy2018}. We show the locations of VR7 and MASOSA on the IRX-$\beta$ plane in Fig. $\ref{fig:IRXbeta}$, together with our compilation of galaxies at $z\approx6-7$, assuming a dust temperature of 45 K. VR7 and MASOSA both lie significantly below the commonly used SMC and/or Calzetti attenuation curves, which could indicate that the dust temperature is higher. Additionally, another explanation could be that dust particles reside relatively closer to the star-forming regions and hence maximise the reddening at fixed IR luminosity \citep[e.g.][]{Ferrara2017}, particularly in compact galaxies.

The constraints on the IR continuum can be converted in an upper limit on the level of obscured star formation. Following \cite{Kennicutt1998}, we constrain SFR$_{\rm IR} <4.6$ M$_{\odot}$ yr$^{-1}$ and  SFR$_{\rm IR} < 4.0$ M$_{\odot}$ yr$^{-1}$ for VR7 and MASOSA at 1$\sigma$ respectively for a dust temperature of 45 K. 
Additionally, we can use the observed UV continuum luminosity as an indicator of the un-obscured SFR. Following again \cite{Kennicutt1998}, we measure SFR$_{\rm UV} = 54^{+3}_{-2}$ M$_{\odot}$ yr$^{-1}$ for VR7 and SFR$_{\rm UV} = 15\pm2$ M$_{\odot}$ yr$^{-1}$ for MASOSA. If we would use the classical method to correct the UV luminosity for dust attenuation based on the observed UV slope \citep[e.g.][]{Meurer1999}, we find SFR$_{\rm dust, Meurer}=279^{+128}_{-118}$ M$_{\odot}$ and SFR$_{\rm dust, Meurer} = 208^{+193}_{-171}$ M$_{\odot}$ yr$^{-1}$ for both galaxies respectively. Such high SFRs are however ruled out by the non-detection of continuum emission in our ALMA observations, which indicate that the fraction of obscured SFR is $\lesssim10-30$ \%.\footnote{We calculate that the obscured SFR would be similar to the unobscured SFR in case the dust temperature is 65 K in MASOSA and 90 K in VR7. Compared to our fiducial calculations assuming T$_{\rm dust}= 45$ K, this would increase the upper limits on the IR luminosity by $+0.4$ and $+1.05$ dex, respectively.} Therefore, combining the limiting obscured SFR as an upper bound to the unobscured SFR, we find SFR$_{\rm UV+IR} = 54^{+5}_{-2}$ M$_{\odot}$ yr$^{-1}$ for VR7 and SFR$_{\rm UV+IR} = 15^{+4}_{-2}$ M$_{\odot}$ yr$^{-1}$ for VR7 and MASOSA, respectively.

\begin{figure*}
\begin{tabular}{cc}
\includegraphics[width=8.7cm]{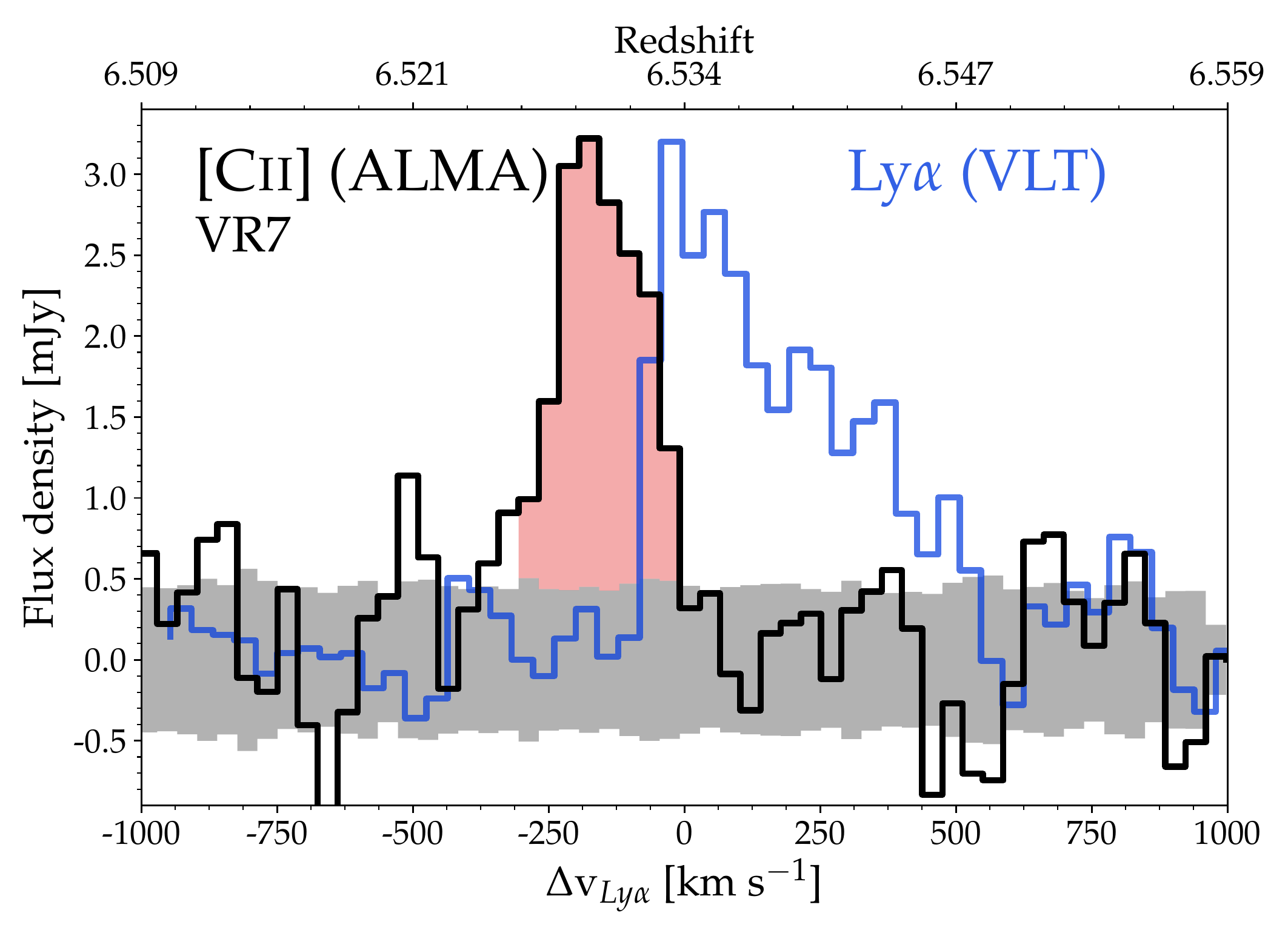}&
\includegraphics[width=8.7cm]{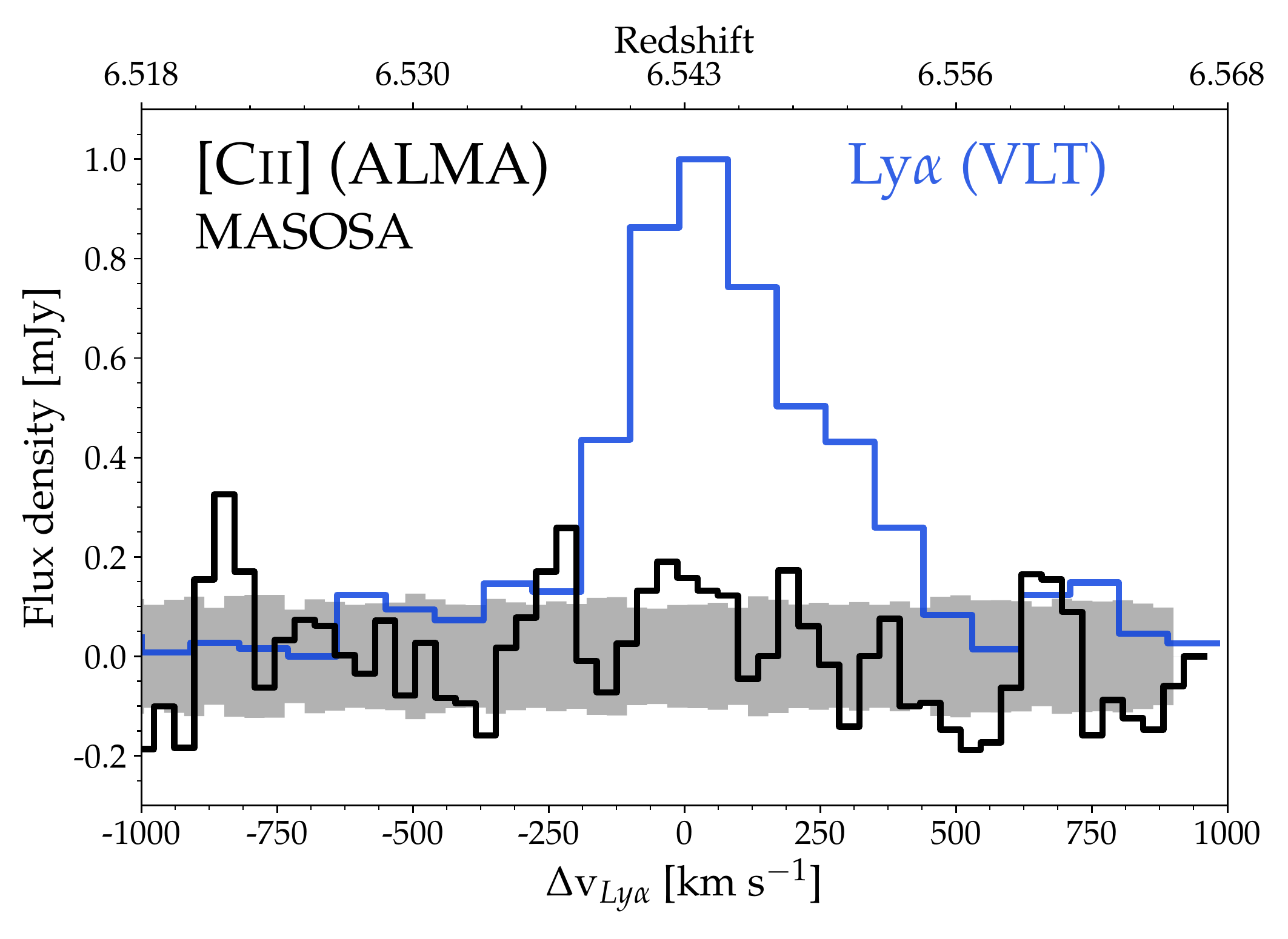}
\end{tabular}
\caption{Extracted 1D spectra centered on redshifted [C{\sc ii}] emission with VR7 on the left and MASOSA on the right. The black line shows the ALMA spectrum that is extracted at the rest-frame UV position (see Fig. $\ref{fig:CII_detection_zooms}$), while the blue line shows the arbitrarily normalised Ly$\alpha$ spectrum. For MASOSA we extract the spectrum within a circular aperture with diameter equal to the beam major axis, while for VR7 we use a aperture that is 1.5 times larger to include extended flux. The noise level of the ALMA data is indicated in grey. Data are binned by a factor two in the frequency direction, resulting in a 38 km s$^{-1}$ velocity resolution. [C{\sc ii}] emission in VR7, detected with integrated S/N$\approx15$, is blue shifted with respect to Ly$\alpha$. The red shaded area highlights the frequency range used to create the integrated [C{\sc ii}] image. We do not detect [C{\sc ii}] emission in MASOSA within 2$''$ and $\pm1000$ km s$^{-1}$ from Ly$\alpha$. \label{fig:CII_detection_spectra}}
\end{figure*}

\section{Integrated [C{\sc ii}] emission properties} \label{sec:CII}
We now focus on the ALMA data around the redshifted [C{\sc ii}]$_{158 \mu {\rm m}}$ frequency. While recent work showed that ALMA resolves [C{\sc ii}] emission in multiple components for a large fraction of high-redshift galaxies \citep[e.g.][]{Matthee2017ALMA,Carniani2018,Hashimoto2018}, we focus first on the integrated [C{\sc ii}] properties by using a low spatial-resolution reduction of our ALMA data (i.e. synthesized beam FWHM=0.7$''$), and analyse a higher resolution reduction in \S $\ref{sec:resolvedVR7}$. 

In order to search for [C{\sc ii}] emission, we visually scanned the data-cube in slices of width 18 km s$^{-1}$ from -1000 to +1000 km s$^{-1}$ with respect to the Ly$\alpha$ redshift, and searched for detections within a radius of 3$''$ (15 pkpc at $z\sim6.5$). We illustrate the results of this scan in Figs. $\ref{fig:CII_detection_zooms}$ and $\ref{fig:CII_detection_spectra}$. Fig. $\ref{fig:CII_detection_zooms}$ shows collapsed [C{\sc ii}] narrow-band images zoomed at the positions of VR7 and MASOSA, while Fig. $\ref{fig:CII_detection_spectra}$ shows 1D spectra extracted at the source location. Normalised Ly$\alpha$ spectra are shown on the same barycentric vacuum velocity scale. We clearly detect [C{\sc ii}] emission in VR7, while no [C{\sc ii}] is detected in MASOSA.

\subsection{Integrated [C{\sc ii}] properties of VR7}
As shown in the left panels in Figs. $\ref{fig:CII_detection_zooms}$ and $\ref{fig:CII_detection_spectra}$, the [C{\sc ii}] emission that is detected in VR7 is roughly co-located with the rest-frame UV emission and is slightly blue shifted with respect to Ly$\alpha$. The peak flux is detected with a S/N=9.4 in the collapsed [C{\sc ii}] image, while the integrated S/N in the 1D extracted spectrum is $\approx15$. 

We measure the redshift of peak [C{\sc ii}] emission and the FWHM of the [C{\sc ii}] line as follows. For 1000 iterations, we perturb each data-point in the 1D extraction with its associated uncertainty and measure the frequency where the flux density is highest, and the (linearly interpolated) frequencies where half of the peak flux density is observed. Then, we obtain the median and the 16, 84nd percentiles of those measurements, and find that the peak [C{\sc ii}] flux is observed at $z_{\rm [CII]} = 6.5285^{+0.0007}_{-0.0004}$. This means that the peak of the Ly$\alpha$ emission-line is redshifted with respect to [C{\sc ii}] by $\Delta v_{\rm Ly\alpha} = 217^{+29}_{-19}$ km s$^{-1}$. The [C{\sc ii}] FWHM is $v_{\rm FWHM, [CII]} = 200^{+48}_{-32}$ km s$^{-1}$, which is a factor $\approx 1.7$ smaller than the Ly$\alpha$ FWHM.

Next, we measure the integrated [C{\sc ii}] luminosity by collapsing the {\it primary-beam corrected} cube over a wider range of frequencies (252.285 GHz - 252.519 GHz; corresponding to -20 to -320 km s$^{-1}$ w.r.t. Ly$\alpha$; illustrated in Fig. $\ref{fig:CII_detection_spectra}$). We use the {\sc imfit} task in {\sc CASA}, which fits an elliptical gaussian profile to the data. This results in a total flux of $447\pm42$ mJy km s$^{-1}$ and a de-convolved source size (major/minor axis) $(1.45\pm0.14'') \times (0.71\pm0.08'')$ with a position angle $96.5\pm5.2 \degree$ ($(1.58\pm0.13'') \times (0.94\pm0.06'')$ before deconvolution).
Following \cite{CarilliWalter2013} this flux translates in a [C{\sc ii}] luminosity $4.76\pm0.44 \times10^8$ L$_{\odot}$, which is emitted over a region of $(7.8\pm0.8) \times (3.8\pm0.4)$ kpc$^2$. The integrated [C{\sc ii}] measurements are summarised in Table $\ref{tab:global_properties}$. The [C{\sc ii}] luminosity, line-width and velocity offset to Ly$\alpha$ are similar to other galaxies at $z\approx6-7$ of comparable UV luminosity, and we will discuss this in more detail in \S $\ref{sec:discuss_CIIUV}$. The major and minor axes of the [C{\sc ii}] emission are larger than the axes of the rest-frame UV emission, which are $(4.6\pm0.2) \times (2.1\pm0.1)$ kpc$^2$ when the UV is modeled as a single component. This indicates [C{\sc ii}] is more extended, similar to observations in quasar hosts \citep{Cicone2015} and a stack of star-forming galaxies \citep{Fujimoto2019}. Moreover, \cite{Rybak2019} find that extended [C{\sc ii}] emission may plausibly dominate the total [C{\sc ii}] emission in high resolution observations of strongly star-forming galaxies at $z\sim3$.

The line-width and size can be converted to an estimate of the total dynamical mass following \cite{Wang2013}, which results in M$_{\rm dyn}$/(sin $i$)$^2 = 1.4^{+0.4}_{-0.3} - 3.0^{+0.7}_{-0.6} \times10^{10}$ M$_{\odot}$ using the semi-minor or semi-major axis, respectively. This mass is a factor $1-2$ times the stellar mass estimated in \cite{Matthee2017SPEC}. 

\begin{table}
\centering
\caption{The integrated properties of VR7 and MASOSA as measured from {\it HST} and ground-based imaging data, optical spectroscopy and ALMA observations. Errors show the 16, 84 percentiles, while upper limits are at the 1$\sigma$ level. See text in \S $\ref{sec:MASOSAlimit}$ that motivates the [C{\sc ii}] luminosity limit for MASOSA.}
\begin{tabular}{lrr}
Property & VR7 & MASOSA \\ \hline
R.A. (J2000) & 22:18:56.36 & 10:01:24.80 \\
Dec.. (J2000) & +00:08:07.32 & +02:31:45.34 \\
M$_{1500}$ & $-22.37^{+0.05}_{-0.05}$ & $-20.94^{+0.14}_{-0.13}$ \\
$\beta_{\rm obs}$ & $-1.49^{+0.26}_{-0.26}$ & $-1.49^{+0.56}_{-0.58}$ \\
$\beta_{\rm Ly\alpha \,\,corr.}$ & $-1.38^{+0.29}_{-0.27}$ & $-1.06^{+0.68}_{-0.72}$ \\
r$_{\rm eff, \rm UV}$/kpc &  $1.56\pm0.05$ & $1.12^{+0.44}_{-0.19}$ \\
L$_{\rm Ly\alpha}$/erg s$^{-1}$ & $2.4^{+0.2}_{-0.2}\times10^{43}$ &  $2.4^{+0.4}_{-0.4}\times10^{43}$ \\
EW$_{\rm Ly\alpha, 0}$/ {\AA} & $34^{+4}_{-4}$ & $145^{+50}_{-43}$ \\   
z$_{\rm Ly\alpha}$ & $6.534\pm0.001$ & $6.543\pm0.003$ \\
$v_{\rm FWHM, Ly\alpha}$/km s$^{-1}$& $340\pm14$ & $386\pm30$ \\
$[3.6]-[4.5]$ & - &  $-0.6\pm0.2$ \\

& & \\ 

{[C{\sc ii}] and IR properties} & & \\ 
S$_{\nu, \rm fit}\Delta$v /mJy km s$^{-1}$& $417.8\pm80.3$ & $<12.7$ \\
L$_{\rm [CII], CASA}$/10$^8$ L$_{\odot}$ & $4.76\pm0.44$ & $<0.22$\\ 
$z_{\rm [CII]}$& $6.5285^{+0.0007}_{-0.0004}$ & - \\
$v_{\rm FWHM, [CII]}$/km s$^{-1}$& $200^{+48}_{-32}$ & -  \\
$\Delta v_{\rm Ly\alpha}$/km s$^{-1}$ & $-217^{+29}_{-19}$ & - \\

r$_{1/2, major, \rm [CII]}$/kpc& $3.9\pm0.4$ & - \\ 
r$_{1/2, minor, \rm [CII]}$/kpc& $1.9\pm0.2$ & - \\
$f_{\nu, \rm 160 \mu m}$/$\mu$Jy beam$^{-1}$ & $<10.6$ & $<9.2$ \\
& & \\ 

Derived properties & & \\
SFR$_{\rm UV, no\,dust}$/M$_{\odot}$ yr$^{-1}$ & $54^{+3}_{-2}$ & $15^{+2}_{-2}$ \\
SFR$_{\rm UV, Meurer}$/M$_{\odot}$ yr$^{-1}$ & $279^{+128}_{-118}$  & $208^{+193}_{-171}$  \\
SFR$_{\rm UV + IR}$/$M_{\odot}$ yr$^{-1}$ & $54^{+5}_{-2}$ & $15^{+4}_{-2}$ \\
EW(H$\beta$+[O{\sc iii}])/{\AA} & - &  $\approx1500$ \\
$Z_{\rm [CII]}/Z_{\odot}$ & $\approx0.2$  & $\lesssim0.07$  \\ 
L$_{\rm IR, T_{d} = 45 K}/10^{10}$L$_{\odot}$& $<2.6$ & $<2.3$\\ 
SFR$_{\rm IR, T_{d} = 45K}$/M$_{\odot}$ yr$^{-1}$& $<4.6$ & $<4.0$ \\
M$_{\rm dyn}$/$(\sin i)^2$ 10$^{10}$ M$_{\odot}$& $3.0^{+0.7}_{-0.6}$  & - \\ 
\hline

\label{tab:global_properties}
\end{tabular}
\end{table}

\subsection{A physically motivated upper limit for the [C{\sc ii}] luminosity in MASOSA} \label{sec:MASOSAlimit}
With the presented observations, no [C{\sc ii}] emission is detected in the LAE MASOSA. Besides the [C{\sc ii}] image collapsed over the frequency range used in Fig. $\ref{fig:CII_detection_zooms}$, we have also inspected collapsed [C{\sc ii}] images ranging from $-1000$ to $+1000$ km s$^{-1}$ with respect to the Ly$\alpha$ redshift and with widths ranging from $50-500$ km s$^{-1}$, but none reveal a detection within 2$''$ from MASOSA. We have also changed the reduction strategy including no UV tapering to a large UV taper of 200k$\lambda$, but none of our attempts resulted in a detection.

We measure the sensitivity of our observations using the r.m.s. value after masking out the position of detected foreground objects. We measure an r.m.s. of 12.7 mJy km s$^{-1}$ beam$^{-1}$ in a collapse of width 200 km s$^{-1}$, which increases to 20.8 mJy km s$^{-1}$ beam$^{-1}$ collapsing channels over 400 km s$^{-1}$. The noise increases by a factor $\approx1.2$ when a UV taper of 300k$\lambda$ is used instead of 650 k$\lambda$. As can be seen in Fig. $\ref{fig:CII_detection_spectra}$, the noise per channel is relatively invariant of frequency, meaning that the sensitivity does not depend on central frequency. These two sensitivities correspond to 1$\sigma$ limiting [C{\sc ii}] luminosities of $0.14-0.22\times10^8$ L$_{\odot}$. 

Alternatively, we assess the significance of the non-detection of [C{\sc ii}] in MASOSA by simulating how fake sources with known luminosity, source-size and width would appear in our data. We simulate point-sources with a range of [C{\sc ii}] luminosities and line widths ranging from $100-400$ km s$^{-1}$ in reductions with different beam sizes. We place these sources in 500 random positions in our collapsed images and measure the median peak S/N. For an object that is unresolved when imaged with a beam FWHM$>0.7''$ (corresponding to 3.8 kpc), we measure a S/N of 3 for a luminosity $(0.45, 0.65, 0.90) \times10^8$ L$_{\odot}$ for widths (100, 200, 400) km s$^{-1}$, respectively. If an object is unresolved when imaged with a beam FWHM $>0.4''$, a S/N of 3 is measured for $(0.38, 0.55, 0.78) \times10^8$ L$_{\odot}$ for each respective velocity width. Therefore, variations in the source-size and line width significantly influence the luminosity for which [C{\sc ii}] emission can be detected. 

\begin{figure}
\hspace{-0.2cm}\includegraphics[width=8.9cm]{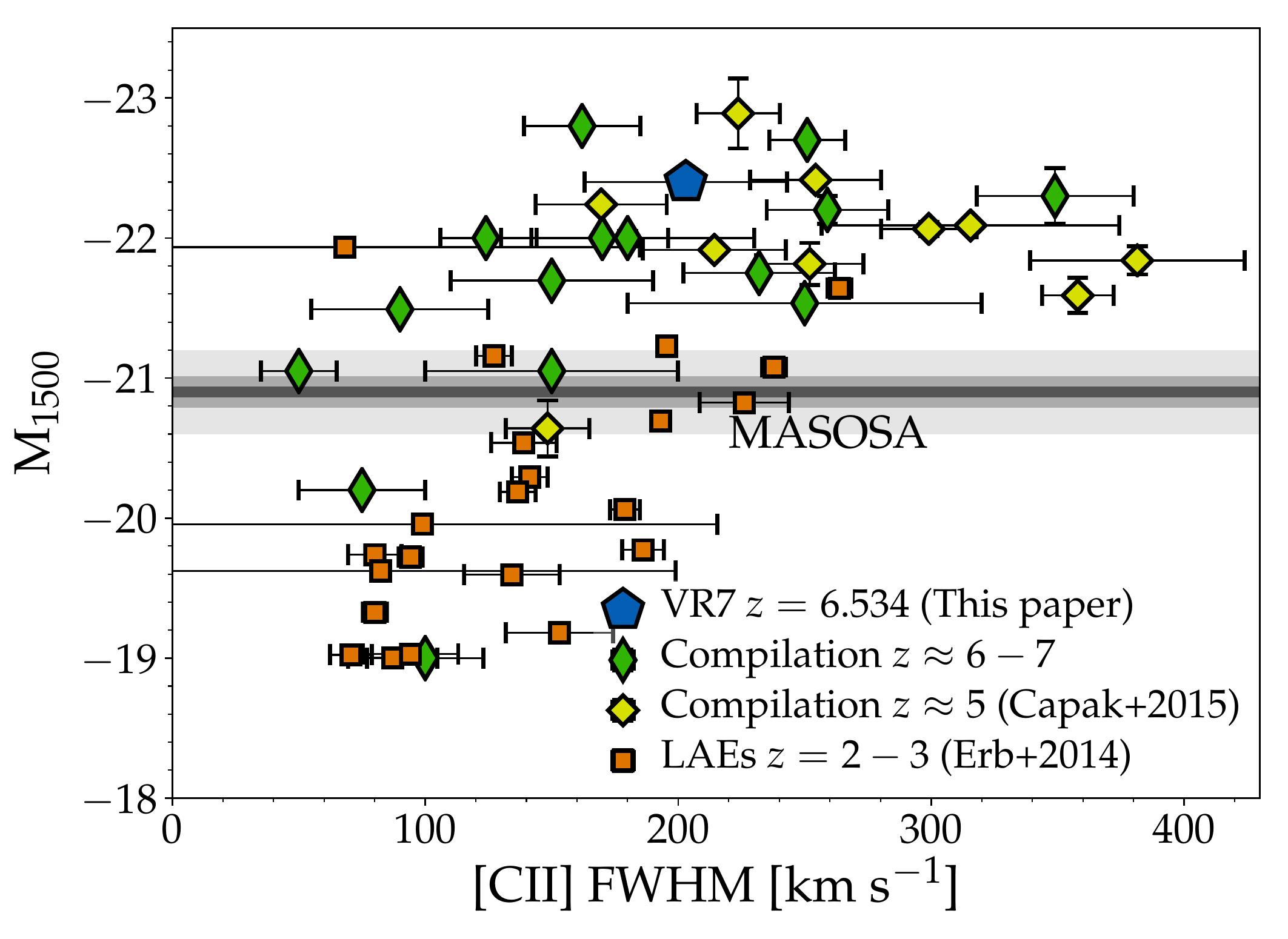}
\caption{The [C{\sc ii}] line FWHM versus UV luminosity for VR7, compilations of galaxies at $z\approx5$ and $z\approx6-7$ and a sample of LAEs at $z\approx2-3$ (where we use the line-width of rest-frame optical nebular lines; \citealt{Erb2014}). UV luminosity, which traces unobscured star formation, is correlated to [C{\sc ii}] width, which traces gravitational potential. Therefore, this relation resembles the `main sequence' between SFR and stellar mass. We highlight the UV luminosity and its uncertainty from MASOSA with a horizontal grey line, indicating that a [C{\sc ii}] FWHM of $\approx120\pm80$ km s$^{-1}$ is expected based on the UV luminosity. \label{fig:motivation_CIIlimits}}
\end{figure}

We show in Fig. $\ref{fig:motivation_CIIlimits}$ that the [C{\sc ii}] line FWHM is correlated with the UV luminosity (with a Spearman rank of 0.46, P-value 98.5 \% for $z\approx5-7$ data on their own), particularly when combined with data at $z=2-3$. This correlation resembles the SFR - stellar mass `main sequence', as the [C{\sc ii}] line-width is expected to be a tracer of gravitational potential and hence mass and the UV luminosity is a proxy for SFR. Increasing scatter in [C{\sc ii}] widths at the brightest UV luminosity could indicate dust attenuation, which could decrease the UV luminosity. MASOSA is among the galaxies with lowest UV luminosity currently targeted by ALMA at $z\approx6-7$ and is thus expected to have a [C{\sc ii}] line width of about $\approx120\pm80$ km s$^{-1}$. Moreover, in galaxies for which both [C{\sc ii}] and Ly$\alpha$ are detected, the [C{\sc ii}] width is typically a factor $\approx0.7\pm0.3$ smaller than the Ly$\alpha$ FWHM, likely because the Ly$\alpha$ line width is affected by radiative transfer effects in the ISM and CGM of galaxies. This would point towards a [C{\sc ii}] width of $\approx250$ km s$^{-1}$ for MASOSA. 

The [C{\sc ii}] half-light radius observed in galaxies at $z\approx6-7$ is typically $\approx2.5\pm1$ kpc, a factor $\approx2$ larger than the UV size of these galaxies measured with {\it HST} \citep{Carniani2018}. As described above,  the UV half-light radius of MASOSA is $\approx1$ kpc. Therefore, a [C{\sc ii}] half-light radius of about $\approx2$ kpc is expected for MASOSA, which corresponds to a major axis of $\approx0.7''$. 

Combining the expected [C{\sc ii}] extent and line width, the most realistic and conservative estimate of sensitivity of our observations is obtained by assuming a line width of 200 km s$^{-1}$ and using a reduction with beam FWHM $\approx0.7''$. This implies a 3$\sigma$ limit for the [C{\sc ii}] luminosity from MASOSA of $0.65\times10^8$ L$_{\odot}$. We compare this upper limit with expectations from the UV luminosity and its implications in \S $\ref{sec:discuss_MASOSA}$. For comparison, this value corresponds to $0.3\times$L$^{\star}_{\rm [CII]}$ in the local Universe \citep{Hemmati2017}.

\begin{figure}
\hspace{-0.3cm}\includegraphics[width=9.1cm]{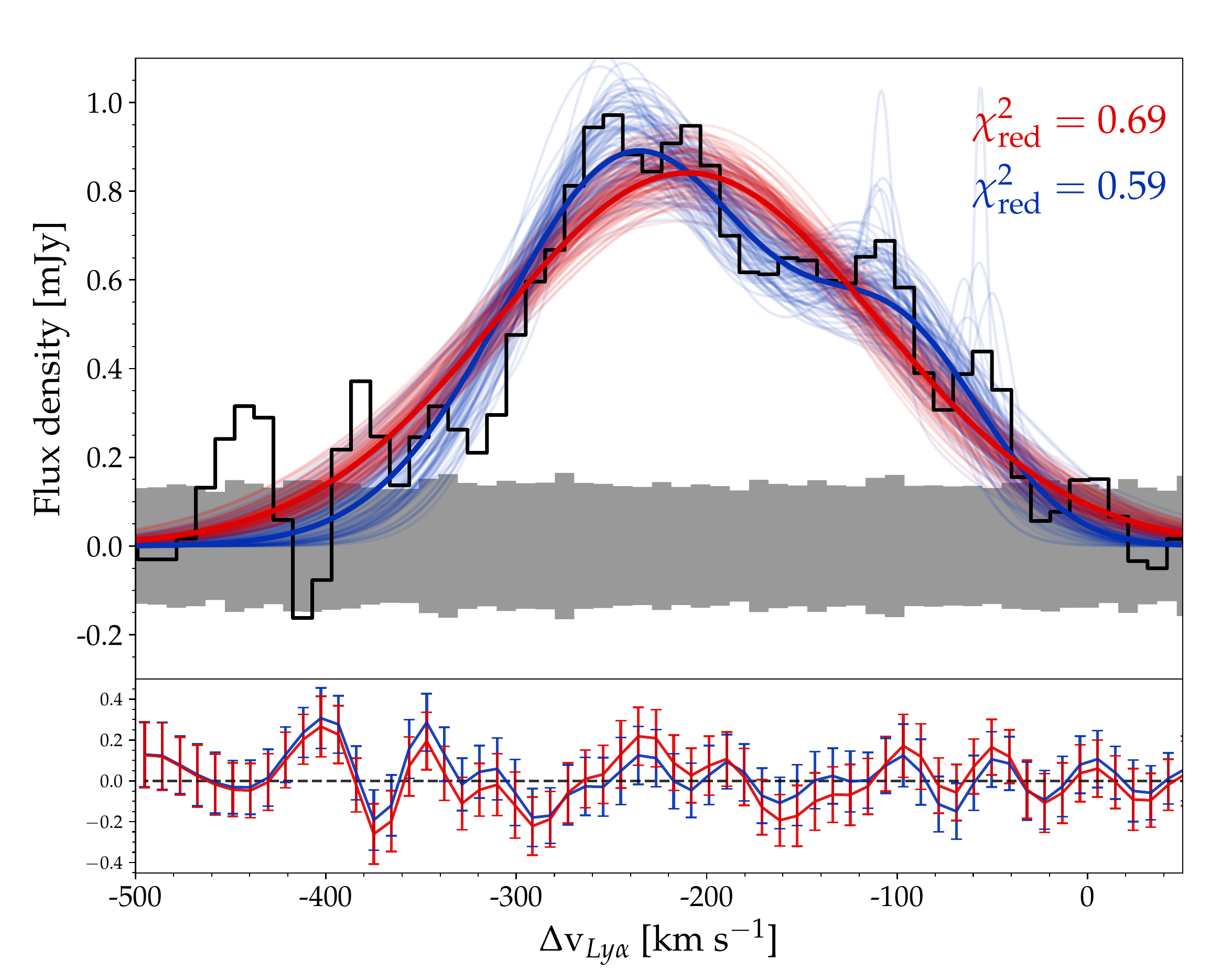}
\caption{1D spectrum of the [C{\sc ii}] line in VR7, as the left panel of Fig. $\ref{fig:CII_detection_spectra}$, but without channel averaging, resulting in a $\approx9$ km s$^{-1}$ resolution. The black line shows the measurement, while the grey region illustrates the noise level. We fit the skewed line-profile with a single gaussian component (red) and a combination of two gaussian components (blue). The bottom panel shows the residuals of the best fits. The data marginally prefer a two-component fit, indicating resolved velocity structure.  \label{fig:VR7_1D_highres}}
\end{figure}

\begin{table}
\centering
\caption{Best-fitted parameters to the two-component gaussian fits of the 1D [C{\sc ii}] line profile of VR7 (shown as the solid blue line in Fig. $\ref{fig:VR7_1D_highres}$). }
\begin{tabular}{lr}
Property ([C{\sc ii}])& Value \\ \hline
Flux$_1$/mJy km s$^{-1}$ beam$^{-1}$& 139.8$^{+29.0}_{-30.3}$\\
Flux$_2$/mJy km s$^{-1}$ beam$^{-1}$ & 43.1$^{+22.9}_{-21.6}$\\
Flux$_2$/(Flux$_1$+Flux$_2$)& 0.24$^{+0.15}_{-0.12}$\\
$\Delta v_{\rm Ly\alpha, 1}$/km s$^{-1}$ & $-211^{+11}_{-9}$\\
$\Delta v_{\rm Ly\alpha, 2}$/km s$^{-1}$& $-87^{+15}_{-15}$\\
FWHM$_{1}$/km s$^{-1}$ & $145^{+41}_{-33}$ \\
FWHM$_{2}$/km s$^{-1}$ & $93^{+26}_{-14}$\\
\label{tab:twogauss_properties}
\end{tabular}
\end{table}

\begin{figure*}
\centering
\includegraphics[width=18.2cm]{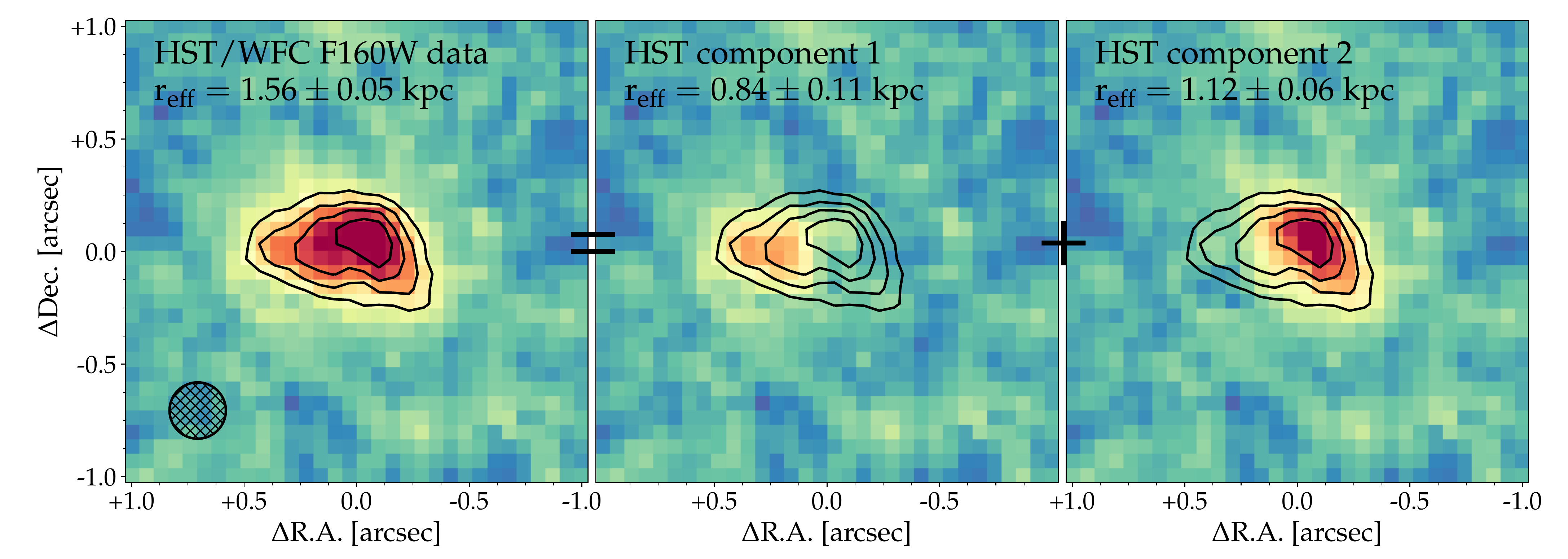}\caption{Zoomed-in {\it HST}/WFC3 image of VR7 in the F160W filter. The left panel shows the data and the size of the PSF-FWHM. The middle and right panels show the individual components of the best-fitted model of two individual exponential components. The contours are drawn in the left panel and shown in the other panels for reference. Each panel lists the effective radius of the respective component, modelled as a single exponential light-profile. \label{fig:VR7_HSTresolved}}
\end{figure*}

 \section{The resolved structure of VR7} \label{sec:resolvedVR7}
Thanks to the high S/N of the detected [C{\sc ii}] line in VR7, we can reduce the ALMA data with a higher spatial resolution and/or relax the velocity-averaging compared to our analysis in \S $\ref{sec:CII}$. In this section, we first investigate the detailed 1D line-profile extracted over the full area over which [C{\sc ii}] emission is detected and then focus on resolving structure spatially in both rest-frame UV and [C{\sc ii}] emission.

\subsection{1D line modelling}
In Fig. $\ref{fig:VR7_1D_highres}$ we show the 1D line-profile of VR7's [C{\sc ii}] line extracted with the native 9 km s$^{-1}$ resolution and extracted over the full [C{\sc ii}] spatial extent. The line-profile appears asymmetrically skewed towards the red, and is slightly better described by a two-component gaussian ($\chi^2_{\rm red}=0.59$) than a single gaussian component ($\chi^2_{\rm red}=0.69$). The best-fitted values of the two-component gaussian fit are listed in Table $\ref{tab:twogauss_properties}$. The line-profile is characterised by a relatively broad (FWHM=$145^{+41}_{-33}$ km s$^{-1}$) luminous component and a narrower (FWHM=$93^{+26}_{-24}$ km s$^{-1}$) component that is a factor $\approx3$ fainter. The narrow component is redshifted by $124^{+19}_{-17}$ compared to the more luminous component, but still blue-shifted by $\approx 90$ km s$^{-1}$ compared to the peak of Ly$\alpha$ emission.

 \begin{figure}
\includegraphics[width=8.6cm]{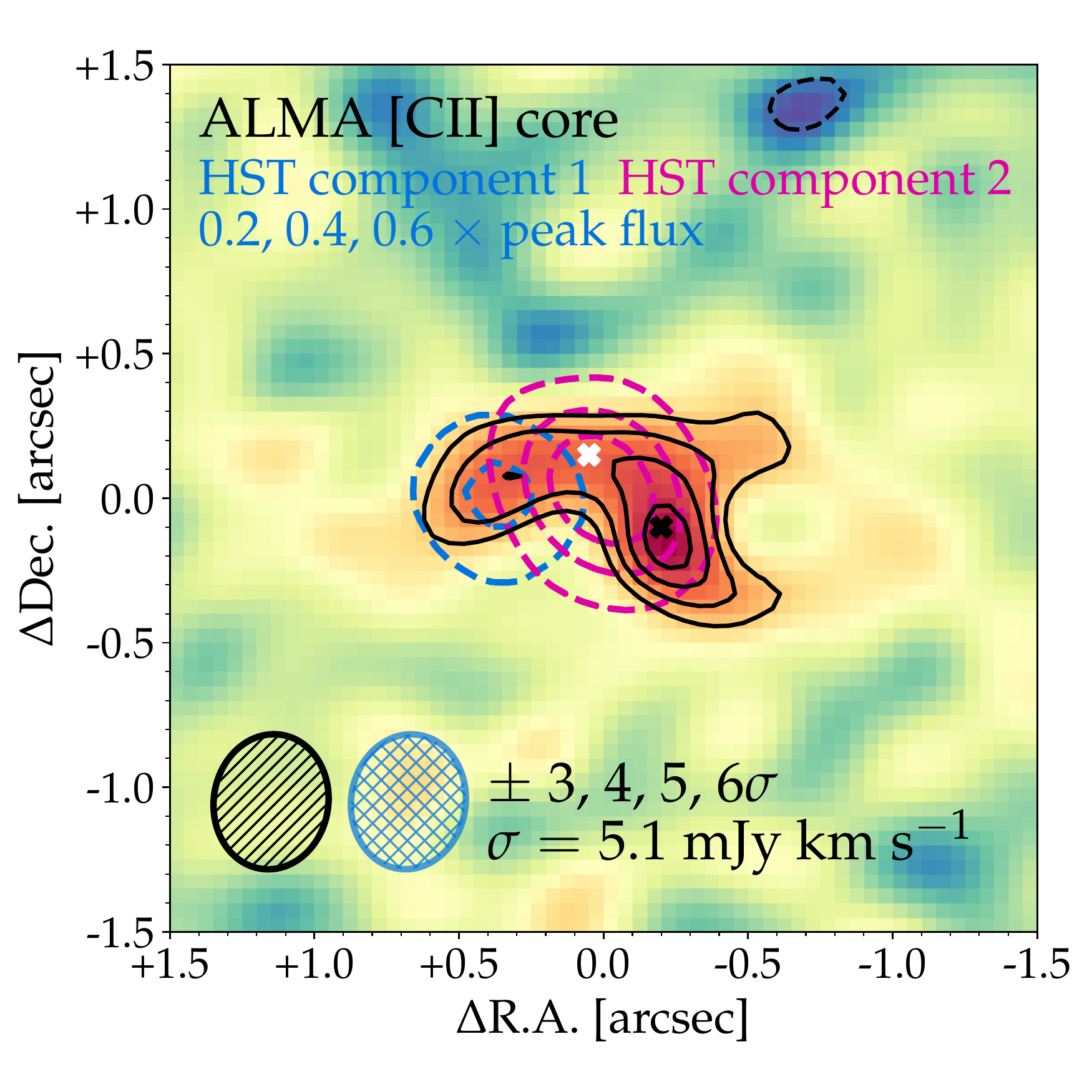}
\caption{Zoomed-in map of the VR7 galaxy. [C{\sc ii}] image from our high resolution ALMA reduction (beam axes $0.47''\times0.40''$), where images are oriented such that east is to the left and north is up. Black lines indicate the $\pm 3,4,5,6\sigma$ contour levels. colored dashed-contours show the individual components in the rest-frame UV identified in {\it HST}/WFC3 imaging, with a PSF-FWHM matched to the ALMA beam by convolving with a 2D asymmetric gaussian. White and black crosses highlight the position of peak [C{\sc ii}] emission at the core velocity($\pm40$ km s$^{-1}$ compared to $z=6.5285$) and at velocity of the redshifted shell ($+100$ to $ +160$ km s$^{-1}$), respectively. Like the UV continuum, [C{\sc ii}] extends in the east-west direction following the centers of both components. Higher surface brightness [C{\sc ii}] emission is extended to the south-west with a position angle $\approx-45\degree$, remarkably similar to component 2 in the rest-frame UV. We note that the relative astrometry may have uncertainties on the order $\approx0.1-0.2''$.  \label{fig:VR7_zooms}}
\end{figure}  

\subsection{Two separate UV components}
VR7's rest-frame UV continuum emission is well resolved in the {\it HST}/WFC3 image, with a maximum extent of about 1.1$''$ ($\approx 6$ kpc) within the 2$\sigma$ contours. While in general the galaxy is elongated in the east-western direction, it also has a tail extending somewhat to the south-west (see the left panel of Fig. $\ref{fig:VR7_HSTresolved}$). Because of this tail, the rest-frame UV light-profile is better fitted by a two-component exponential model than a single component model (with reduced $\chi^2_r=1.10$ versus $\chi^2_r=2.83$, detailed in Appendix $\ref{app:size}$). The best-fitted double exponential model contains two components separated by $0.35\pm0.01''$ ($1.88\pm0.07$ kpc). We show the individual components in the middle and right panel of Fig. $\ref{fig:VR7_HSTresolved}$. The image for component 1 is created by subtracting the PSF-convolved model of component 2, and vice versa. We list their properties in Table $\ref{tab:twocomp_properties}$. Component 1 lies in the east and is relatively compact, while component 2 coincides with the peak UV emission and is elongated in the south-western direction. Component 2 is slightly larger and a factor $1.8^{+0.7}_{-0.5}$ more luminous, implying that both clumps have very similar SFR surface density. We note that we refrain from over-interpreting the colors of modelled components because of Ly$\alpha$ emission possibly affecting the morphology in the F110W filter, but refer the reader to Appendix $\ref{app:size}$ for more details.

\begin{table}
\centering
\caption{Best-fitted parameters for the two-component exponential models of the {\it HST}/WFC3 F160W image of VR7 (shown in Fig. $\ref{fig:VR7_HSTresolved}$). $\Delta$R.A. and $\Delta$Dec. are the relative positions with respect to the center listed in Table $\ref{tab:global_properties}$.}
\begin{tabular}{lrr}
Property & Component 1 & Component 2 \\ \hline
$\Delta$R.A. & $+0.263''$ &  $-0.079''$ \\
$\Delta$Dec. & $+0.104''$ & $+0.126''$\\
r$_{\rm eff}$ & $0.84\pm0.11$ kpc & $1.12\pm0.06$ kpc \\
PA & $76\pm9 \degree$  & $48\pm4 \degree$ \\
Ellipticity & $0.52\pm0.12$ & $0.62\pm0.08$ \\
Fraction of total flux & $36\pm7$ \% & $64\pm7$ \% \\

\label{tab:twocomp_properties}
\end{tabular}
\end{table}

\subsection{Spatially resolved [C{\sc ii}] and UV emission}  
Now, we focus on spatially resolving the [C{\sc ii}] emission using a high-resolution reduction ($0.47''\times0.40''$ beam; $2.5$kpc$\times2.2$kpc), see Fig. $\ref{fig:VR7_zooms}$. Extended [C{\sc ii}] emission is observed over an area $\approx1.3''\times0.6''$, with the largest elongation in the east-west direction, similar to what was found above in a lower resolution reduction. Interestingly, higher surface brightness contours are rotated by about $45\degree$ clockwise, resembling component 2 in the rest-frame UV. Besides the peak [C{\sc ii}] emission in the west, a hint of a second peak is seen in the east at $\approx-0.5''$ from the image centre. Fig. $\ref{fig:VR7_zooms}$ also compares the [C{\sc ii}] map with the rest-frame UV components identified in the {\it HST}/WFC3 F160W imaging (Fig. $\ref{fig:VR7_HSTresolved}$). In general, the [C{\sc ii}] morphology is similar to the rest-frame UV, with the major elongation in the east-western direction following the centers of both components. Intriguingly, the peak [C{\sc ii}] emission extends towards the south-west, which is similar to the major direction of elongation of component 2 in the {\it HST} data.\footnote{The relative astrometry of the ALMA and {\it HST} data have been verified with only a single foreground object, meaning uncertainties on the order $\approx0.1-0.2''$ exist. However, the result that the position angles of high surface brightness regions of [C{\sc ii}] and UV emission do not align is robust to these uncertainties.}

 \begin{figure}
\includegraphics[width=8.9cm]{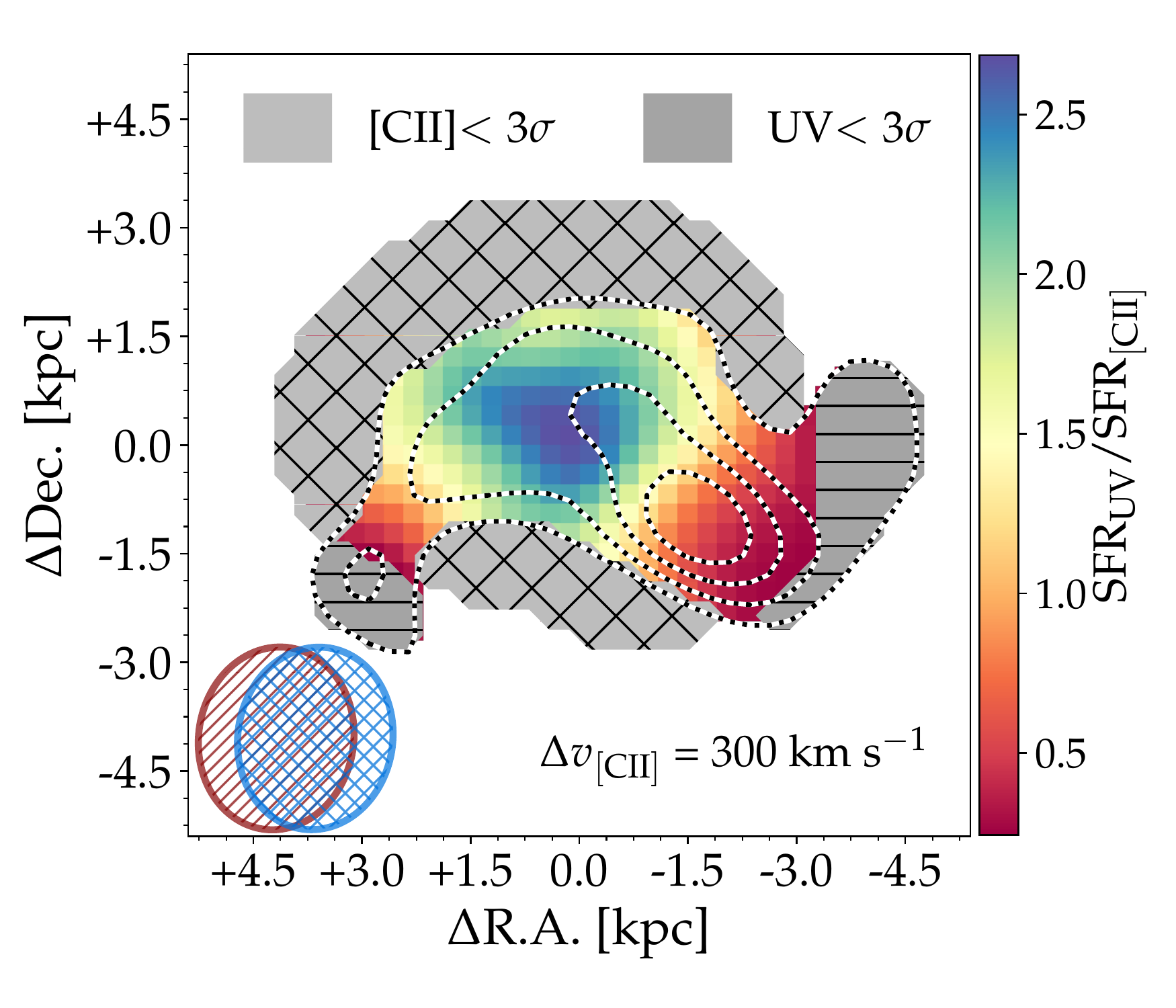}
\caption{The spatial variation of the SFR$_{\rm UV}$ to SFR$_{\rm [CII]}$ ratio, where the first is not dust corrected and the latter estimated from the scaling relation in the local Universe by \citet{DeLooze2014}. {\it HST} data have been smoothed to match the ALMA beam. The grey horizontally-hatched region shows where no flux is detected in the rest-UV at $>3\sigma$, while the cross-hatches highlight where no [C{\sc ii}] flux is detected at $>3 \sigma$. The dotted black-and-white line shows the [C{\sc ii}] level as in Fig. $\ref{fig:VR7_zooms}$.} 
 \label{fig:VR7_SFRUVCII}
\end{figure}

 \begin{figure*}
\begin{tabular}{ccc}
\hspace{-0.3cm}\includegraphics[width=6.1cm]{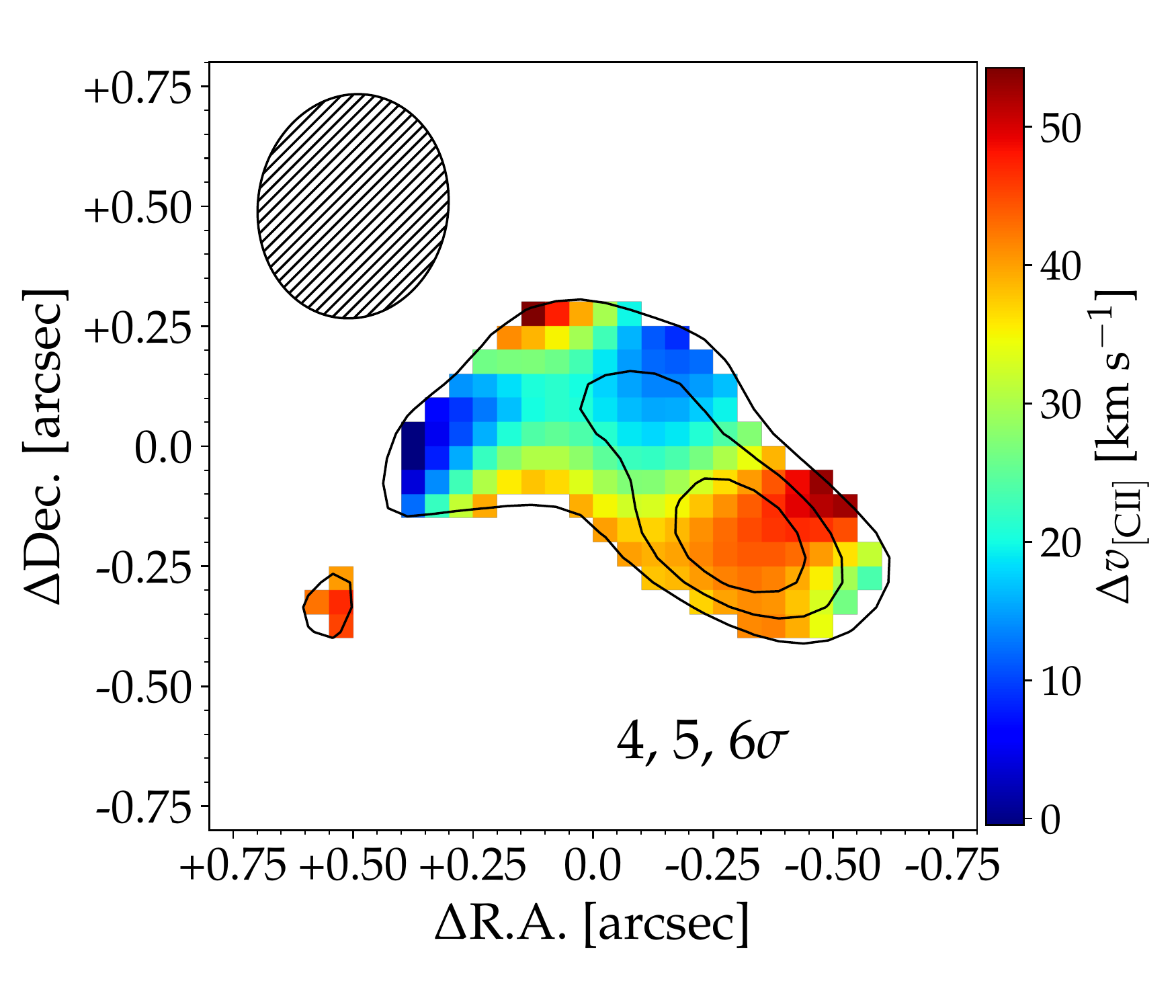}&
\hspace{-0.3cm}\includegraphics[width=6.1cm]{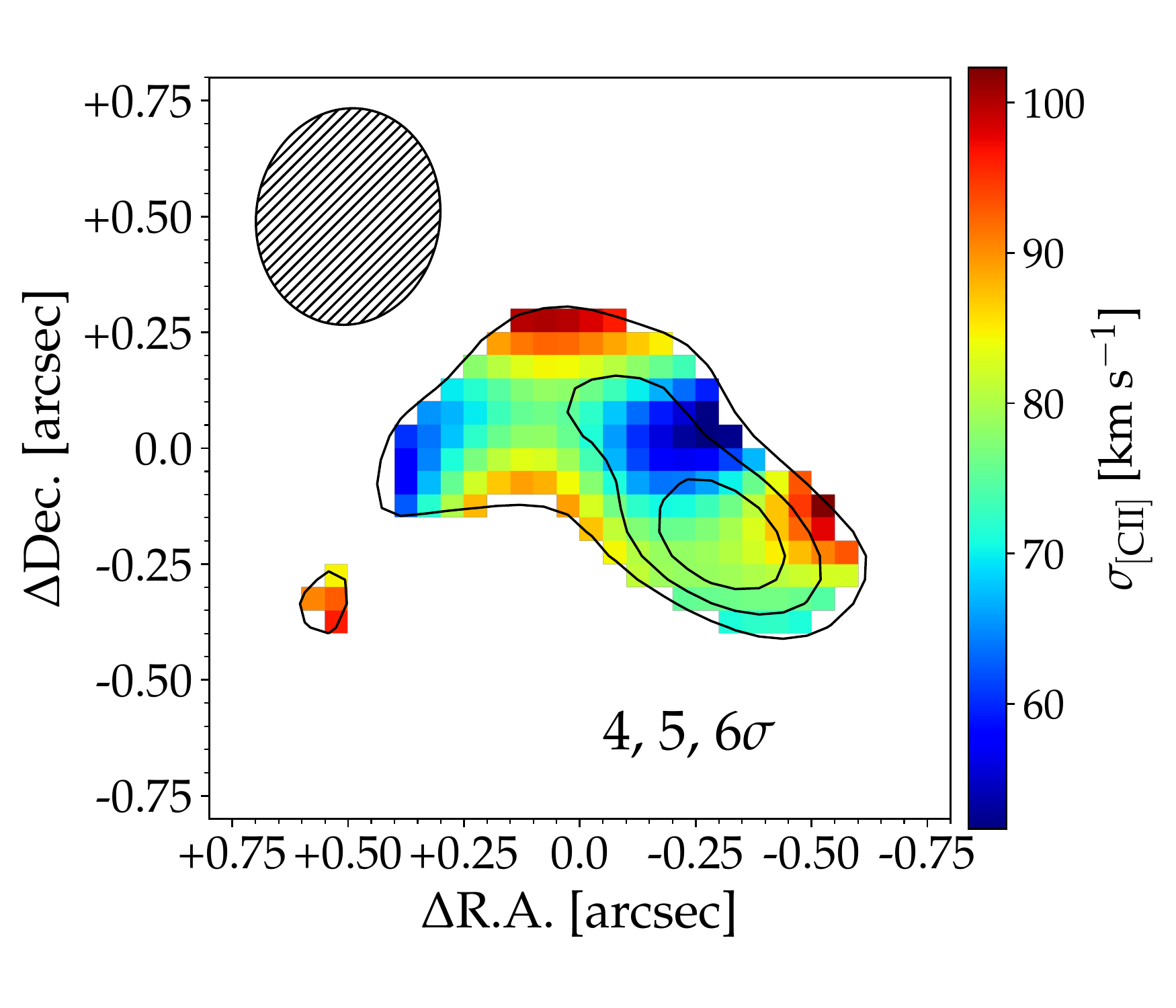}&
\hspace{-0.3cm}\includegraphics[width=6.1cm]{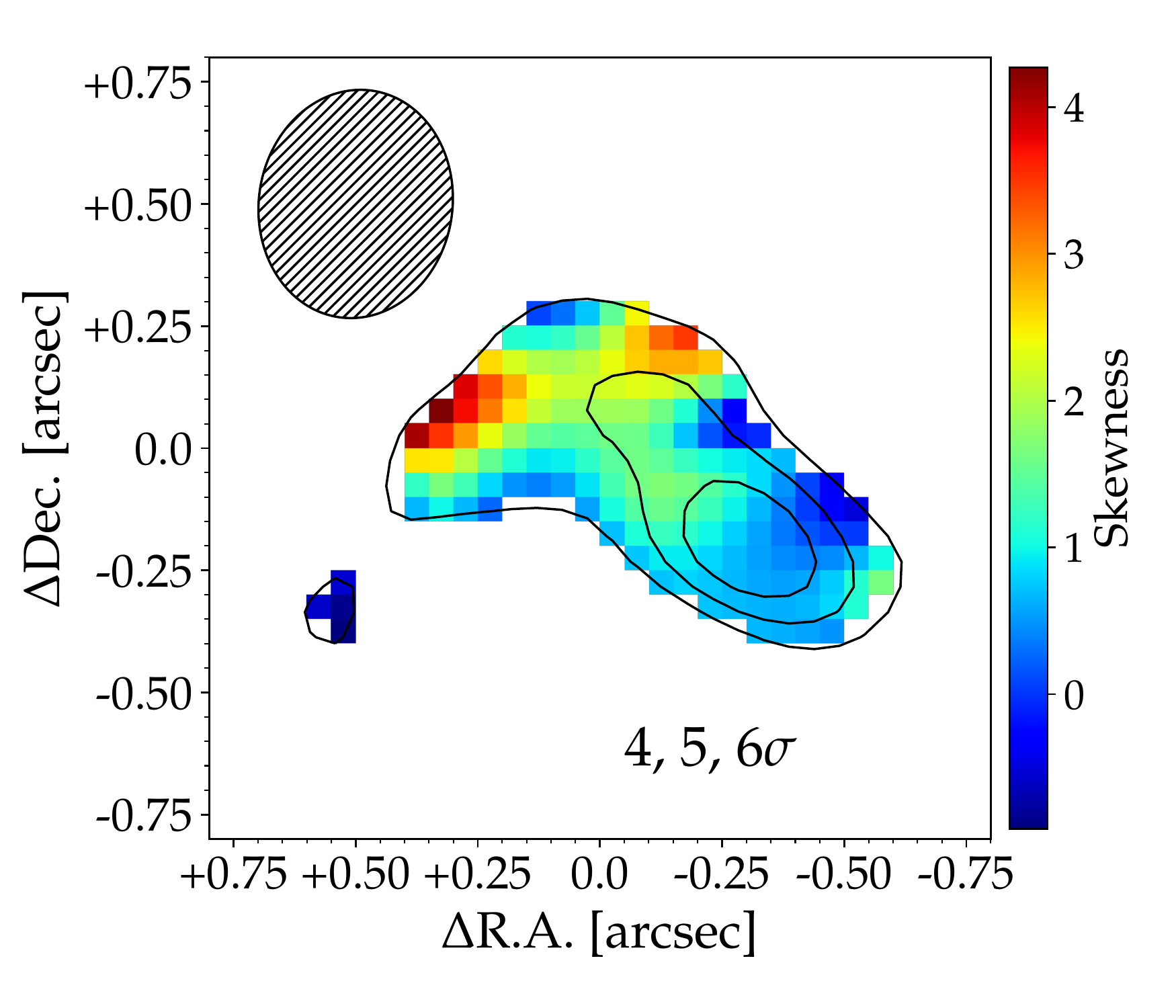}\\
\end{tabular}
\caption{First (left), second (middle) and third (right) moment maps of VR7, based on the highest spatial resolution ALMA reduction and using frequencies within [-100,+200] km s$^{-1}$ with respect to the [C{\sc ii}] redshift $z=6.5285$. These moments correspond to the velocity map, the dispersion map and the skewness map, respectively. The moment maps show complex structure and suggest VR7 is dispersion-dominated with a typical $\Delta v/2\sigma \approx 0.3$ in the highest S/N regions.  \label{fig:VR7_velocity}} 
\end{figure*}

\subsection{Resolved UV-[C{\sc ii}] ratio}
As a result of the differences in the UV and [C{\sc ii}] morphologies, the UV-[C{\sc ii}] ratio differs within the galaxy. We quantify the variation of the UV-[C{\sc ii}] ratio by converting the luminosities to (unobscured) SFR surface densities following \cite{Kennicutt1998} for the UV and the observed relation between [C{\sc ii}] luminosity and SFR in the local Universe \citep{DeLooze2014} and illustrate the results in Fig. $\ref{fig:VR7_SFRUVCII}$.

The average SFR$_{\rm UV}$/SFR$_{\rm [CII]}$ ratio in the region where both UV and [C{\sc ii}] are detected at $3\sigma$ is normal, with a mean ratio $1.3\pm0.1$, but the scatter on $\sim2$kpc scales is large ($0.7$). For example, the lowest SFR$_{\rm UV}$/SFR$_{[CII]}$ ratio, observed in the south-western part of the galaxy, is $\approx0.4$, while the highest ratio of about $\approx2.8$ is only 2.8 kpc away. The region with peak [C{\sc ii}] emission has a relatively low UV luminosity, potentially indicating obscured star formation (although no dust is detected in our ALMA observations, see \S $\ref{sec:FIR}$) and/or a higher metallicity. We note that the UV colors of component 2 also indicate that it is older and more evolved (see Appendix $\ref{app:size}$), although we stress these colors need to be interpreted cautiously. The central-eastern part of the galaxy has a relatively high UV-[C{\sc ii}] ratio, indicating little obscuration and a low metallicity. No UV emission is detected at the position of the tentative [C{\sc ii}] emitting component in the south-east. The outskirts of the galaxy tend to have a relatively high [C{\sc ii}] luminosity, indicating a strong contribution from PDR emission instead [C{\sc ii}] emission originating from H{\sc ii} regions. This is consistent with measurements in local low-metallicity galaxies \citep{Cormier2019} and the Milky Way \citep{Pineda2013}. Most observed [C{\sc ii}] emission originates from PDR regions ($\sim 50$ \%; \citealt{Pineda2013}), but with other significant contributions from molecular hydrogen and cold atomic gas. Only a small fraction  of the [C{\sc ii}] (which decreases with decreasing metallicity; \citealt{Cormier2019}) originates from H{\sc ii} regions. However, note that local galaxies are not affected by a strong CMB background that is present at high redshift.

\subsection{Velocity structure} 
\subsubsection{Moment maps}
Here we explore the velocity structure of VR7 using the first, second and third moment maps of the [C{\sc ii}] emission in the high resolution reduction, which can be interpreted as maps showing the peak velocity, velocity width and line skewness, respectively, see Fig. $\ref{fig:VR7_velocity}$.\footnote{We note that the simple interpretation of the first and second moments may be somewhat confused by strong line skewness. For example, the first moment of a line-emitting region with fixed central velocity throughout the system, but with a red asymmetry in only part of the system, will show a slight gradient following the skewness gradient, as a red skewed line will result in a higher first moment.} 

The first moment map (left panel in Fig. $\ref{fig:VR7_velocity}$) reveals that the peak velocity varies along the $45\degree$ SW axis, the same position angle as component 2. While this pattern somewhat resembles rotation, the maximum velocity difference is only $\Delta v \approx 40$ km s$^{-1}$, which combined with the velocity dispersion of $\sigma \approx65$ km s$^{-1}$ results in a kinematic ratio $\Delta v/2\sigma \approx 0.3$. This indicates a dispersion dominated system \citep[e.g.][]{ForsterSchreiber2009}, unlike two luminous LBGs observed with ALMA in lower spatial resolution by \cite{Smit2017}. 

The second moment map (center panel in Fig. $\ref{fig:VR7_velocity}$) shows variations in the velocity dispersion within a factor $\approx1.5$. The region with highest flux density has relatively low dispersion, while the north-east region has slightly broader line emission. It should be noted that the lowest dispersions in the second moment map are on the order of $\sigma\approx60$ km s$^{-1}$, which are similar to the line-width of the dominant (broader) component identified in our fit to the 1D spectrum. This could indicate that the second, narrow and slightly redshifted component identified from the 1D spectrum (Fig. $\ref{fig:VR7_1D_highres}$) originates mostly from the north-eastern region of the galaxy. This is supported by the line skewness map (right panel of Fig. $\ref{fig:VR7_velocity}$). While the highest skewness values that are found in regions where [C{\sc ii}] is detected at $\approx4\sigma$ significance should be interpreted with caution, the skewness still increases towards the north-east in the regions with high S/N. 

 \subsubsection{Position-velocity diagrams} \label{sec:PV}
\begin{figure}
\includegraphics[width=8.9cm]{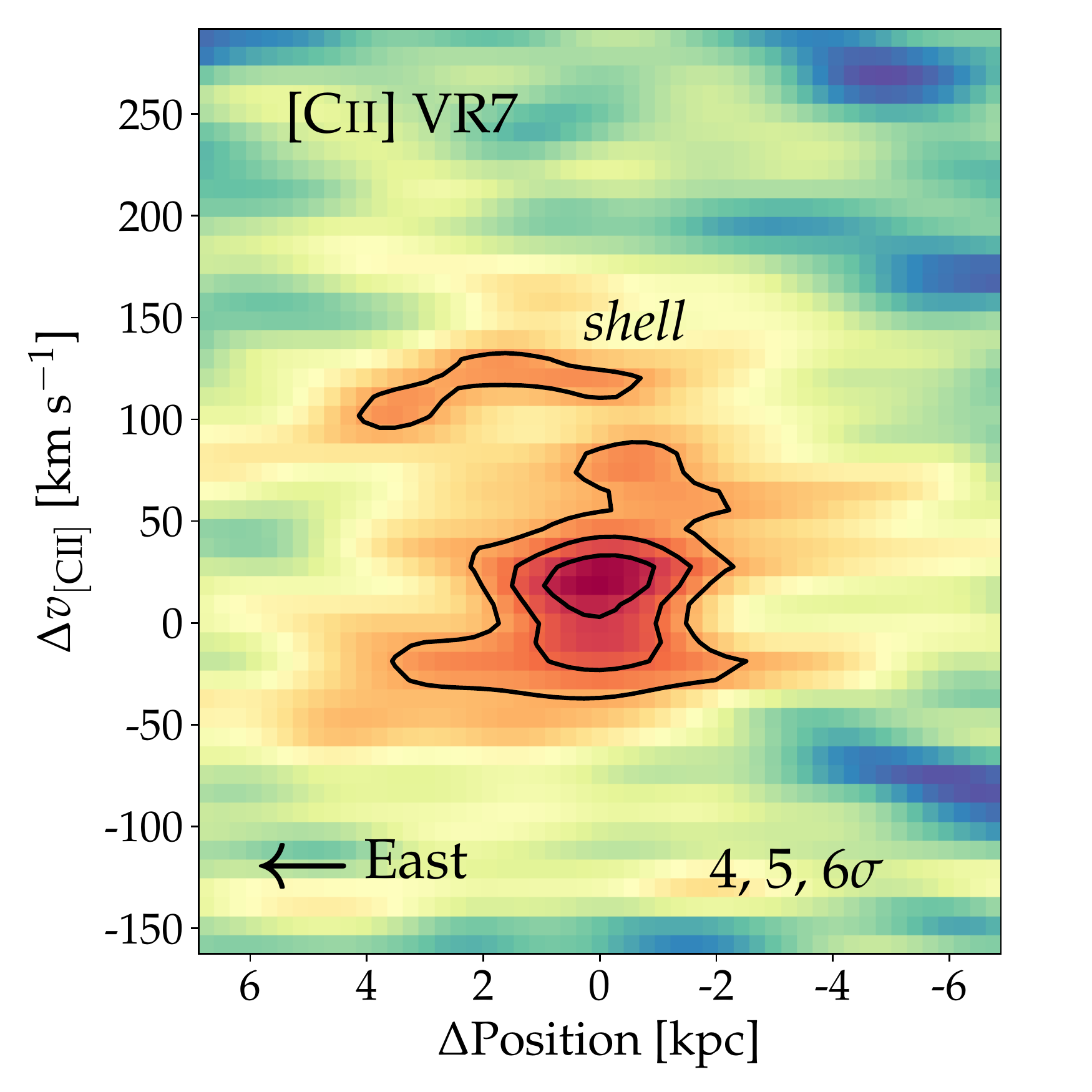}\\
\includegraphics[width=8.9cm]{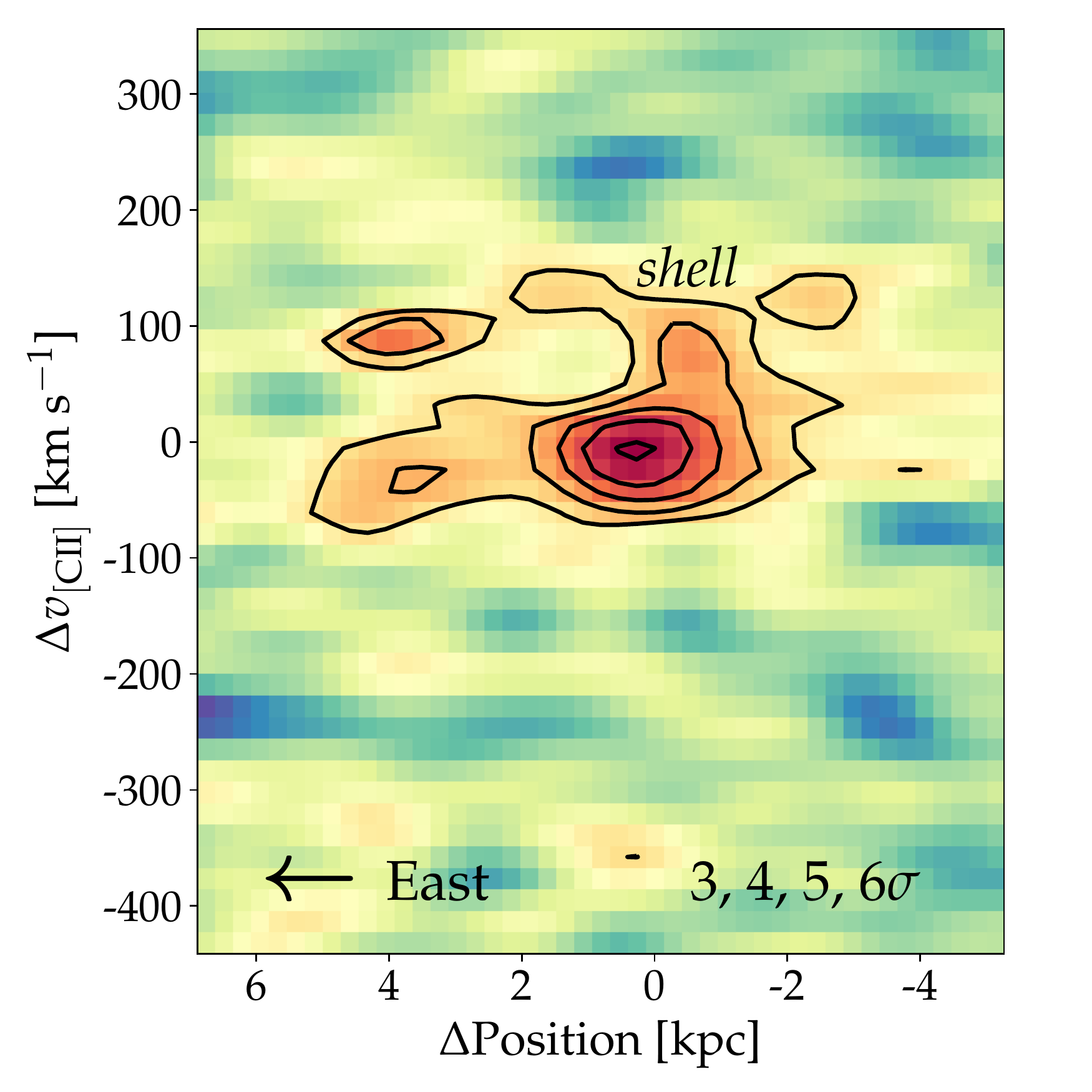}\\
\caption{Position velocity diagrams of VR7. The top row uses an ALMA reduction optimised for high spectral resolution but with low spatial resolution (natural weight, UV taper; as Fig. $\ref{fig:CII_detection_zooms}$). The bottom row uses a high spatial resolution reduction (briggs weight R=0.5; as Fig. $\ref{fig:VR7_zooms}$), but averages over two spectral elements. Flux is averaged over the north-south direction over a 1.5$''$ width. Note that the y-axis scale differs between top and bottom rows. The top-row shows that the [C{\sc ii}] line profile can be resolved in a dominant central component and a faint, diffuse component {\it shell} at a redshift of $\approx100$ km s$^{-1}$. This diffuse component is likely resolved out and at lower significance in the high resolution reduction shown in the bottom row. } 
 \label{fig:VR7_PV}
\end{figure}

We also analyse position-velocity (PV) diagrams in order to study how the [C{\sc ii}] line varies within VR7. We construct PV diagrams by averaging over a slit-width 1.5$''$ around the center in the NS direction (PA=$90\degree$) using the {\sc CASA} task {\sc impv}. Fig. $\ref{fig:VR7_PV}$ shows the results for a reduction optimised for high spectral resolution (top row) and high spatial resolution (bottom row), where the first is achieved by reducing the data with normal weighting and UV tapering, while the latter is achieved with briggs weighting, but averaging over two velocity channels. 

Fig. $\ref{fig:VR7_PV}$ confirms the result from the 1D analysis, which is that the [C{\sc ii}] line profile is composed of (at least) two components: a central dominant component that is relatively compact and a redshifted extended component that has a narrower line-width, that we name the {\it shell} in the rest of the paper. The significance of the shell is $>4\sigma$ in the reduction with low spatial resolution, but lower in the reduction with higher spatial resolution, which indicates that the [C{\sc ii}] emission is somewhat diffuse and resolved out in the latter reduction. The shell is redshifted by $\approx100$ km s$^{-1}$ compared to the central component and is therefore consistent with the narrow gaussian component identified from the integrated 1D spectrum (see Table $\ref{tab:twogauss_properties}$). As the shell is chemically enriched, it is more likely to be outflowing than inflowing. While the flux from the shell extends over roughly the same spatial scales as the central component, the center is shifted slightly towards the north-east, consistent with the skewness map in Fig. $\ref{fig:VR7_velocity}$. It is challenging to simultaneously show collapsed images of the core [C{\sc ii}] emission and the shell because of the small separation and the extended nature of the shell. The positions of peak [C{\sc ii}] flux are therefore indicated with crosses in Fig. $\ref{fig:VR7_zooms}$, which shows that the separation is $\approx0.35''$, or $\approx2$ kpc.

Isolating the dominant component in the bottom panel of Fig. $\ref{fig:VR7_PV}$, we measure a size FWHM ranging from 0.35-0.50$''$. As the beam major-axis is 0.46$''$, this means that the central component is unresolved and has half-light radius $\lesssim1.3$ kpc. Combined with the line-width of FWHM=$145^{+41}_{-33}$ km s$^{-1}$, this results in an estimate of the dynamical mass M$_{\rm dyn}$/(sin $i$)$^2 \lesssim5.5\pm3\times10^{9}$ M$_{\odot}$ \citep{Wang2013,Matthee2017ALMA}. Besides the dominant component, the bottom panel of Fig. $\ref{fig:VR7_PV}$ also hints that there is a second fainter compact [C{\sc ii}] emitting component at an offset of about 4 kpc $\approx 0.7''$ to the east (see also Fig. $\ref{fig:VR7_zooms}$, which shows that this component overlaps with one of the UV components). The peak flux of this component is a factor two lower than the dominant component and is detected at only $\approx3-4\sigma$ significance. 

\section{Discussion} \label{sec:discussion}
We now focus on interpreting the ALMA and {\it HST} observations and place them into the context of other observations. We first address the non-detections of [C{\sc ii}] and IR continuum in MASOSA and discuss ways to reconcile this with the potentially red color in the UV (\S $\ref{sec:discuss_MASOSA}$). Secondly, we discuss the nature of VR7 based on current data in \S $\ref{discuss:VR7}$ and we finally discuss the relation between [C{\sc ii}] and UV emission in \S $\ref{sec:discuss_CIIUV}$. 

\subsection{The nature of MASOSA: dusty? Metal poor?} \label{sec:discuss_MASOSA}
MASOSA is not detected in [C{\sc ii}] or FIR continuum emission. This may be explained by a higher dust temperature combined with a low dust mass and a relatively low metallicity. Such properties are indeed expected given MASOSA's high-redshift and very high Ly$\alpha$ EW (typically found in young, metal poor and dust-free galaxies; e.g. \citealt{Sobral2015,Nakajima2016,Trainor2016,Sobral2018}). 

\begin{figure*}
\begin{tabular}{cc}
\includegraphics[width=8.9cm]{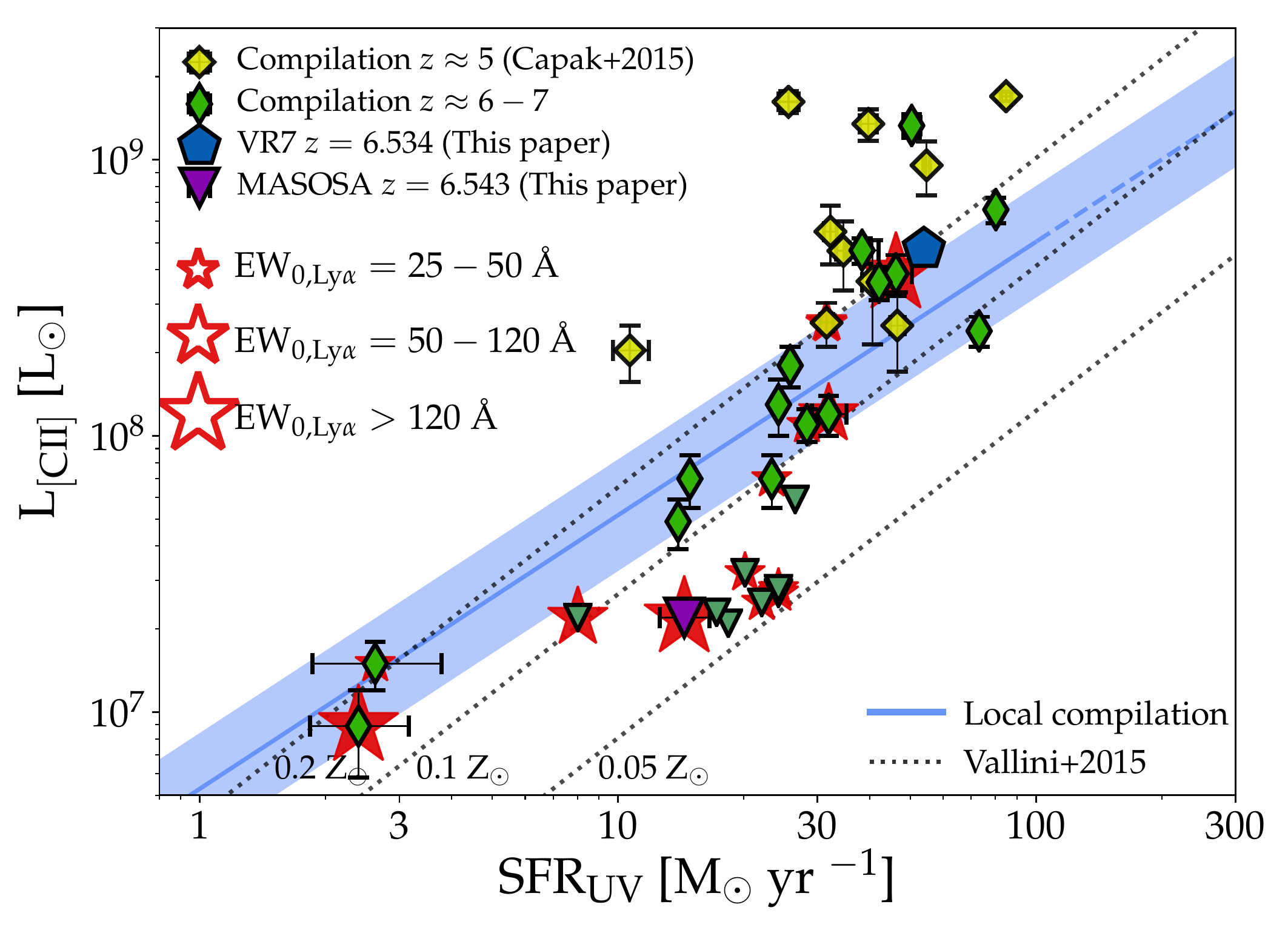}&
\hspace{-0.4cm}\includegraphics[width=8.9cm]{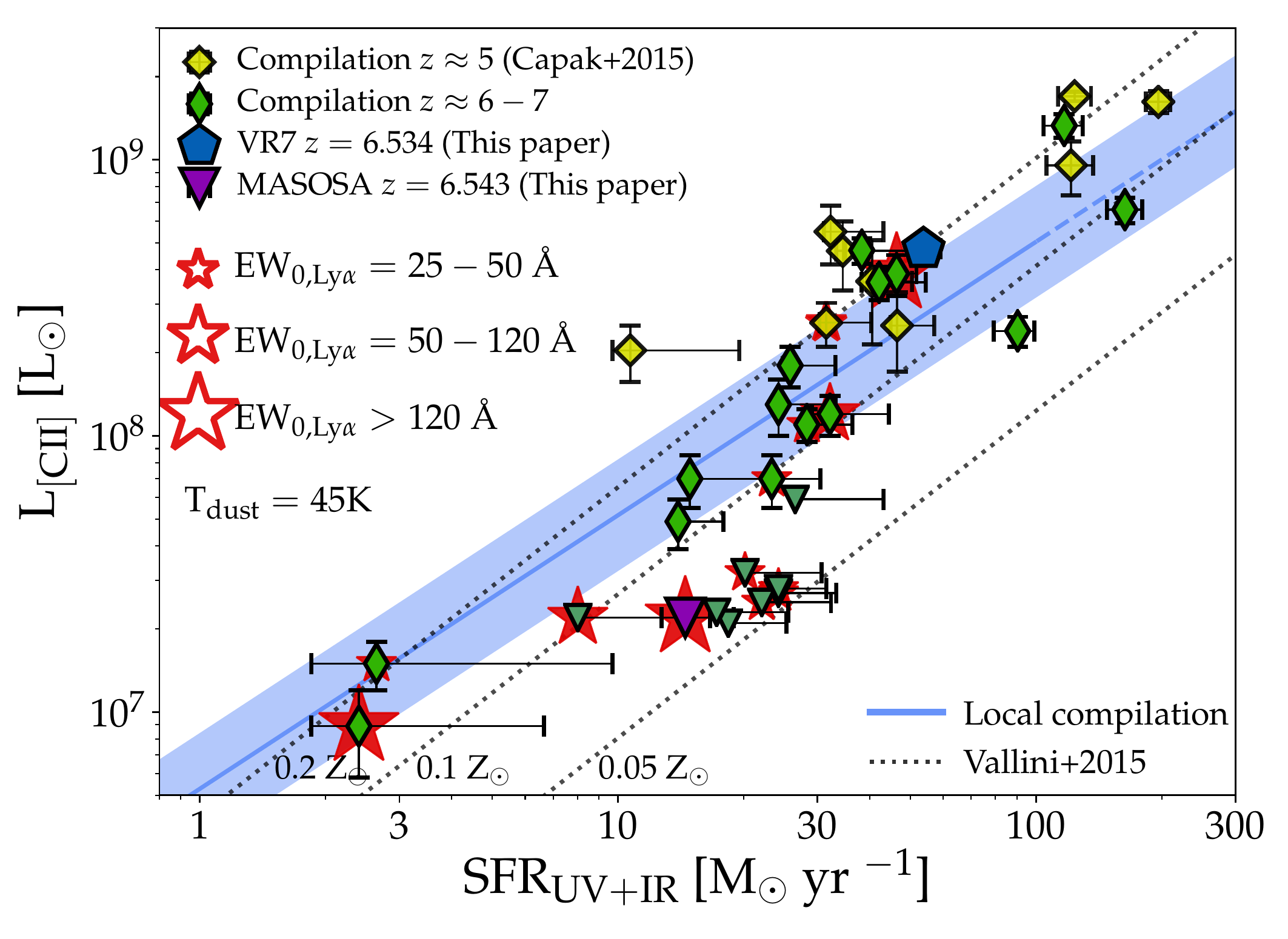}
\end{tabular}
\caption{The relation between SFR$_{\rm UV}$ and [C{\sc ii}] luminosity at $z\approx6-7$. The {\it left} panel shows SFR$_{\rm UV}$ estimated from the rest-frame UV luminosity following \citet{Kennicutt1998}. The {\it right} panel shows SFR$_{\rm UV+IR}$ where we have converted IR continuum detections to SFR$_{\rm IR}$ assuming a dust temperature of 45 K and assign the 1$\sigma$ limiting SFR$_{\rm IR}$ as an upper error on the SFR as in \S $\ref{sec:IR_SFR}$. We compare our integrated luminosity measurement for VR7 (blue pentagon) and the upper limit for MASOSA (purple triangle) to other galaxies observed at $z\approx5$ and $z\approx6-7$ (yellow green diamonds; all computed consistently, see Appendix $\ref{appendix:compilation}$) and to the relation observed in the local Universe (\citealt{DeLooze2014}; blue band; where the shaded region shows the dispersion, the line-style changes to dashed at the luminosities were relations are extrapolated and the relation is shifted to the same IMF as used here). Upper limits are at the 1$\sigma$ level. We indicate galaxies with strong Ly$\alpha$ lines with red stars. Dotted lines show the relation between [C{\sc ii}] luminosity and SFR at different gas-phase metallicities inferred from hydrodynamical simulations by \citet{Vallini2015}, which used the same IMF as our analysis.} 
 \label{fig:SFRUV_LCII}
\end{figure*}

As illustrated in Fig. $\ref{fig:SFRUV_LCII}$, the [C{\sc ii}] luminosity limits implies that MASOSA is significantly [C{\sc ii}]-deficient given its SFR, compared to the typical [C{\sc ii}] luminosity found for star-forming galaxies in the local Universe \citep[e.g.][]{DeLooze2014}. This can be explained by a lower gas-phase metallicity \citep[e.g.][]{Vallini2015,Olsen2017}. Following the relations derived by \cite{Vallini2015} based on radiative hydrodynamical simulations, the [C{\sc ii}] upper limit in MASOSA implies a gas-phase metallicity $\lesssim0.07$ Z$_{\odot}$ (see Fig. $\ref{fig:SFRUV_LCII}$) providing further evidence for the very low metallicity of this source as discussed in \cite{Sobral2015}. An additional rough metallicity-estimate can be derived based on {\it Spitzer}/IRAC photometry. For the redshift of MASOSA, the [3.6] band is contaminated by H$\beta$+[O{\sc iii}] line-emission, while the [4.5] band includes H$\alpha$ emission \citep[e.g.][]{SchaererBarros2009,Raiter2010IRAC}. Assuming a flat continuum in the rest-frame optical, MASOSA blue IRAC color ($[3.6]-[4.5]=-0.6\pm0.2$) indicates an EW(H$\beta$+[O{\sc iii}])$\approx1500$ \citep[e.g.][]{Smit2014} and hence a gas-phase metallicity $\approx0.01-0.4$ Z$_{\odot}$ \citep{Castellano2017}. More detailed constraints on the gas-phase metal abundances in the ISM of MASOSA can be obtained from future deep (near-)infrared spectroscopy.

MASOSA's possible red UV slope, $\beta=-1.06^{+0.68}_{-0.72}$, is relatively unexpected given the UV luminosity ($\beta\approx-2.1$; e.g. \citealt{Bouwens2012} and Fig. $\ref{fig:MUV_BETA}$), the high Ly$\alpha$ EW \citep{Hashimoto2016}, the blue IRAC colors and the non-detection of dust continuum. Besides dust attenuation that we discussed in \S $\ref{sec:FIR}$\footnote{Any significant amount of dust-obscuration in MASOSA would also imply a lower metallicity in order to explain the non-detection of the [C{\sc ii}] line.}, a red UV slope could be reconciled with a young age in case of a combination of strong nebular continuum emission, strong UV emission lines besides and/or multiple clumps.

For example, \cite{Raiter2010} show that strong nebular continuum free-free emission can redden the observed UV slope of low metallicity galaxies with hard ionising spectra up to $\beta\approx-1.5$ at $\lambda_0=1500$ {\AA}. The most likely emission line boosting the F160W magnitude (and hence the observed $\beta$) is C{\sc iii}]$_{1909}$. C{\sc iii}]$_{1909}$ is the strongest UV emission line in star-forming galaxies besides Ly$\alpha$ \citep[e.g.][]{Shapley2003} and can have EW$_{0}\approx10-20$ {\AA} \citep[e.g.][]{Rigby2015,Stark2015_CIII,Maseda2017}, particularly in galaxies with strong Ly$\alpha$ emission \citep[e.g.][]{LeFevre2017}. However, we calculate that correcting for a C{\sc iii}] EW of $25$ {\AA} would imply a slope to $\beta=-1.2\pm0.7$, meaning that exceptionally strong lines would be required \citep[see also][]{Stroe2017}. Moreover, MASOSA has a slightly different morphology in the reddest {\it HST} band (e.g. Fig. $\ref{fig:thumbs}$ and Fig. $\ref{fig:app_MASOSA}$), indicating the presence of a second, relatively faint red clump. Such a clump may be physically disconnected from the youngest stellar population. By fitting the F160W data with a combination of two point sources\footnote{The F160W data of MASOSA is described equally well by a single component within the uncertainties, see Appendix $\ref{app:size}$, but we use the results from a two component fit to address the implication of this hypothesis.}, we find that the red clump contributes $\approx20$ \% of the flux. Reducing the F160W flux by 20 \% results in $\beta=-1.5\pm0.6$, meaning that the main component of MASOSA can plausibly be `normally' blue. 

Concluding, MASOSA has common properties for galaxies with strong Ly$\alpha$ emission (relatively compact, low metallicity and dust content) and is likely a young star bursting galaxy, with little evidence for being powered by an AGN. While MASOSA's red UV color needs to be verified with deeper observations, a combination of a possible second component, relatively strong rest-frame UV lines such as C{\sc iii}] and strong nebular continuum emission could reconcile MASOSA's relatively red color with being a young and relatively dust free galaxy. 

\subsection{The nature of VR7: hot dust? merger?} \label{discuss:VR7}
VR7 is a large UV luminous star-forming galaxy with relatively strong Ly$\alpha$ emission, relatively normal [C{\sc ii}] emission. Remarkably, VR7 is the most UV-luminous SFG at $z\approx6-7$ for which no dust emission is detected at 160$\mu$m. The limiting 160$\mu$m/1500{\AA} flux density ratio is extremely low and in the local Universe only seen for extremely metal poor systems as I Zw 18 (Fig. $\ref{fig:IRUV}$). The Ly$\alpha$ EW implies a Ly$\alpha$ escape fraction of $\approx15-20$ \% \citep{SM2019}, which indicates that a significant fraction of Ly$\alpha$ emission is destroyed by dust. Moreover, the moderate color $\beta\approx-1.4$ also indicates dust attenuation. Therefore, it is likely that dust is present in VR7 at high temperatures (T$>60$ K), which would also reconcile the limits on the 160$\mu$m continuum flux density with common IRX-$\beta$ relations (Fig. $\ref{fig:IRXbeta}$ and e.g. \citealt{Faisst2018}).
The non-detection of IR continuum emission, the narrowness of the [C{\sc ii}] line and the UV morphology all indicate that VR7 is not powered by an AGN \citep[e.g.][]{Ota2014}.

In \S $\ref{sec:resolvedVR7}$, we showed that the [C{\sc ii}] and rest-frame UV images clearly resolve the light from VR7, with clear indications of multiple components even though they are separated by only $\approx0.3-0.4''$. This separation is a factor 2-3 smaller than the separation of individual components in other luminous Ly$\alpha$ emitter at $z\approx6.5$ \citep[e.g.][]{Matthee2017ALMA,Carniani2018}. Whether the components that are building up the VR7 galaxy are already at a smaller physical separation compared to other luminous LAEs (i.e. VR7 is in a more advanced stage of a merger), or whether this is a viewing angle/inclination effect, can not be addressed with the current data. Interestingly, among the luminous LAEs targeted with ALMA so far, VR7 has the lowest Ly$\alpha$ EW, which could indicate it is in a slightly more evolved stage than LAEs with higher Ly$\alpha$ EW. A slightly more evolved stage is also in line with its moderate UV slope, particularly since the other luminous LAEs have $\beta\approx-2$.

\subsection{What determines the [C{\sc ii}]/UV ratio} \label{sec:discuss_CIIUV}
Here, we address the relative [C{\sc ii}] and UV luminosities of galaxies at $z\approx6-7$, compared to galaxies in the local Universe. Earlier work either found relatively high \citep{Capak2015}, moderately low \citep{Pentericci2016} or extremely low [C{\sc ii}] luminosities at fixed SFR$_{\rm UV}$ \citep[e.g.][]{Ota2014}. As detailed in Appendix $\ref{appendix:compilation}$, we homogenise the measurements by converting all UV luminosities to SFR$_{\rm UV}$ consistently and we revise upper limits following the method applied to MASOSA. Our results are compared to relations between [C{\sc ii}]  luminosity and SFR based on observations of galaxies in the local Universe \citep{DeLooze2014}. Like our observations, we rescale the relation to the common SFR calibration of \cite{Kennicutt1998}. The results are shown in Fig. $\ref{fig:SFRUV_LCII}$, where the left panel uses SFR estimated from only the UV, while the right panel uses the information on the dust continuum provided by ALMA to correct for obscured SFR.
 
While VR7 has an integrated [C{\sc ii}]-SFR$_{\rm UV}$ ratio that is very similar of other galaxies with similar SFR\footnote{We note that the global SFR$_{\rm UV}$ and [C{\sc ii}] surface densities of VR7 also agree with galaxies in the local Universe observed by \cite{DeLooze2014}.}, MASOSA is [C{\sc ii}] deficient, similar to several other galaxies with SFR$_{\rm UV} \approx10-20$ M$_{\odot}$ yr$^{-1}$. There is significant dispersion in [C{\sc ii}] luminosities at fixed SFR$_{\rm UV}$, particularly in the range SFR$\approx30-60$ M$_{\odot}$ yr$^{-1}$ that includes most of the objects currently observed. A part of this dispersion is related to variation in relative strengths of Ly$\alpha$ emission, since the SFR-[C{\sc ii}] ratio is anti-correlated with the observed Ly$\alpha$ EW \citep[e.g.][]{Carniani2018,Harikane2018}. Indeed, a large fraction of objects with [C{\sc ii}] deficits are strong Ly$\alpha$ emitters (Fig. $\ref{fig:SFRUV_LCII}$). Furthermore, among the sample of luminous LAEs known at $z\approx6-7$, MASOSA is at the higher end of the Ly$\alpha$ EW distribution (e.g. Fig. $\ref{fig:MUV_LYA}$), in agreement with this picture. On the other hand, Fig. $\ref{fig:SFRUV_LCII}$ shows that several other strong Ly$\alpha$ emitters, such as CR7 \citep{Matthee2017ALMA}, have rather typical [C{\sc ii}]  luminosities given the SFR. 

In the range SFR$_{\rm UV+IR}\approx30-200$ M$_{\odot}$ yr$^{-1}$, the detected galaxies at $z\approx6-7$ are roughly co-located with galaxies in the local Universe and follow a similar slope log$_{10}$(L$_{\rm [CII]})\propto$ log$_{10}$(SFR) once obscured SFR is accounted for. At SFRs below 30 M$_{\odot}$ yr$^{-1}$ the results are less clear, as the [C{\sc ii}]  line is undetected in a large fraction of galaxies, with a 1$\sigma$ limiting luminosity well below the local relation. 

If relatively faint [C{\sc ii}] lines are confirmed in these systems, it would imply that the relation between [C{\sc ii}]  luminosity and SFR steepens at low SFRs and that a larger fraction of the ISM is ionised. This potentially indicates a stronger correlation between gas-phase metallicity and SFR in lower mass galaxies. Alternatively, it could also indicate more bursty star formation in fainter galaxies (leading to a higher ionisation parameter) or other differences in ISM properties. Such a steepening is observed in the local Universe using a sample of low-metallicity dwarf galaxies \citep{DeLooze2014}. Furthermore, such steepening would also explain why relatively low [C{\sc ii}] luminosities are found among most strong Ly$\alpha$ emitters, as the fraction of galaxies with strong Ly$\alpha$ emission is higher at low SFRs in the observed sample (Fig. $\ref{fig:SFRUV_LCII}$). This picture simultaneously allows higher [C{\sc ii}] luminosities in the most luminous Ly$\alpha$ emitters found in the large wide-field surveys \citep[e.g.][]{Matthee2017ALMA,Carniani2018}.

\section{Conclusions} \label{sec:conclusions}
In this paper, we have presented new deep follow-up observations of two luminous Ly$\alpha$ emitters at $z\approx6.5$ with {\it HST}/WFC3 and ALMA with a resolution of 1.5-2 kpc. The {\it HST} data are used to characterise the strength and morphology of rest-frame UV emission, constraining the un-obscured SFR. ALMA data are used to constrain the level of FIR dust continuum emission and measure the strength and dynamics of the atomic [C{\sc ii}]$_{158\mu \rm m}$ fine-structure line. The targeted galaxies, VR7 and MASOSA \citep{Sobral2015,Matthee2017SPEC}, have a high $\approx 2\times L^{\star}$ Ly$\alpha$ luminosity, while their UV luminosity differs by a factor four. Table $\ref{tab:global_properties}$ summarises our measurements. Our main results are the following:

\begin{enumerate}
\item VR7 is among the most luminous and largest star-forming galaxies known at $z\approx6-7$ with an unobscured SFR=54$^{+3}_{-2}$ M$_{\odot}$ yr$^{-1}$ and an extent of $\approx 6$ kpc at the 2$\sigma$ level in the UV. VR7 has a typical moderate color given its high UV luminosity, $\beta=-1.38^{+0.29}_{-0.27}$, and moderate Ly$\alpha$ EW$_0=34\pm4$ {\AA}. The {\it HST}/WFC3 rest-frame UV data of VR7 strongly prefers a two-component exponential model over a single component model. The individual components are separated by $1.88\pm0.07$ kpc and have sizes r$_{\rm eff} = 0.84\pm0.11$ kpc and r$_{\rm eff} = 1.12\pm0.06$ kpc. The larger component is a factor $1.8^{+0.7}_{-0.5}$ more luminous and is elongated in the south-western direction. 

\item MASOSA, which has a $\approx$L$^{\star}$ UV  luminosity (unobscured SFR=15$^{+2}_{-2}$ M$_{\odot}$ yr$^{-1}$) and a normal size (r$_{\rm eff}=1.12^{+0.44}_{-0.19}$ kpc), is among the galaxies with highest Ly$\alpha$ EW known at $z\approx6-7$ (EW$=145^{+50}_{-43}$ {\AA}). Surprisingly, MASOSA seems to have a relatively red UV continuum color $\beta=-1.06^{+0.68}_{-0.72}$, which could indicate strong C{\sc iii}] line-emission, nebular continuum emission and/or a multiple component nature. 

\item No dust continuum emission is detected in either VR7 or MASOSA (\S $\ref{sec:IR_SFR}$). Unless the dust temperature is very high (T$_{\rm dust}\gtrsim60$K), this indicates little amounts of dust present in both systems. VR7 has an extremely low L$_{\rm IR}$/L$_{\rm UV}$ ratio, similar to the ratio of the extremely metal poor dwarf galaxy I Zw 18 and the luminous LAE CR7 \citep{Matthee2017ALMA}. Assuming T$_{\rm dust}=45$K, we find that the obscured SFR in both galaxies is a small fraction of the unobscured SFR ($\lesssim10$ \% and $\lesssim30$ \%, respectively). 

\item Our ALMA data reveals a strong [C{\sc ii}] emission detection in VR7 (S/N=15), but none in MASOSA (\S $\ref{sec:CII}$). VR7 has a typical integrated [C{\sc ii}] luminosity relative to its SFR, indicating a metallicity $\approx0.2$ Z$_{\odot}$. For MASOSA, we derive a physically motivated [C{\sc ii}] upper limit that results in a [C{\sc ii}] deficit given the SFR$_{\rm UV}$ and that implies a very low metallicity $<0.07$ Z$_{\odot}$. Combining our measurements with a large compilation, we find indications that the relation between [C{\sc ii}] luminosity and SFR steepens at SFRs $\lesssim30$ M$_{\odot}$ yr$^{-1}$, which could be explained due to a tighter correlation between SFR and metallicity in these galaxies compared to more luminous systems. Such a correlation naturally explains why most galaxies with high Ly$\alpha$ EW have relatively low [C{\sc ii}] luminosity.

\item {[C{\sc ii}]} emission in VR7 is extended (major/minor axes $(7.8\pm0.8) \times (3.8\pm0.4)$ kpc$^2$), a factor $\approx1.8$ larger than the UV emission. The faint and low surface brightness [C{\sc ii}] emission extends mostly in the east-west direction. The high surface brightness [C{\sc ii}] emission also extends 45$\degree$ towards to south-west (\S $\ref{sec:resolvedVR7}$). We find that the UV-[C{\sc ii}] ratio varies significantly within the galaxy (up to factors of $\approx7$ on $\approx 3$ kpc scales). The dominant [C{\sc ii}] emitting component is compact, with an estimated half-light radius $\lesssim 1.3$ kpc and being slightly offset from the peak UV emission. We find that some faint, diffuse [C{\sc ii}] emission is redshifted by $+100$ km s$^{-1}$ with respect to the central component, possibly indicating an outflow and explaining the skewness and asymmetry seen in the 1D spectrum. Finally, we find indications for a second compact component at a 4 kpc separation.
\end{enumerate}

Our observations show that there is little dust present in luminous Ly$\alpha$ emitters at $z\approx6-7$, unless the dust has very high temperatures ($\gtrsim60$ K). Future ALMA surveys should therefore aim at observationally constraining these temperatures by performing deep observations at higher frequencies. There are large variations in the [C{\sc ii}] luminosities of galaxies at $z\approx6-7$ which are strongly related to the SFR$_{\rm UV}$, albeit with significant scatter. Deep ALMA observations of galaxies with SFR$_{\rm UV}<10$ M$_{\odot}$ yr$^{-1}$ (M$_{1500}\gtrsim-20$) should test whether the relation between [C{\sc ii}] luminosity and SFR$_{\rm UV}$ indeed steepens at low SFRs. ALMA and {\it HST} are capable of resolving the most luminous systems at $z\approx6-7$ and reveal that they are actively assembling from multiple components with strong variations in the SFR$_{\rm UV}$-[C{\sc ii}] ratio, while their environment is already being influenced by galactic outflows. In the future, resolved rest-frame optical spectroscopy using integral field unit spectroscopy on {\it JWST} will provide crucial complementary measurements of the ISM metallicity, the contribution from non-thermal emission and the masses, ages and metallicities of stellar populations.

\acknowledgments
We thank an anonymous referee for constructive comments and suggestions. We thank Max Gronke for comments on an earlier version of this paper. LV acknowledges funding from the European Union's Horizon 2020 research and innovation program under the Marie Sk\l{}odowska-Curie Grant agreement No. 746119. This paper makes use of the following ALMA data: ADS/JAO.ALMA\#2017.1.01451.S. ALMA is a partnership of ESO (representing its member states), NSF (USA) and NINS (Japan), together with NRC (Canada) and NSC and ASIAA (Taiwan) and KASI (Republic of Korea), in cooperation with the Republic of Chile. The Joint ALMA Observatory is operated by ESO, AUI/NRAO and NAOJ. Based on observations obtained with the Very Large Telescope, programs: 294.A-5018, 097.A-0943 \& 99.A-0462. Based on observations made with the NASA/ESA Hubble Space Telescope, obtained [from the Data Archive] at the Space Telescope Science Institute, which is operated by the Association of Universities for Research in Astronomy, Inc., under NASA contract NAS 5-26555. These observations are associated with program \#14699.

\clearpage
%

\vspace{5mm}
\facilities{ALMA, {\it HST} (WFC3), VLT (FORS2, MUSE, X-SHOOTER).}

\software{astropy, \citep{Astropy}, imfit \citep{Erwin2015}, matplotlib \citep{Hunter2007}, Topcat \citep{Topcat}, Scipy \citep{Scipy}, Swarp \citep{Bertin2010}}




\appendix

\setcounter{figure}{0} \renewcommand{\thefigure}{A.\arabic{figure}}
\setcounter{table}{0} \renewcommand{\thetable}{A.\arabic{table}}

\section{Continuum detections in the ALMA data}
\begin{figure*}
\begin{tabular}{cc}
\includegraphics[width=8.3cm]{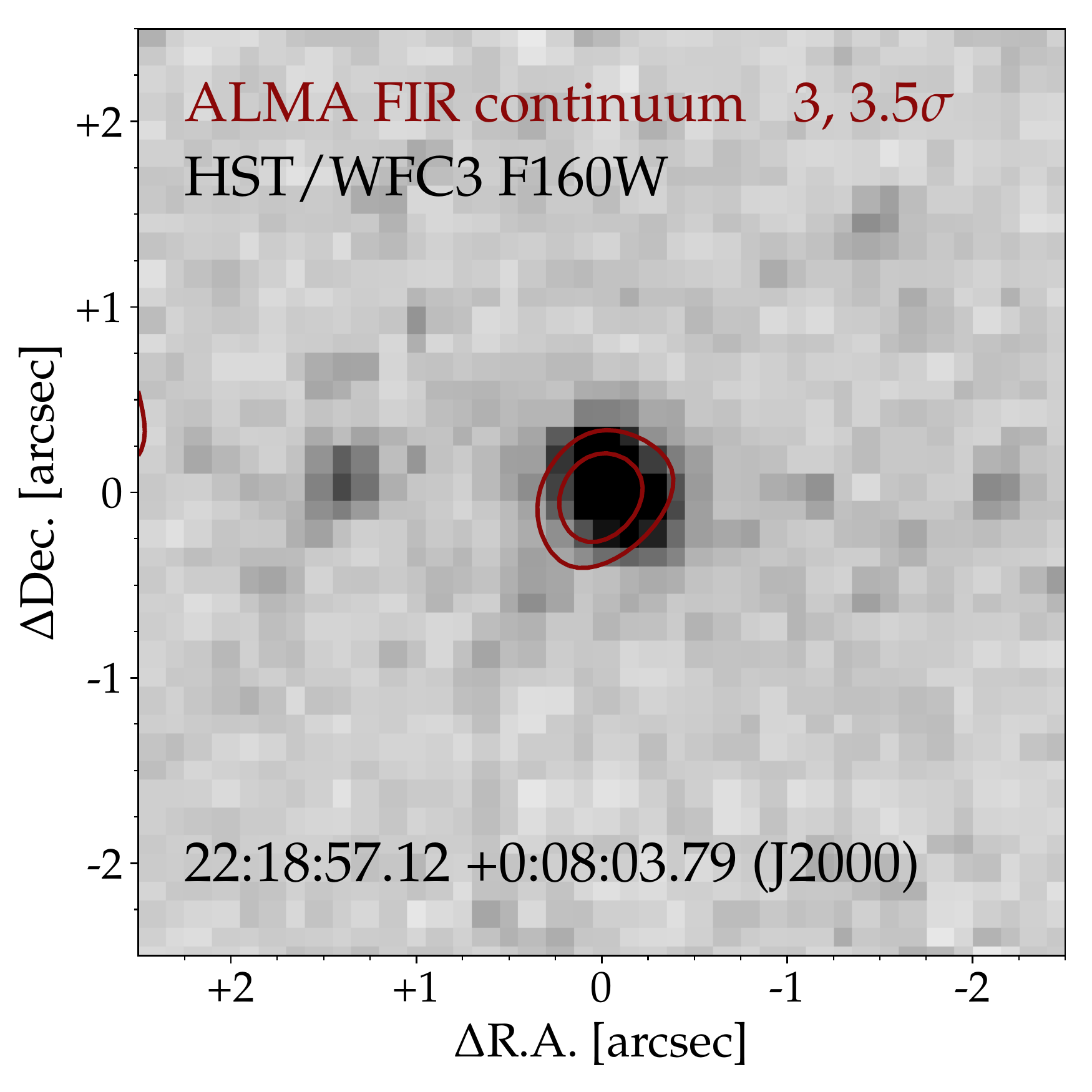} &
\includegraphics[width=8.3cm]{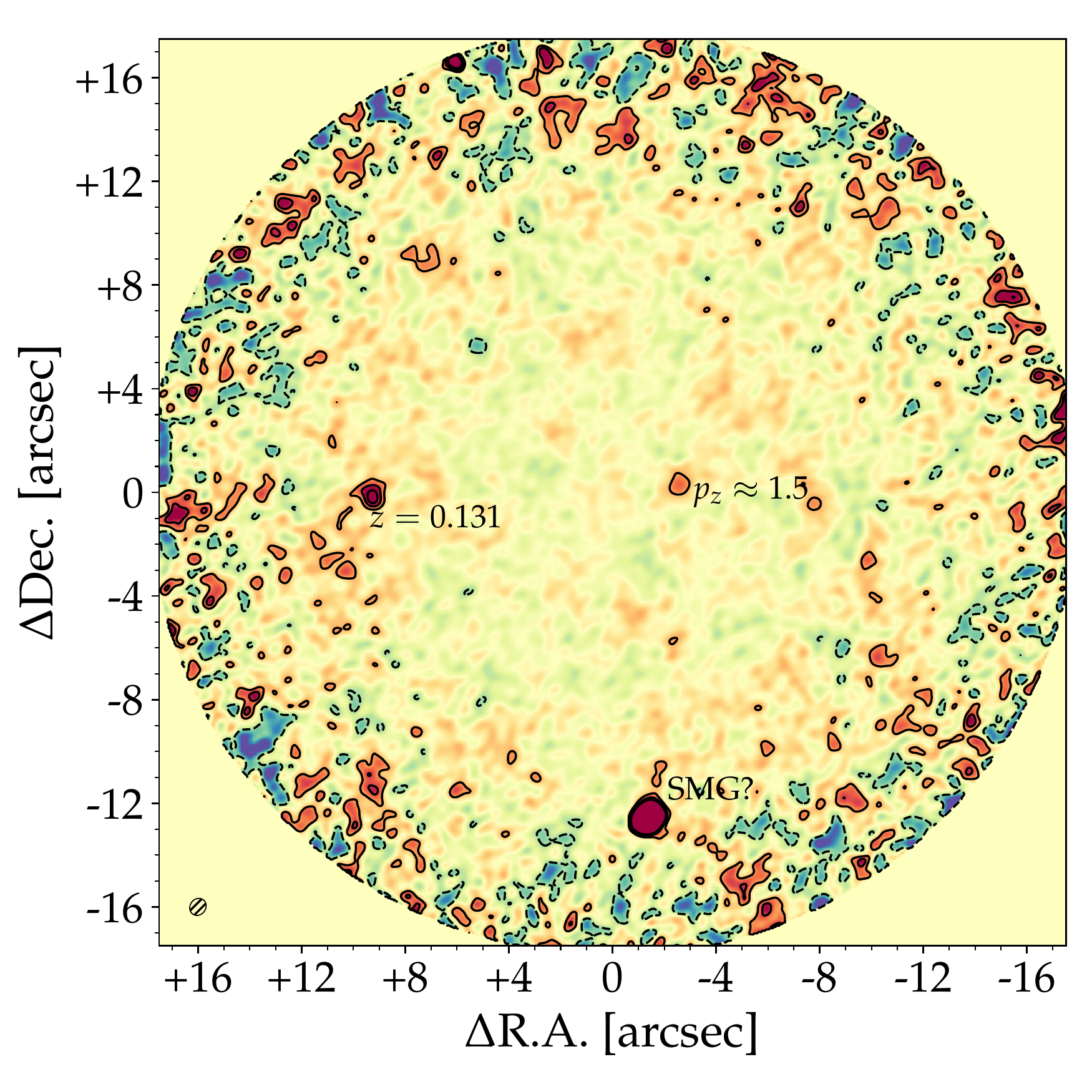}
\end{tabular}
\caption{Continuum maps of foreground sources in the field of VR7 (left) and MASOSA (right). For VR7 we zoom in on the foreground source detected with S/N=4.2 and show that the IR continuum contours coincide well with the {\it HST}/WFC3 position. For MASOSA we show our full primary-beam corrected image and highlight the positions of the three foreground sources, where contour levels are at the 4, 8, 12$\sigma$ level. \label{fig:foregrounds}}
\end{figure*}

Besides our targets, our continuum maps reveal a number of detections, see Fig. $\ref{fig:foregrounds}$. Three objects are detected in the MASOSA pointing with S/N ranging from 5-56, while two objects with S/N 3.5-4.2 are (tentatively) detected in the VR7 pointing, see Table $\ref{tab:foreground_properties}$ for details. 
In the MASOSA field, the highest S/N detection is likely a dusty and highly obscured star-forming galaxy, as it is not detected in optical imaging (limiting AB magnitudes $\sim28$) nor in the {\it HST}/WFC3 data presented here. The object is detected in the MIR {\it Spitzer}/IRAC [3.6] and [4.5] bands, showing it is very red. No [C{\sc ii}] or any other emission line is detected in the ALMA data, indicating it is not at similar redshift as MASOSA. The second most luminous continuum detection coincides with the centre of a face-on spiral galaxy with an SDSS-redshift of $z=0.131$. The third continuum detection (with a S/N=5.0) is relatively near MASOSA (at $\approx2.5''$, see also the right panel in Fig. $\ref{fig:FIR_zooms}$), but coincides with a galaxy with photometric-redshift $z=1.46\pm0.05$ in the \cite{Laigle2016} catalogue.
The continuum detections in the VR7 field are of lower significance. The detection with S/N=4.2 is shown in the left panel of Fig. $\ref{fig:foregrounds}$ and is co-located with a faint galaxy in the {\it HST} data. No emission-line is detected in our MUSE data (Matthee et al. in prep), suggesting that the galaxy is below $z<3$ (otherwise Ly$\alpha$ could have been detected) and above $z>1.5$ (otherwise [O{\sc ii}] could have been detected). The other continuum detection may be associated to a LAE at $z=3.062$, although it is offset by 0.25$''$ to the north-west.

\begin{table*}
\centering
\caption{Global properties of foreground objects. Continuum flux is measured using the primary beam corrected map, while the S/N is measured before the primary beam correction. }
\begin{tabular}{llrrp{9.8cm}} \hline
R.A. & Dec. &$f_{\nu, \rm 250 GHz}$ & S/N & Comment  \\ 
(J2000) & (J2000) & ($\mu$Jy beam$^{-1}$) & & \\ \hline
10:01:24.705 & +2:31:32.73 &  1212   & 56.0 & Dusty, highly obscured star forming galaxy with unknown photo$-z$. Detected in the {\it Spitzer}/IRAC [3.6] and [4.5] bands, but not detected in optical bands or our {\it HST}/WFC3 imaging data.\\
10:01:25.427 & +2:31:45.23 & 138 & 8.2 & In the centre of a relatively nearby face-on spiral galaxy at $z=0.131$ with identifier SDSS J100125.41+023145.2. \\
10:01:24.629& +2:31:45.62 & 64 & 5.0 & Galaxy with photo$-z=1.46\pm0.05$ and M$_{\rm star}=10^{10.37\pm0.08}$ M$_{\odot}$, ID 810182 in \cite{Laigle2016} $\approx2''$ away from MASOSA. \\ \hline

22:18:57.117 &  +0:08:03.79 & 55 & 4.2 & Detected in {\it HST}/WFC3 with F110W=24.4. Marginally detected in a combined $griz$ image. No emission-line detected in MUSE (Matthee et al. in prep). Likely photo-$z\approx1.5-2.8$. \\
22:18:57.143 &  +0:08:18.29 & 79 & 3.5 & Associated(?) with a LAE identified in MUSE $z_{Ly\alpha}=3.062$, offset 0.25$''$ to the north-west \\
\hline

\label{tab:foreground_properties}
\end{tabular}
\end{table*}

\setcounter{figure}{0} \renewcommand{\thefigure}{B.\arabic{figure}}

\section{Fitting rest-frame UV sizes}\label{app:size}
Here we provide more details on the fitting of the {\it HST}/WFC3 data, in particular on fitting VR7 as a single component or a two component model and the fitting results for MASOSA.

\subsection{Single component versus two component in VR7's {\it HST} data} 
\begin{figure*}
\centering
\includegraphics[width=18.2cm]{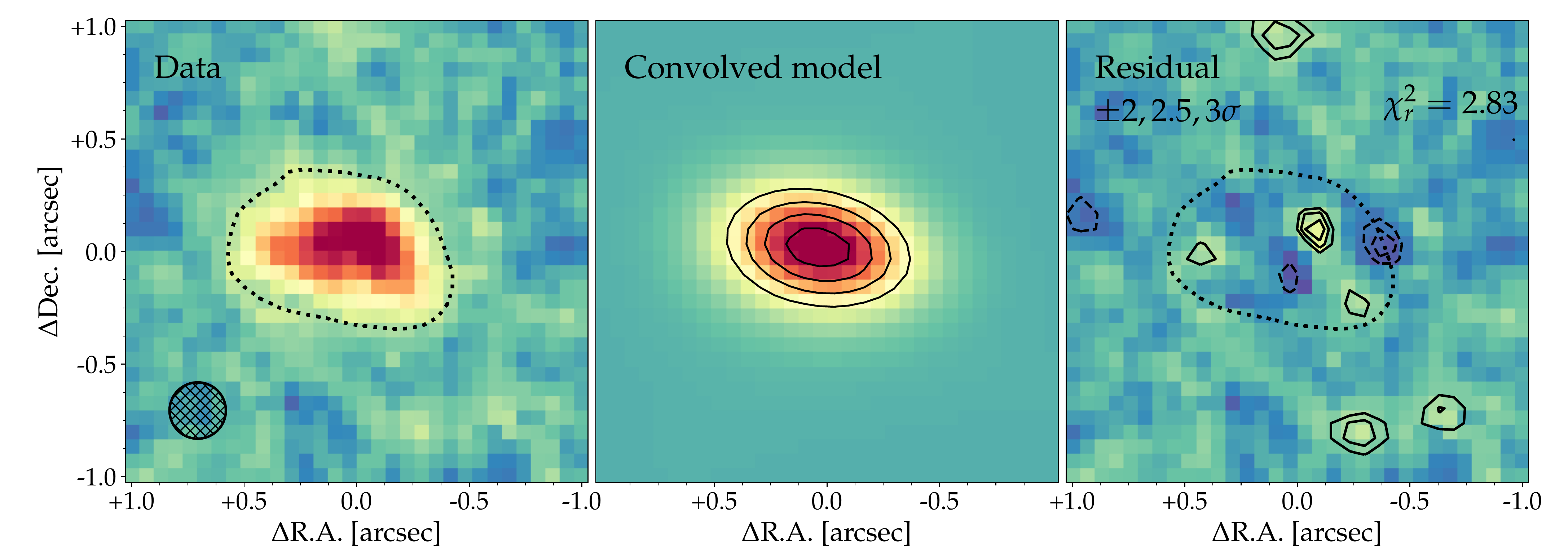}\\
\includegraphics[width=18.2cm]{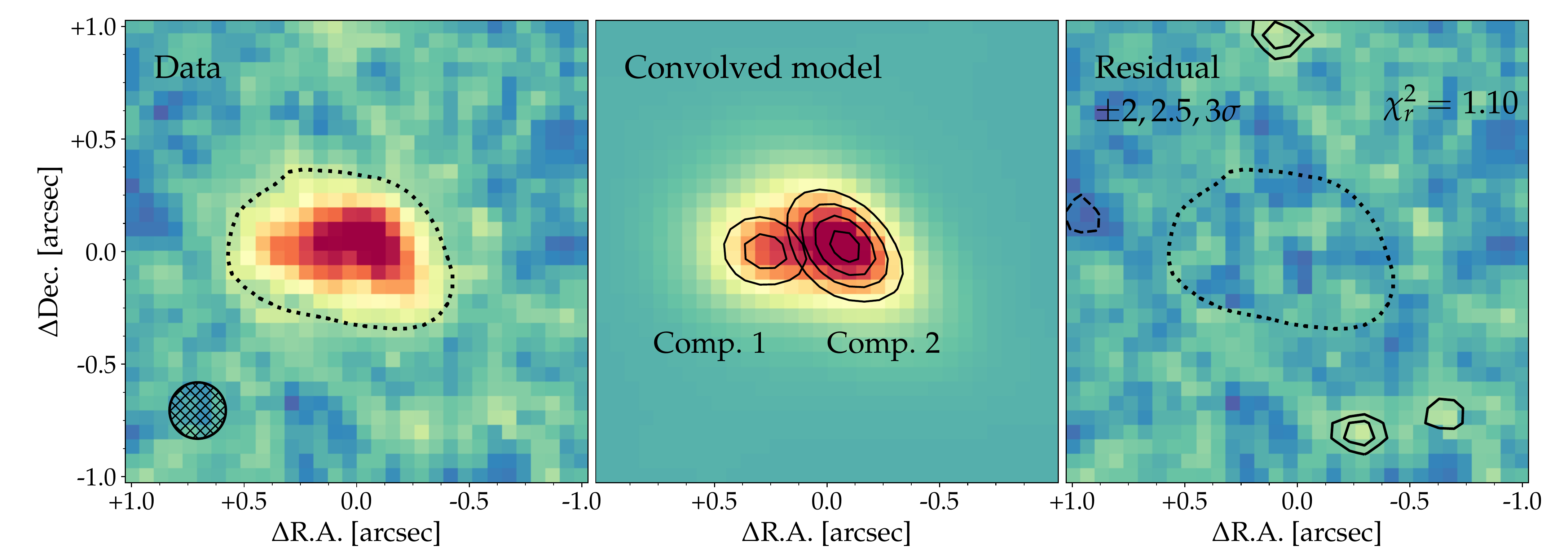}\\
\caption{Zoomed-in {\it HST}/WFC3 image of VR7 in the F160W filter. The left panel shows the data and the size of the PSF-FWHM. The middle panels show the models and the right panels show the residuals. The top row shows a single exponential light-profile, while the bottom row shows a combination of two exponential light-profiles. For reference, the dotted contour illustrate the outline of the galaxy. Reduced $\chi^2$ values are computed within the dotted contour-level. The rest-frame UV light-profile of VR7 is better described by two individual components. \label{fig:VR7_HST_single_vs_double}}
\end{figure*}

As described in the main text (\S $\ref{sec:UVsize}$), we find that the light profile of VR7 in the {\it HST}/WFC3 F160W data is better described by a combination of two exponential profiles than by a single exponential (or S\'ersic) profile. The result of the fitting using {\sc imfit} \citep{Erwin2015} is shown in Fig. $\ref{fig:VR7_HST_single_vs_double}$. The residual map of the single-component fit shows clear structure at the $\approx3\sigma$ level, while residuals of such strength are not seen in the more flexible two-component fit. We note that the F110W data (which morphology is possibly affected by Ly$\alpha$ emission) is also better described by a two-component fit, although the relative flux and sizes of the two component reverse (i.e. a larger and more luminous component in the east and a smaller fainter component south-west). While this implies a color gradient, we note that these (and particularly their uncertainties) are challenging to quantify without strong priors such as assuming that the relative sizes, position angles and ellipticities are the same and assuming a homogeneously distributed Ly$\alpha$ surface brightness. If we use the morphology of the individual components in the F160W data as a prior and assume Ly$\alpha$ emission is homogeneously distributed, we measure that the flux fraction in component 1 (Fig. $\ref{fig:VR7_HSTresolved}$) is $42\pm6$ \% and $36\pm7$ \% in the F110W and F160W data, respectively. This implies component 1 has a UV luminosity and color M$_{1500}=-21.41\pm0.17$ and $\beta=-2.04^{+0.89}_{-0.95}$, while component 2 has M$_{1500}=-21.77\pm0.12$ and $\beta=-1.14\pm0.62$. Without subtracting Ly$\alpha$ emission from the F110W flux density the UV slopes would be bluer, with $\Delta \beta \approx-0.3$.

\subsection{Single component in MASOSA} 
\begin{figure*}
\centering
\includegraphics[width=18.2cm]{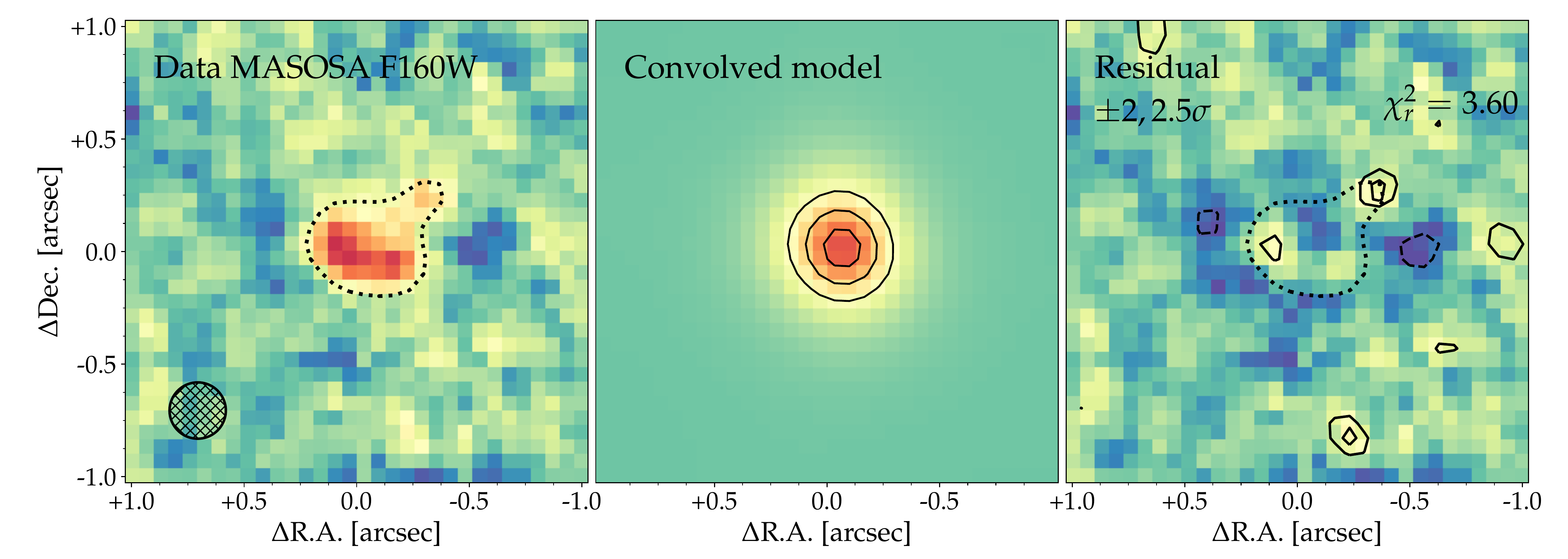}
\caption{As Fig. $\ref{fig:VR7_HST_single_vs_double}$, but showing the single component exponential fit to MASOSA. \label{fig:app_MASOSA}}
\end{figure*}

We show our best fit single exponential model to the F160W from MASOSA in Fig. $\ref{fig:app_MASOSA}$. Due to the limited S/N of the galaxy the data can not reliably distinguish between a single component or a combination of two components, although visually a hint of a second component can be seen in the north-west of the galaxy. Deeper {\it HST} data is required to confirm this component.

\section{Compilation and homogenisation of measurements from galaxies at $z\approx6-7$} \label{appendix:compilation}
\setcounter{figure}{0} \renewcommand{\thefigure}{C.\arabic{figure}}
\setcounter{table}{0} \renewcommand{\thetable}{C.\arabic{table}}

In order to compare our measurements to earlier observations of galaxies at $z\approx6-7$, we compile a sample of spectroscopically confirmed galaxies at $z\approx6-7$ for which the [C{\sc ii}] line and adjacent continuum has been observed with ALMA. This compilation includes LAEs \citep{Ota2014,Inoue2016,Matthee2017ALMA,Carniani2018Himiko}, UV-selected galaxies confirmed through Ly$\alpha$ emission \citep{Maiolino2015,Willott2015,Pentericci2016,Carniani2018,Hashimoto2018Dragons}, lensed UV-selected galaxies \citep{Knudsen2016,Bradac2017} and UV-selected galaxies confirmed through [C{\sc ii}] emission \citep{Smit2017}. We also include the sample of LBGs, LAEs and galaxies selected through their photometric redshift at $z\approx5$ from \cite{Capak2015} with updated measurements from \cite{Barisic2017}. For a proper comparison, we collect observed properties (i.e. UV, [C{\sc ii}] and Ly$\alpha$ luminosity, line-widths and IR flux densities) and self-consistently convert these to derived properties (i.e. star formation rates and IR luminosities). Table $\ref{tab:compilation}$ shows the observed properties of the galaxies at $z\approx6-7$, while Table $\ref{tab:derived}$ lists derived properties. The Ly$\alpha$ properties are listed in Table $\ref{tab:Lya}$.

We use published values of M$_{1500}$ and $\beta$ wherever possible, but for several objects we calculated M$_{1500}$ based on published photometry. When available, Ly$\alpha$ luminosities have been computed using published line-flux measurements. For galaxies in which no [C{\sc ii}] is detected, we measure the physically motivated upper limit similar to the procedure in \S $\ref{sec:MASOSAlimit}$. This means that we estimate the expected line-width based on the UV luminosity (typically 150-200 km s$^{-1}$) and convert published noise values to this width assuming rms$_{\rm [CII]} \propto \sqrt{\Delta v_{\rm channel}}$, where $\Delta v_{\rm channel}$ is the channel width over which the noise is measured. Measurements of/upper limits on the continuum flux at $\lambda_0=160\mu$m (typically $\lambda_{obs}=1.2$mm) have been converted to an FIR ($\lambda=42.5-122.5\mu$m) and IR ($\lambda=8-1000 \mu$m) luminosity following \cite{Ota2014}, after correcting the contribution from CMB heating following \cite{daCunha2013}. We assume a modified blackbody of the form:
B$_{\nu}=\frac{M_{\rm dust} \kappa}{D_L^2}  \frac{2 h \nu^3}{c^2 e^{\frac{h \nu}{k_B T_d} -1}} $, where $D_L$ is the luminosity distance, $k_B$ the Boltzmann constant, $T_d$ the dust temperature, $M_{\rm dust}$ the dust mass, $\kappa = \kappa_{850} (\frac{\nu}{\nu_{850}})^{\beta}$, where $\kappa_{850}$ is the dust opacity at 850 $\mu$m, which we assume is 0.077 m$^2$ kg$^{-1}$ similar to \cite{Ota2014} and $\beta$ is the grey-body power-law exponent, which we assume to be $\beta=1.5$ following \cite{Ota2014}. Then, the obscured SFR is calculated following \cite{Kennicutt1998}:  log$_{10}$(SFR$_{\rm IR}$/M$_{\odot}$ yr$^{-1}$) = log$_{10}$ ($L_{\rm IR}$/erg s$^{-1}$) - 43.48. M$_{1500}$ is  converted to an un-obscured SFR$_{\rm UV}$ following \cite{Kennicutt1998}: log$_{10}$(SFR$_{\rm UV}$/M$_{\odot}$ yr$^{-1}$) = -0.4 (M$_{1500}$+18.03). The SFRs can be converted to the calibrations in \cite{KennicuttEvans2012} that follow a Kroupa IMF by multiplying SFR$_{\rm UV}$ by a factor 0.63 and multiplying SFR$_{\rm IR}$ by a factor 0.86.

\begin{table*}
\caption{Observed properties of the galaxy compilation at $z\approx6-7$. Upper limits are at the 1$\sigma$ level and in case of [C{\sc ii}] based on the expected line-width from the UV luminosity (Fig. $\ref{fig:motivation_CIIlimits}$).}
\begin{tabular}{lccccccc}
ID & $z$ & M$_{1500}$ & $\beta$ & f$_{160 \mu m}$/$\mu$Jy & L$_{[\rm CII]}/10^8$ L$_{\odot}$ & $v_{\rm FWHM, [CII]}$/km s$^{-1}$ & Reference \\ \hline
A383-5.1       & 6.029 & $-19\pm0.3$ & $-2\pm0.7$ & $<11$ & $0.089\pm0.031$ & $100\pm23$ & K16           \\
NTTDF2313      & 6.07 & $-21.2$ & & $<18$ & $<0.21$ &  & C18        \\
WMH5           & 6.075 & $-22.8$ & & $218\pm41$ & $6.6\pm0.7$ & $251\pm15$ & W15             \\
BDF2203        & 6.12 & $-21.5$ & & $<23$ & $1.3\pm0.3$ & $150\pm50$ & C18              \\
CLM1           & 6.176 & $-22.7$ & & $44\pm26$ & $2.4\pm0.3$ & $162\pm23$ &  W15              \\
GOODS3203      & 6.27 & $-21.6$ & & $<41$ & $<0.59$ &  & C18             \\
COSMOS20521    & 6.36 & $-21.13$ & $-2.17$ & $<20$ & $<0.23$ &  & C18           \\
VR7            & 6.534 & $-22.4\pm0.05$ & $-1.4\pm0.3$ & $<10.6$ & $4.76\pm0.14$ & $203\pm40$ & This paper             \\
MASOSA         & 6.543 & $-20.9\pm0.1$ & $-1.1\pm0.7$ & $<9.2$ & $<0.22$ &  & This paper             \\
UDS4821        & 6.56 & $-21.3$ & & $<24$ & $<0.32$ &  & C18          \\
Himiko         & 6.595 & $-21.8\pm0.1$ & $-2\pm0.4$ & $<27$ & $1.2\pm0.2$ & $180\pm50$ & C18           \\
CR7            & 6.604 & $-22.2\pm0.1$ & $-2.2\pm0.4$ & $<7$ & $3.87\pm0.65$ & $259\pm24$ & M17           \\
COSMOS24108    & 6.629 & $-21.67$ & $-1.76\pm0.2$ & $<18$ & $1.1\pm0.15$ & $150\pm40$ & P16             \\
UDS16291       & 6.638 & $-20.97$ & $-2.4\pm0.4$ & $<20$ & $0.7\pm0.15$ & $50\pm15$ & P16       \\
NTTDF6345      & 6.701 & $-21.57$ & $-1.3\pm0.5$ & $<16$ & $1.8\pm0.3$ & $250\pm70$ & P16           \\
RXJ13471216    & 6.765 & $-19.1\pm0.4$ & $-2.5\pm0.7$ & $<15$ & $0.15\pm0.03$ & $75\pm25$ & B17             \\
COS-2987030247 & 6.807 & $-22.1\pm0.1$ & $-1.22\pm0.5$ & $<25$ & $3.6\pm0.5$ & $124\pm18$ & L17 \\
SDF-46975      & 6.844 & $-21.5$ & & $<19.2$ & $<0.27$ &  & M15         \\
COS-3018555981 & 6.854 & $-22\pm0.1$ & $-1.18\pm0.5$ & $<29$ & $4.7\pm0.5$ & $232\pm30$ & S17               \\
IOK-1          & 6.96 & $-21.4$ & $-2.07\pm0.26$ & $<21$ & $<0.25$ &  & O14                \\
BDF-521        & 7.008 & $-20.3$ & & $<17.4$ & $<0.22$ &  & M15           \\
BDF-3299       & 7.109 & $-20.9$ & & $<7.8$ & $0.49\pm0.1$ & $75\pm20$ &M15,C17 \\
COSMOS13679    & 7.145 & $-21.46$ & $-1.54\pm0.4$ & $<14$ & $0.7\pm0.15$ & $90\pm35$ & P16               \\
B14-65666      & 7.152 & $-22.3\pm0.05$ & $-1.85\pm0.5$ & $130\pm25$ & $13.3\pm1.3$ & $349\pm31$ & B18,H18  \\
SXDF-NB1006-2  & 7.212 & $-21.5$ & $-2.6$ & $<14$ & $<0.28$ &  & I16            \\ \hline
\end{tabular}

References: B17 \citep{Bradac2017}; B18 \citep{Bowler2018}; C17 \citep{Carniani2017}; C18 \citep{Carniani2018}; H18 \citep{Hashimoto2018Dragons}; I16 \citep{Inoue2016}; K16 \citep{Knudsen2016}; L17 \citep{Laporte2017}; M15 \citep{Maiolino2015}; M17 \citep{Matthee2017ALMA}; O14 \citep{Ota2014};  P16 \citep{Pentericci2016}; S17\citep{Smit2017}; W15 \citep{Willott2015}   

\label{tab:compilation}
\end{table*}

\begin{table}
\caption{Derived properties of the galaxy compilation at $z\approx6-7$, based on the measurements listed in Table $\ref{tab:compilation}$. SFRs are based on the calibrations presented in \cite{Kennicutt1998}. Infrared luminosities have been corrected for CMB-heating following \cite{daCunha2013}. Upper limits are at the 1$\sigma$ level. }
\centering
\begin{tabular}{lcccc}
ID & SFR$_{\rm UV}$/M$_{\odot}$ yr$^{-1}$ & L$_{\rm IR, Td=35K}/10^{10}$ L$_{\odot}$ & L$_{\rm IR, Td=45K}/10^{10}$ L$_{\odot}$ & SFR$_{\rm IR, Td=45K}$/M$_{\odot}$ yr$^{-1}$   \\ \hline
A383-5.1       & $2.4\pm0.7$ & $<1.2$ & $<2.4$ & $<4.2$ \\
NTTDF2313      & $18.6$ & $<2.0$ & $<4.0$ & $<6.9$ \\
WMH5           & $81.3$ & $24.0\pm4.5$ & $48.2\pm8.9$ & $83.9\pm15.6$ \\
BDF2203        & $24.5$ & $<2.6$ & $<5.1$ & $<9.0$ \\
CLM1           & $74.1$ & $5.0\pm2.7$ & $10.0\pm5.6$ & $17.4\pm9.8$ \\
GOODS3203      & $26.9$ & $<4.8$ & $<9.6$ & $<16.6$ \\
COSMOS20521    & $17.5$ & $<2.4$ & $<4.8$ & $<8.3$ \\
VR7            & $54.4\pm2.5$ & $<1.3$ & $<2.7$ & $<4.6$ \\
MASOSA         & $14.5\pm2.1$ & $<1.2$ & $<2.3$ & $<4.0$ \\
UDS4821        & $20.4$ & $<3.1$ & $<6.1$ & $<10.5$ \\
Himiko         & $32.7\pm3.1$ & $<3.5$ & $<6.9$ & $<12.0$ \\
CR7            & $47.0\pm4.1$ & $<0.9$ & $<1.8$ & $<3.1$ \\
COSMOS24108    & $28.7$ & $<2.3$ & $<4.6$ & $<8.1$ \\
UDS16291       & $15.1$ & $<2.6$ & $<5.2$ & $<9.0$ \\
NTTDF6345      & $26.2$ & $<2.1$ & $<4.2$ & $<7.3$ \\
RXJ13471216    & $2.7\pm1.0$ & $<2.0$ & $<4.0$ & $<7.0$ \\
COS-2987030247 & $43.0\pm3.7$ & $<3.4$ & $<6.7$ & $<11.7$ \\
SDF-46975      & $24.5$ & $<2.7$ & $<5.2$ & $<9.1$ \\
COS-3018555981 & $39.1\pm3.5$ & $<4.0$ & $<7.9$ & $<13.8$ \\
IOK-1          & $22.4$ & $<3.0$ & $<5.9$ & $<10.3$ \\
BDF-521        & $8.1$ & $<2.5$ & $<4.9$ & $<8.6$ \\
BDF-3299       & $14.1$ & $<1.2$ & $<2.3$ & $<4.0$ \\
COSMOS13679    & $23.7$ & $<2.1$ & $<4.1$ & $<7.2$ \\
B14-65666      & $51.2\pm2.5$ & $19.8\pm3.9$ & $38.3\pm7.6$ & $66.7\pm13.3$ \\
SXDF-NB1006-2  & $24.5$ & $<2.2$ & $<4.2$ & $<7.3$ \\ \hline

\end{tabular}
\label{tab:derived}

\end{table}

\begin{table}
\caption{Lyman-$\alpha$ properties of the galaxy compilation at $z\approx6-7$.}
\centering
\begin{tabular}{lccc}
ID & L$_{\rm Ly\alpha}/10^{42}$ erg s$^{-1}$ & EW$_{\rm 0, Ly\alpha}$/{\AA} & $\Delta v_{\rm Ly\alpha}$/km s$^{-1}$  \\ \hline
A383-5.1       & 7 & 138 & 90$\pm$20 \\ 
NTTDF2313      & & &  \\ 
WMH5           & & 13 & 265$\pm$52 \\ 
BDF2203        & & 3 &  \\ 
CLM1           & & 50 & 430$\pm$69 \\ 
GOODS3203      & & 5 &  \\ 
COSMOS20521    & 3 & 10 &  \\ 
VR7            & 24 & 34 & 217$\pm$25 \\ 
MASOSA         & 24 & 145 &  \\ 
UDS4821        & & 48 &  \\ 
Himiko         & 43 & 65 & 145$\pm$15 \\ 
CR7            & 83 & 211 & 167$\pm$27 \\ 
COSMOS24108    & 10 & 27 & 240$\pm$0 \\ 
UDS16291       & 0.8 & 6 & 110$\pm$0 \\ 
NTTDF6345      & 4 & 15 & 110$\pm$0 \\ 
RXJ13471216    & 6 & 26 & 20$\pm$90 \\ 
COS-2987030247 & 5 & 16 & 326$\pm$30 \\ 
SDF-46975      & 15 & 43 &  \\ 
COS-3018555981 & & &  \\ 
IOK-1          & & 42 &  \\ 
BDF-521        & & 64 &  \\ 
BDF-3299       & 7 & 50 & 64$\pm$10 \\ 
COSMOS13679    & 13 & 28 & 135$\pm$0 \\ 
B14-65666      & 2.6 & 4 & 772$\pm$45 \\ 
SXDF-NB1006-2  & & 33 &  \\ \hline

\end{tabular}
\label{tab:Lya}

\end{table}



\bibliography{almavr7.bib}



\end{document}